\documentclass[11pt,a4paper]{article} %11pt?
\usepackage{jheppub,multirow}
\usepackage{amsthm}
\usepackage{xcolor}
\usepackage{float}
\usepackage{url}
\usepackage{microtype}
\usepackage{subcaption}
\usepackage{empheq}
\usepackage{booktabs}
\usepackage[makeroom]{cancel}
\usepackage{dsfont} %Nice \1
\usepackage{ytableau} %Make Young tableaux
\usepackage{setspace}%Set line spacing to 1.15
\usepackage[skins]{tcolorbox}

\newcommand{\beq}{\begin{equation}}
\newcommand{\eeq}{\end{equation}}

\newcommand{\bR}{\mathbb{R}}

\newcommand{\fg}{\mathfrak{g}}

\newcommand{\fq}{\mathfrak{q}}

\newcommand{\cN}{\mathcal{N}}

\newcommand{\cW}{\mathcal{W}}

\newcommand{\cT}{\mathcal{T}}

\newcommand{\E}{{\epsilon}}

\newcommand{\tabref}[1]{Table~\ref{#1}}
\newcommand{\sh}{\mbox{sh}}
\newcommand{\ch}{\mbox{ch}}
\def\ie{\begin{equation}\begin{aligned}}
\def\fe{\end{aligned}\end{equation}}
\def\xTB{$\times$} % inside table
\newcommand*\xbar[1]{%
	\hbox{%
		\vbox{%
			\hrule height 0.5pt % The actual bar
			\kern0.3ex%         % Distance between bar and symbol
			\hbox{%
				\kern-0.1em%      % Shortening on the left side
				\ensuremath{#1}%
				\kern-0.05em%      % Shortening on the right side
			}%
		}
	}
}

\newtheoremstyle{fullit}
{\topsep}      % ABOVESPACE
{\topsep}      % BELOWSPACE
{\normalfont}  % BODYFONT % \itshape
{0pt}          % INDENT (empty value is the same as 0pt)
{\itshape}     % HEADFONT
{.\ }          % HEADPUNCT
{0pt}          % HEADSPACE. `plain` default: {5pt plus 1pt minus 1pt}
{\thmname{#1} \thmnumber{#2}}             % CUSTOM-HEAD-SPEC

\theoremstyle{definition}

\theoremstyle{fullit}

\DeclareCaptionSubType*[alph]{figure}
\tikzset{cross/.style={cross out, draw=black, minimum size=2*(#1-\pgflinewidth), inner sep=0pt, outer sep=0pt},
	cross/.default={1pt}}
\definecolor{urlssc}{HTML}{161680}
\definecolor{linksc}{HTML}{80164B}
\definecolor{citesc}{HTML}{16804B}

\title{On the Quantization of Seiberg-Witten Geometry}
\author{Nathan Haouzi$^a$,}
\author{Jihwan Oh$^b$}
\affiliation{$^a$Simons Center for Geometry and Physics, State University of New York,
	Stony Brook, NY 11794}
\affiliation{$^b$Department of Physics,
	University of California, Berkeley, CA 94720, U.S.A.}
\emailAdd{nhaouzi@scgp.stonybrook.edu, jihwanoh@berkeley.edu}

\abstract{We propose a double quantization of four-dimensional $\cN=2$ Seiberg-Witten geometry, for all classical gauge groups and a wide variety of matter content. This can be understood as a set of certain non-perturbative Schwinger-Dyson identities, following the program initiated by Nekrasov \cite{Nekrasov:2015wsu}. The construction relies on the computation of the instanton partition function of the gauge theory on the so-called $\Omega$-background on $\mathbb{R}^4$, in the presence of half-BPS codimension 4 defects. The two quantization parameters are identified as the two parameters of this background. The Seiberg-Witten curve of each theory is recovered in the flat space limit. Whenever possible, we motivate our construction from type IIA string theory.}
\begin{document}
	\maketitle
	\setlength{\parindent}{0pt}
	\clearpage
	
	%%%%%%%%%%%%%%%%%%%%%%%%%%%%%%%%%%%%%%%%%%%%%%%
	%%%%%%%%%%%%%%%%%%%%%%%%%%%%%%%%%%%%%%%%%%%%%%%
	%%%%%%%%%%%%%%%%%%%%%%%%%%%%%%%%%%%%%%%%%%%%%%%
	%%%%%%%%%%%%%%%%%%%%%%%%%%%%%%%%%%%%%%%%%%%%%%%
	%%%%%%%%%%%%%%%%%%%%%%%%%%%%%%%%%%%%%%%%%%%%%%%
	%%%%%%%%%%%%%%%%%%%%%%%%%%%%%%%%%%%%%%%%%%%%%%%
	%%%%%%%%%%%%%%%%%%%%%%%%%%%%%%%%%%%%%%%%%%%%%%%
	%%%%%%%%%%%%%%%%%%%%%%%%%%%%%%%%%%%%%%%%%%%%%%%

	\newpage
	
	\section{Introduction}
	
	$\cN=2$ supersymmetric Yang-Mills holds a special place in the realm of quantum field theories. Non-perturbative effects play a crucial role in defining the effective action, yet these effects are tractable. In particular, the low energy physics can be determined with the help of a holomorphic function known as the prepotential. Indeed, this function encodes both the perturbative  one loop physics, as well as the  non-perturbative instanton corrections to the effective action. In the seminal work \cite{Seiberg:1994rs}, Seiberg and Witten put forward a formidable proposal to solve the pure $SU(2)$ gauge theory. In short, they showed that the prepotential can be determined via the symplectic geometry of an algebraic curve, referred to nowadays as the Seiberg-Witten curve. In this way, a large class of examples was solved by means of a hyperelliptic curve. As the gauge group and matter content is varied, it is also not uncommon to encounter non-hyperelliptic curves, of finite or infinite order.\\
	
	The present paper deals with the quantization of such geometries, a notion that comes about in the following way: 
	As an illustrative example, let $T^{4d}$ be the pure four-dimensional  $\cN=2$  $SU(N)$ gauge theory. The Seiberg-Witten curve can be written as\footnote{The choice of the variable name $M$ is not innocent here;  we will later identify it as the mass of a one-dimensional fermion that couples to the bulk gauge theory.}
	\begin{equation}
	\label{A1SW}
	y(M) + \frac{\Lambda^{2 N}}{y(M)} = \cT(M; \{e_i\}) \; ,
	\end{equation}
	where $(M, y)\in \mathbb{C}\times \mathbb{C}^*$, $\Lambda$ is the QCD dynamical scale, and $\cT(M; \{e_i\})$ is a polynomial of degree $N$ in $M$ whose roots define the vacuum of the theory, and depend implicitly on the  Coulomb vevs of $T^{4d}$.

Nekrasov and Okounkov developed a powerful method to derive such a curve, relying on the direct evaluation of $T^{4d}$ 's  prepotential   \cite{Nekrasov:2002qd, Nekrasov:2003rj}. Crucially, the construction relies on the fact that the four-dimensional spacetime admits a two-parameter deformation known as the $\Omega$-background, which can be thought of as a weak $\cN=2$ supergravity background. We denote it as $\mathbb{C}^2_{\epsilon_1, \epsilon_2}$, where $\epsilon_i$ rotates the $i$-th $\mathbb{C}$-line, for $i=1, 2$. In this background, the prepotential, plus an infinite number of corrections which vanish in the flat space limit, are equal to the logarithm of a sum of instanton integrals. By exploiting equivariant localization, the computation simplifies drastically to become a mere multiple contour integral. It can then be shown that the Seiberg-Witten curve is nothing but the saddle point equation of such an instanton integral, not unlike the large $N$ analysis in matrix models. The saddle point analysis can be carried out by simply taking the limit $\epsilon_1, \epsilon_2 \rightarrow 0$. In this picture, the complex variable $y$ which appears on the left-hand side of the Seiberg-Witten curve \eqref{A1SW} is realized as the vev of an operator,
	\beq
	y(M) \equiv \left\langle Y(M)\right\rangle \; .
	\eeq
	This $Y$-operator is a generating function of the chiral ring of $T^{4d}$. So-called ``i-Weyl" reflections \cite{Nekrasov:2012xe} are defined to act on the operator $Y$ as $Y \rightarrow\Lambda^{2N}\, Y^{-1}$. Starting with a ``highest weight"  $Y$ and acting on it with such a reflection, we generate a representation whose character is the left-hand side of \eqref{A1SW}\footnote{The general statement is as follows: consider a (conformal or asymptotically free) four-dimensional $\cN=2$ quiver gauge theory, where the gauge group is a product of special unitary groups, resulting in a quiver whose graph $\Gamma$ is a Dynkin diagram of type $\fg= A, D, E$. There exists a relation between the Seiberg-Witten geometry of the theory and the representation theory of the quiver itself. In particular, one can always express the Seiberg-Witten curve of the gauge theory in terms of the characters of the fundamental representations of $\fg$. In the case at hand, $T^{4d}$ consists of a single gauge node, so the associated algebra is simply $\fg= A_1$, and the left-hand side of \eqref{A1SW} is the character of the fundamental (or spin 1/2) representation.}.\\

	In fact, sending both $\epsilon_1$ and $\epsilon_2$ to zero is overkill to carry out a saddle point analysis, as only one parameter really needs vanish, say $\epsilon_2\rightarrow 0$, while the other parameter $\epsilon_1$ can be kept arbitrary; this is sometimes referred to as the Nekrasov-Shatashvili limit. Therefore, a natural question to ask is what type of saddle point equations one obtains with $\epsilon_1$ still present. This problem was addressed in \cite{Nekrasov:2013xda}. The authors showed that the saddle point equations can be understood as a $q$-deformation of the Seiberg-Witten curve of $T^{4d}$, where the role of Planck's constant $\log(q)\equiv \hbar$ is played by the surviving  $\Omega$-background parameter $\epsilon_1\equiv \hbar$ (see also the work \cite{Bullimore:2014awa}). The saddle point equation no longer defines an algebraic curve, but is now instead a difference equation\footnote{Physicists sometimes short-handedly refer to it as a quantum curve. The terminology can be confusing as the usual algebraic Seiberg-Witten curve already contains the information about the ``quantum" corrections to the Coulomb branch, but from a geometric perspective, the Seiberg-Witten curve is ``classical".}.
	In our $SU(N)$ example, the saddle-point equation reads:
	\begin{equation}
	\label{A1examplecurveq}
	 \left\langle Y(M)\right\rangle + \Lambda^{2 N} \left\langle \frac{1}{Y(M+\hbar)}\right\rangle = \cT(M; \{e_i^{\hbar}\}) \; .
	\end{equation}
	In this context, $Y$ and $M$ should be understood as two operators satisfying an uncertainty principle $[M, Y]=i\, \hbar$, and the notation $\frac{1}{Y(M)}$ is a shorthand for the inverse of the operator $Y(M)$.
	The function $\cT(M; \{e_i^{\hbar}\})$ is still a polynomial of degree $N$ in the variable $M$, whose roots  $\{e_i^{\hbar}\}$ define the vacuum of the theory, depending explicitly on $\hbar$. By analogy with the previous construction of the curve \eqref{A1SW}, the left-hand side of \eqref{A1examplecurveq} is sometimes called the fundamental $q$-character of  $A_1$. More precisely, it is a (twisted) Yangian $q$-character of the fundamental representation of $Y(A_1)$. This character had already been constructed back in the 90's \cite{Frenkel:qch}, in the context of two-dimensional conformal field theories. The fact that such an object makes an appearance in a four-dimensional gauge theory context is a striking example of what is sometimes called the BPS/CFT correspondence.\\
	
	It turns out to be possible to obtain a similar equation in the full $\Omega$-background, with both $\epsilon_1$ and $\epsilon_2$ present. In particular, a saddle point analysis of instanton integrals is no longer possible, since the two parameters are now kept arbitrary. This is the program carried out in \cite{Nekrasov:2015wsu}. Namely, one first defines a half-BPS point-defect in $T^{4d}$ and an associated operator $Y(M)$, with $M\in\mathbb{C}$ a coordinate on an auxiliary complex line. Second, one computes the instanton partition function of $T^{4d}$ on the $\Omega$-background, in the presence of the defect.
	
	This defect partition function can be expressed in two ways:
	First, it is a sum of $Y$-operator correlators:
	\begin{equation}
	\label{A1examplecurveqqleft}
	Z(M) = \left\langle Y(M)\right\rangle + \Lambda^{2 N}\left\langle\frac{1}{Y(M+\epsilon_1+\epsilon_2)}\right\rangle \; .
	\end{equation}
	Second, if expanded in the defect fugacity $M$, the (suitably normalized) partition function turns out to be a finite polyonomial of degree $N$ in the variable $M$:
	\begin{equation}
	\label{A1examplecurveqqright}
	Z(M) = \cT(M; e_i^{\epsilon_1, \epsilon_2}) \; .
	\end{equation}
	The roots $\{e_i^{\epsilon_1, \epsilon_2}\}$ of the polynomial once again define the vacuum of the theory, this time in the full $\Omega$-background.\\
	
	Putting \eqref{A1examplecurveqqleft} and \eqref{A1examplecurveqqright} together, we obtain:
	\begin{equation}
	\label{A1examplecurveqq}
	\left\langle Y(M)\right\rangle + \Lambda^{2 N}\left\langle\frac{1}{Y(M+\epsilon_1+\epsilon_2)}\right\rangle  = \cT(M; e_i^{\epsilon_1, \epsilon_2}) \; .
	\end{equation}
	By analogy with the previous constructions, the left-hand side of this equation is sometimes called the fundamental $qq$-character of  $A_1$. More precisely, it is a deformation (by $\epsilon_2$) of the Yangian $q$-character of the fundamental representation of $Y(A_1)$, with $\epsilon_1$ playing the role of the first ``$q$" here. This deformed character had once again already appeared in the 90's  \cite{Shiraishi:1995rp,Awata:1995zk,Frenkel:1998}, in the study of two-dimensional conformal field theories.

	The difference equation \eqref{A1examplecurveq} is recovered in the limit $\epsilon_2 \rightarrow 0$, and  the $SU(N)$ Seiberg-Witten curve \eqref{A1SW} is further recovered in the limit $\epsilon_1, \epsilon_2 \rightarrow 0$.\\
	
	Though it is a priori far from obvious, the above equation has a beautiful interpretation as a non-perturbative Schwinger-Dyson identity for the theory $T^{4d}$ \cite{Nekrasov:2015wsu}. Roughly, the idea is as follows: Given a correlator defined by a path integral in quantum field theory, the Schwinger-Dyson equations can be understood as constraints that must be satisfied by such a correlator. This comes about from demanding that the path integral remain invariant under a slight shift of the  contour (provided that the measure is left invariant by such a shift). In particular, we could ask about a contour modification that takes us from a given topological sector of $T^{4d}$ to another distinct topological sector, related to the first by a large gauge transformation. In other words, what symmetries of $T^{4d}$ are made manifest when we change the instanton number? The above $Y$-operator is a natural observable to answer this question: as a codimension 4 defect operator in $T^{4d}$, it can mediate the change in instanton number of the theory. The equation \eqref{A1examplecurveqq} can then be understood as a regularity condition on the correlator  $\left\langle Y(M)\right\rangle$. Namely, the correlator has poles in the fugacity $M$, but the Schwinger-Dyson equation tells us that there exists a precise combination of $Y$-operator vevs (the left-hand side of the equation) which is pole-free in $M$.

In this fashion, non-perturbative Schwinger-Dyson identities can be derived for $SU(N)$ gauge theories with or without fundamental and adjoint matter, and for quiver gauge theories made up of such special unitary groups\footnote{$\fg=ADE$-type  quivers made up of special unitary groups were considered in the original paper \cite{Nekrasov:2015wsu}; a formal construction for ``fractional" quivers, which have an arbitrary lacing number, was given in \cite{Kimura:2017hez} using a five-dimensional setup, and in \cite{Haouzi:2019jzk} using a one-dimensional modified ADHM quantum mechanics.}. These identities are what we will refer to as ``quantized Seiberg-Witten geometry" in this paper.\\

A natural question is whether it is possible to extend this analysis to $SO(N)$ and $Sp(N)$ gauge groups, as well as  $SU(N)$ gauge theories with different matter content, such as symmetric and antisymmetric matter. The Seiberg-Witten geometry for these models has been known since the 90's (see \cite{Ennes:1999fb} for a summary). However, no attempt has been made to quantize it. 
In this paper, we address this point, by constructing non-perturbative Schwinger-Dyson identities for all the above models. Whenever possible, we provide a string theory derivation of our results, based on a type IIA brane realization, possibly with orientifolds. For example, the quantized geometry of the $SO(2N)$ and $Sp(N)$ theories will be defined using a D4/D$4'$/O8 brane system. In some cases, such as the $SO(2N+1)$ theory, our construction will rely on field theoretic arguments only, with no underlying brane setup.\\

It will turn out that all the non-perturbative Schwinger-Dyson identities we derive take the form:
\begin{equation}
\label{generalexamplecurveqq}
Z(\left\langle Y(M)\right\rangle) = \cT(M, \{e^{\E_1, \E_2}_i\}) \; .
\end{equation}
Above, $Z(\left\langle Y(M)\right\rangle)= \left\langle Y(M)\right\rangle + \ldots$ will be an infinite Laurent series in the point defect operator vev $\left\langle Y(M)\right\rangle$, organized as an instanton expansion. The series is finite only in the case of a pure $SU(N)$ gauge theory, that is the left-hand side of \eqref{A1examplecurveqq}, or in the presence of fundamental matter.
	
We give an explicit algorithm to write this Laurent series to arbitrary high order in the instanton counting parameter. 
We then conjecture the highly nontrivial fact that the right-hand side of the equation, $\cT(M, \{e^{\E_1, \E_2}_i\})$, is a \emph{finite} polyonimal in $M$. Turning off the $\Omega$-background, we further argue that \eqref{generalexamplecurveqq} becomes the known Seiberg-Witten curve of the corresponding gauge theory.\\

The paper is organized as follows: in Section 2, we describe the general method to construct non-perturbative Schwinger-Dyson identities for gauge theories with a classical gauge group. The theories will be five-dimensional on a circle, and four-dimensional results will be recovered when shrinking the circle size to zero. We also describe how to extract the Seiberg-Witten geometry of the models. In section 3, we test our methods on a wide variety of examples, with a type IIA string theory picture whenever we can provide one. In section 4, we comment on whether our construction of the quantized geometry is expected to be unique.

\section{Constructing Non-Perturbative Schwinger-Dyson Identities}

Though our discussion so far has been purely phrased in a four-dimensional context, it will be worthwhile for us to start our analysis in five dimensions.
We claim that the construction of quantized Seiberg-Witten geometry is intimately related to the following problem: consider 5d SYM with gauge group $G$, defined on the manifold $S^1 \times \mathbb{R}^4$. How does a supersymmetric Wilson loop wrapping the circle $S^1$ interact with the instantons wrapping that same $S^1$ ? Our goal in this section is to make the connection between these two ideas explicit.

\subsection{The 5d Gauge theory and Instantons}

Consider 5d maximal SYM theory, with classical gauge group $G=SU(N), SO(N), Sp(N)$, defined on $S^1(R) \times \mathbb{R}^4$, where $R$ is the radius of the circle $S^1(R)$. The theory has $SO(1, 4)$ Lorentz symmetry and $SO(5)$ $R$-symmetry.
We give a non-zero vev to a vector multiplet scalar, $\langle \Phi \rangle\neq 0$. This forces the theory to go on the Coulomb branch, where the gauge symmetry $G$ breaks to a maximal abelian subgroup. Correspondingly, the $SO(5)$ $R$-symmetry is broken to $SO(4)_R=SU(2)^R_+\times SU(2)^R_-$.\\
	
In this paper, we will be interested in counting instantons, which  are solutions of the self-dual Yang-Mills equations on $\mathbb{R}^4$. In five dimensions, these  instantons are massive BPS particles wrapping $S^1(R)$.  As such, they preserve the little group of $SO(1, 4)$, which we write as $SO(4)_L=SU(2)_+\times SU(2)_-$.\\
	
	Let us then denote the doublet indices of $SU(2)_+$, $SU(2)_-$, $SU(2)^R_+$, $SU(2)^R_-$, by $\alpha, \dot{\alpha}, a, a'$, respectively. The 5d maximal SYM theory has 16 supercharges, which we  write as $Q_{\alpha a}, Q_{\dot{\alpha} a}, Q_{\alpha a'}$ and $Q_{\dot{\alpha} a'}$. The self-dual instantons will preserve $Q_{\dot{\alpha} a}$ and $Q_{\dot{\alpha} a'}$, and carry a $U(1)$  topological charge $k=\frac{1}{8\pi^2}\int_{\mathbb{R}^4}\mbox{Tr}(F\wedge F)$. The mass of such an instanton is proportional to $8 \pi^2/g_{5d}^2$, where $g_{5d}$ is the 5d gauge coupling. The other type of 1/2-BPS massive elementary particle is the W-boson. It preserves the supercharges $Q_{\alpha a}$ and $Q_{\dot{\alpha} a'}$, and has a mass proportional to $\langle \Phi \rangle$.\\

We now want to ask what happens when we introduce a Wilson loop in the 5d theory.

	\subsection{Wilson Loops and Instantons}

Recall that a Wilson loop is formulated as the trace of a holonomy matrix, where a quark is parallel-transported along a closed curve in spacetime, and the trace is evaluated in some irreducible representation of the gauge group $G$. The vev of such a loop is the phase shift of the quark wavefunction.

	Going back to the 5d maximal SYM theory, we introduce a supersymmetric Wilson loop wrapping the $S^1(R)$ and sitting at the origin of $\mathbb{R}^4$. This can be done with the help of a one-dimensional fermion field $\chi$, transforming in the fundamental representation of $G$ and in the fundamental representation of another background gauge group $G'$ (we will refer to it as the defect gauge group), coupled to the 5d gauge field in the bulk as \cite{Gomis:2006sb}
	\begin{equation}
	\label{1dfermion}
	S^{5d/1d}=\int dt\; \chi_{i,\rho}^\dagger\, \left( \delta_{\rho\sigma}(\delta_{ij} \, \partial_t - i\,A^{[5d]}_{t, ij}   - \Phi^{[5d]}_{ij})   + \delta_{ij}\, \widetilde{A}_{t,\rho \sigma} \right)\, \chi_{j,\sigma} \; .
	\end{equation}
	Above, $A^{[5d]}_t$ and $\Phi^{[5d]}$ are the pullback of the 5d gauge field and the adjoint scalar of the vector multiplet, respectively.  $\widetilde{A}_{t}$ is the (nondynamical) gauge field the 1d fermions couple to. $i$ and $j$ are indices for the fundamental representation of $G$, while $\rho$ and $\sigma$ are indices for the fundamental representation of $G'$. The variable $t$ is periodic, with period $R/(2\pi)$. The eigenvalues $M_\rho$ of the background gauge field $\widetilde{A}_{t}$ are (large) masses for the fermions. Those parameters set the energy scale for the excitation of the fermions. Such a loop is indeed 1/2-BPS, as it preserves the supercharges $Q_{\alpha a}$ and $Q_{\dot{\alpha} a'}$.

	Then, evaluating the path integral of the coupled 5d/1d system amounts to computing
	\beq\label{5d1dpathintegral}
	Z(M) =\int D\psi D\chi \, e^{i(S^{5d}[\psi]+ S^{5d/1d}[\psi, \chi, M])} \; .
	\eeq
	Here, $\psi$ denotes collectively all the fields of the bulk 5d theory, written as $S^{5d}$, while $S^{5d/1d}$ denotes the coupling  term \eqref{1dfermion}.\\

Since instantons are particles in 5 dimensions, counting them amounts to computing the partition function of their quantum mechanics, which is essentially a Witten index. Crucially, one needs to treat carefully the contribution of coincident zero-size instantons, as the moduli space is singular there.  In general, the so-called ADHM \cite{Atiyah:1978ri} construction is a powerful way to resolve such singularities. In our case, the instantons are also coincident with the Wilson loop, so regularizing their contribution must be done with extra care. Naively, one could try to simply localize the Wilson loop at ADHM solutions in the absence of a loop\footnote{For recent work following this approach, see for instance \cite{Gaiotto:2015una}.}, but that is not the way to proceed here. Instead, one should generalize the ADHM construction altogether and study how the presence of the loop modifies the instanton background from the onset. The instantons are then described by a $\cN=(0,4)$ gauged quantum mechanics\footnote{In this paper, when we talk about $\cN=(0,2)$ or $\cN=(0,4)$ supersymmetry in the context of a quantum mechanics, what we really mean is the reduction of two-dimensional $\cN=(0,2)$ or $\cN=(0,4)$ supersymmetry to one dimension.}, where the preserved supercharges are $Q_{\dot{\alpha} a'}$.\\

	\subsection{The Partition Function as a Witten Index}

	Let $T^{5d}$ be a  5d $\cN=1$ gauge theory with classical gauge group $G$ and flavor symmetry group $K$,  on the manifold  $S^1(R) \times \mathbb{R}^4$. In the rest of this paper, the 5d bare Chern-Simons term (when it exists) is set to zero. If we view $\mathbb{R}^4$ as $\mathbb{C}\times \mathbb{C}$, then we can denote the coordinate on the first $\mathbb{C}$-line as $z_1$, and the coordinate on the second $\mathbb{C}$-line as $z_2$. We introduce an $\Omega$-background by viewing the 5d spacetime as a $\mathbb{C}\times\mathbb{C}$ bundle over $S^1(R)$, where as we go around the circle,  we make the identification
	\begin{align}\label{omega}
	(z_1, z_2) \sim  (z_1\, e^{R \epsilon_1}, z_2\, e^{R \epsilon_2})\; ,
	\end{align}
	with $\epsilon_1$ and $\epsilon_2$ real.
	We write the partition function of $T^{5d}$ with Wilson loop as the Witten index of the following $\cN=(0,2)$ ADHM gauged quantum mechanics:
	\ie
	\label{index}
	Z=\text{Tr}_{{\cal H}_{\text{\tiny QM}}}\left[(-1)^F\, \widetilde{\fq}^k\, e^{-R\{Q^\dagger,Q\}}\, e^{2R\E_+ (J_+-J^R_+)}\,e^{2R\E_- J_-}\,e^{R\sum_{i} a_i\Pi_i}\,e^{R\sum_{\rho} M_{\rho}\Lambda_{\rho}}\,e^{R\sum_{d} m_d L_d}\right],
	\fe
	${\cal{H}}_{QM}$ is the Hilbert space of the five-dimensional field theory on $\mathbb{R}^4$. The trace is taken over all BPS states of the ADHM quantum mechanics. These are the BPS states annihilated by the supercharges $Q\equiv Q_{1\dot{1}}$ and $Q^\dagger\equiv Q_{2\dot{2}}$, with Hamiltonian $\{Q^\dagger,Q\}$; it therefore counts states in $Q-$cohomology.\\
	
	$F$ is the fermion number. $k$ is the topological $U(1)$ charge, conjugate to the instanton counting fugacity  $\widetilde{\fq}\equiv \exp\left(-8\pi^2 R/g^2_{5d}\right)$, with $R$ the radius of $S^1$. We have also defined
	$J_+$, $J_-$, and $J^R_+$ as the Cartan generators of $SU(2)_+$, $SU(2)_-$, and $SU(2)^R_+$, respectively.  $\Pi_i$, $\Lambda_{\rho}$ and $L_{d}$ are Cartan generators of the 5d gauge group $G$, the Wilson loop defect group $G'$, and the additional 5d flavor symmetry group $K$, respectively. They are all flavor symmetry groups from the one-dimensional perspective. As far as the conjugate variables are concerned, we have introduced the 5d Coulomb parameters $\{a_i\}$, the Wilson loop defect fugacities $\{M_{\rho}\}$, additional masses as $\{m_d\}$, and redefined the $\Omega$-background parameters as  
	\beq
	\label{epsilons}
	\epsilon_+\equiv \frac{\epsilon_1+\epsilon_2}{2}\; , ~~~~~ \epsilon_-\equiv \frac{\epsilon_1-\epsilon_2}{2}\; .
	\eeq

	The index is the grand canonical ensemble of instanton BPS states. It is a product of a perturbative factor involving the classical and 1-loop contributions, and of a factor capturing the instanton corrections. The perturbative part will drop out after normalization, so we will safely ignore it in the rest of this work. Meanwhile, we organize the instanton part as a sum over all instanton sectors $k$.\\
	
	We can evaluate the gauge theory index in the weak coupling regime of the  UV quantum mechanics, where it reduces to Gaussian integrals around saddle points.  These saddle points are parameterized by  $\phi=R\, \varphi^{(QM)}+ i\, R\, A^{(QM)}_t$, with $A^{(QM)}_t$ the gauge field and $\varphi^{QM}$ the scalar in the vector multiplet of the quantum mechanics. We denote the gauge group of this quantum mechanics as $\widehat{G}$, and the (complexified) eigenvalues of $\phi$ as $\phi_1, \ldots, \phi_k$. Performing the Gaussian integrals over massive fluctuations, the index reduces to a zero mode integral of various 1-loop determinants, which we write schematically for now in a five-dimensional language as: 
	\begin{align}
	\label{5dintegral}
	&Z_{inst}  =\sum_{k=0}^{\infty}\; \frac{\widetilde{\fq}^k}{|W(\widehat{G})|} \oint  \left[ \prod_{I=1}^{k}\frac{d\phi_I}{2\pi i}\right]Z^{(k)}_{vec}\cdot Z^{(k)}_{matter}\cdot Z^{(k)}_{defect}  \; .
	\end{align}
	The five-dimensional gauge coupling is implicitly written in the instanton counting fugacity  $\widetilde{\fq}\equiv \exp\left(-8\pi^2 R/g^2_{5d}\right)$. $|W(\widehat{G})|$ is the Weyl group order of the quantum mechanics gauge group $\widehat{G}$. If $G=U(N)$, then $\widehat{G}=U(k)$. If $G=SO(N)$, then $\widehat{G}=Sp(k)$. If $G=Sp(N)$, then $\widehat{G}=O(k)$. In that last case, the group $\widehat{G}=O(k)$ is disconnected, and one should turn on additional holonomies that will slightly modify the above formula, as we will see later in detail.\\ 
	
	The factor $Z^{(k)}_{vec}(\{\phi_I\}, \{a_i\}, \epsilon_1, \epsilon_2)$ contains all the physics of the 5d vector multiplet. We have made explicit the dependence on the $N$ Coulomb parameters $\{a_i\}$ of $G$, the $k$ integration variables $\phi_I$ of $\widehat{G}$, and the $\Omega$-background parameters $\epsilon_1$ and $\epsilon_2$.\\  
	
	The factor $Z^{(k)}_{matter}(\{\phi_I\}, \{m_d\}, \epsilon_1, \epsilon_2)$ contains all the physics of the 5d matter multiplets, where $\{m_d\}$  are the corresponding masses. In this paper, $Z^{(k)}_{matter}$ stands for matter in symmetric, antisymmetric, adjoint, or fundamental representations of $G$. In the case of a massive hypermultiplet in the adjoint representation of $G$, the gauge theory is sometimes referred to as having $\cN=1^*$ supersymmetry.\\

	The factor $Z^{(k)}_{defect}(\{\phi_I\}, \{M_\rho\}, \{a_i\}, \epsilon_1, \epsilon_2)$ contains all the physics of the Wilson loop. It introduces the $N'$ defect fermion masses $\{M_\rho\}$ in the instanton partition function. Note that it depends on the 5d Coulomb moduli $\{a_i\}$, as is clear from the coupled 5d/1d action term \eqref{1dfermion}.\\

	Since the partition function is a multi-dimensional integral in the 1d Coulomb moduli $\phi_I$'s, we need a robust method to compute the residues. We adopt the so-called Jeffrey-Kirwan (JK) residue prescription \cite{Jeffrey:1993}. The prescription was first popularized in our context in a two-dimensional setup \cite{Benini:2013xpa}, and in the works \cite{Hwang:2014uwa, Cordova:2014oxa, Hori:2014tda}. It has been used widely in the high energy community since then, so we will be brief in reviewing it. Each factor $Z^{(k)}_{vec}$, $Z^{(k)}_{matter}$, and $Z^{(k)}_{defect}$ has the following general form:
	\ie\label{integRand}
	\frac{\prod_{i=1}^{n_1}\sinh(\vec\rho_i\vec\phi_i+\ldots)}{\prod_{j=1}^{n_2}\sinh(\vec\rho_j\vec\phi_j+\ldots)}\; ,
	\fe
	where $\vec\rho$ is a $k$-tuple vector, which consists exclusively of entries in the set $\{0,\pm1,\pm\frac{1}{2}\}$. $n_1$ and $n_2$ are positive integers specified by the details of the ADHM quantum mechanics. Finally, ``$\ldots$" stands for a linear function of the spacetime fugacities $\epsilon_1$, $\epsilon_2$, as well as 1d flavor fugacities. Since $\sinh(i\pi)=\sinh(0)=0$, there can be many poles in \eqref{integRand} at a specific value of $\vec\phi=\vec\phi_0$. 
	
	On general ground, if the gauge group $\widehat{G}$ has $k$ abelian factors, the index will depend on the choice of a $k$-vector, the Fayet-Iliopoulos (F.I.) parameter $\zeta$ of the quantum mechanics, due to nonzero $\phi$-poles at $\pm \infty$. In particular, the index will be computed in a certain chamber, corresponding to a distinct choice of F.I. parameter, and one should expect wall-crossing phenomena when $\zeta=0$. 
	The JK prescription instructs us to pick yet another $k$-vector $\eta$, not a priori related to the F.I. parameter, but by choosing $\eta=-\zeta$, it can be argued that the contributions of the $\phi$-residues at $\pm \infty$ vanish. In this paper, we find the choice $(1,3,5,\ldots,2k-1)$ convenient, and will therefore work in the corresponding $\zeta$-chamber. When the gauge group does not contain any abelian factor, so when $\widehat{G}=Sp(k), O(k)$, there are no wall-crossing phenomena, and the index will not depend on the choice of $\eta$. 
	
	Having defined $\eta$, we are to choose $k$ hyperplanes from the arguments of $\sinh$ functions in the denominator of \eqref{integRand}. Those hyperplanes will take the following form:
	\ie
	\label{linearsystem}
	\vec\rho_j\cdot\vec\phi_j+\ldots=0\;,\;\;\;\text{where }j=1,\ldots,k.
	\fe
	The contours of the partition function are then chosen to enclose poles which are solutions of the above linear system of equation, but only if the vector  $\eta$ also happens to lie in the cone spanned by the vectors $\vec\rho_l$. One practical way to test this condition is to construct a $k\times k$ matrix ${\bf{Q}}=Q_{ji}=(\rho_j)_i$, where $\vec\rho_j=((\rho_j)_1,\ldots,(\rho_j)_k)$, and test if all the components of $\eta\, {\bf{Q}}^{-1}$ are positive.\\
	
	Sometimes, it can happen that a solution of the system of equations \eqref{linearsystem} yields additional zeroes in the denominator of \eqref{integRand}. This typically results in non-simple poles, and we discuss them in the appendix.

	\subsection{The Wilson Loop Defect Operator $Y$}
	\label{ssec:Yoperator}
	
	Before we can make contact with Seiberg-Witten geometry, we need to discuss in some detail the contours of the partition function.\\
		
	First, we specialize the Wilson loop factor $Z^{(k)}_{defect}$ to be made up of only one fermion, $N'=1$\footnote{The case $N'>1$ will not be relevant to our discussion, as will be clear a fortiori. It corresponds to considering Wilson loops in higher irreducible representations of $G$ \cite{Gomis:2006sb}. We will comment on it later.}.

	Second, we organize the poles in the partition function into two sets: for a given instanton number $k$,  let $\mathcal{M}_k$ be the set of poles selected by the JK-residue prescription in the  partition function \eqref{5dintegral}  (and should therefore all be enclosed by the contours). Meanwhile, let $\mathcal{M}^{pure}_k$ be the set poles selected by the JK-residue prescription in the pure partition function, in the absence of the factor $Z^{(k)}_{defect}$ in the integrand.  We denote the instanton partition function in the absence of defects  as $\left\langle 1\right\rangle$.\\
	
	For a given instanton number $k$, the sets $\mathcal{M}_k$ and  $\mathcal{M}^{pure}_k$ have finite, but different sizes: as we will see explicitly in the next section, $Z^{(k)}_{defect}$ always contains at least one pole depending on the defect fugacity $M$\footnote{There is one notable exception, which is $k=1$ for a $Sp(N)$ gauge theory. We will come back to this in the next section.}, so for a given instanton number $k$, the former set  is always strictly larger than the latter set:  $|\mathcal{M}_k|>|\mathcal{M}^{pure}_k|$. This simple observation has important consequences, as we can derive non-perturbative Schwinger-Dyson equations for Wilson loop vevs from it. We first define the expectation value of a defect operator:
	\begin{align}
	\label{Yoperator}
	\left\langle \left[Y(M)\right]^{\pm 1}\right\rangle \equiv
	\sum_{k=0}^{\infty}\frac{\widetilde{\fq}^{k}}{|W(\widehat{G})|} \, \oint_{\mathcal{M}^{pure}_k}  \left[\frac{d\phi_I}{2\pi i}\right]Z^{(k)}_{vec}\cdot Z^{(k)}_{matter} \cdot \left[Z^{(k)}_{defect}(M)\right]^{\pm 1}\, . 
	\end{align}
	At first sight, it may seem like $\left\langle Y(M)\right\rangle$ is just the partition function \eqref{5dintegral}, but we still need to specify the contours of the above integral. Namely, we \emph{choose} the contours in \eqref{Yoperator} to only enclose poles in the set $\mathcal{M}^{pure}_k$, meaning no enclosed pole will depend on $M$. Put differently, we ignore all the poles originating from  $Z^{(k)}_{defect}$. This differs from the contour prescription used in defining the partition function $Z(M)$, where all the poles are taken in the set $\mathcal{M}_k$.\\ 
	
	Our first main result in this paper will be to show that given a classical gauge group $G$, the partition function of $T^{5d}$ can be written as
	\beq
	\label{manyterms}
	Z(M) = \left\langle Y(M) \right\rangle + \sum_{k=1}^{\infty}\,  \sum_{j=1}^{\mathcal{M}_k\setminus\mathcal{M}^{pure}_k} \widetilde{\fq}^{j}\, F_{k,j}\big(\left\langle Y(M)\right\rangle, \{a_i\} \{m_d\}, M,\epsilon_1, \epsilon_2, R\big) \, ,
	\eeq
	where $F_{k,j}$ is a rational function of the defect $Y$-operator vev, the spacetime fugacities $\epsilon_1$, $\epsilon_2$, the circle radius $R$ and the various masses. In using the notation $Z(M)$, we made the dependence of the partition function on the defect fugacity $M$ explicit, while all the other fugacities are kept implicit. For instance, the dependence on the 5d Coulomb parameters is implicitly encoded in the vev $\left\langle Y(M)\right\rangle$ and the function $F$.\\
	
	The meaning of this expression is as follows: as we have noted, the partition function has the same integrand as the defect operator vev,  the first term $\left\langle Y(M) \right\rangle$ on the right-hand side. However, the contours on the left-hand side enclose more poles than those of  $\left\langle Y(M) \right\rangle$, since $|\mathcal{M}_k|>|\mathcal{M}^{pure}_k|$, for each instanton number $k$. The sum is there to make up for that deficit of $M$-dependent poles. Each term in the sum on the right-hand side stands for a residue of the integral \eqref{5dintegral}, evaluated at a pole in $\mathcal{M}_k\setminus \mathcal{M}^{pure}_k$, that is to say a pole in the set $\mathcal{M}_k$ not present in the set $\mathcal{M}^{pure}_k$. Quite nontrivially, we find that the residues turn out to be rational functions of $Y$-operator vevs. After summing over all the terms in the right-hand side, the partition function is recovered.
	
	\begin{figure}[h!]
		%\begin{center}
		\emph{}
		%\hspace{-20ex}
		\centering
		\includegraphics[trim={0 0 0 3cm},clip,width=0.8\textwidth]{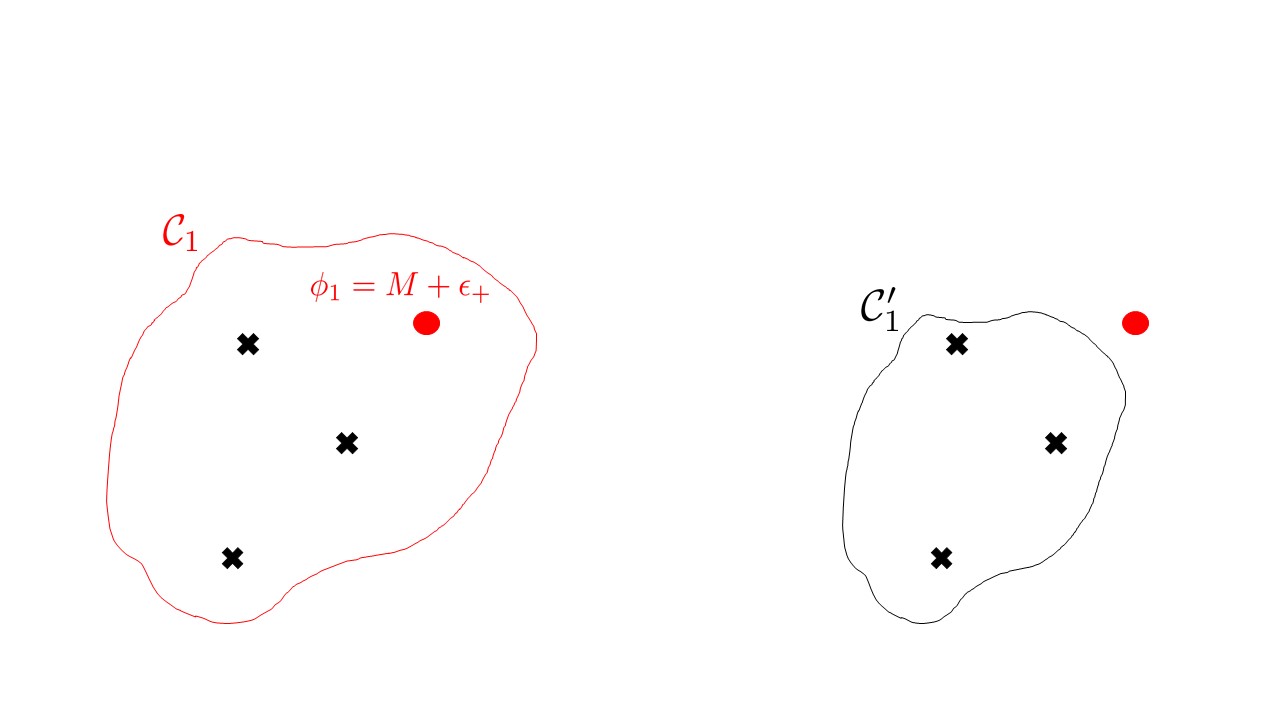}
		\vspace{-12pt}
		\caption{The black crosses denote poles in the set $\mathcal{M}^{pure}_k$, from the pure partition function, while the red dot denotes a pole in the set $\mathcal{M}_k\setminus \mathcal{M}^{pure}_k$. Such a pole is due to the factor $Z^{(k)}_{defect}$ in the integrand. On the left, we show a possible contour for the computation of the 5d partition function at $k=1$, say for a $SU(3)$ theory. Note that by the JK prescription, we must in particular enclose the new pole in red. Remarkably, it is equivalent to trade this contour for the one on the right, which now only encloses the poles in the set $\mathcal{M}^{pure}_1$, but with a modified integrand; for the contour on the right, the integrand will now contain insertions of new $Y$-operators, with an instanton shift of one unit to account for the missing pole.} 
		\label{fig:contours}
		%\end{center}
	\end{figure}
	
	We emphasize here that at no point in the discussion did we need to know how to compute the instanton partition function of the gauge theory in the absence of the defect, that is to say the content of the set of poles $\mathcal{M}^{pure}_k$. What is instead relevant here is the set of poles $\mathcal{M}_k\setminus \mathcal{M}^{pure}_k$, due entirely to the insertion of the defect.\\

	We normalize the partition function,
	\beq
	\label{Znormalized}
	\widetilde{Z}(M)\equiv \frac{Z(M)}{\widetilde{\left\langle 1 \right\rangle}} \; ,
	\eeq
	where $\widetilde{\left\langle 1 \right\rangle}\equiv\left\langle 1\right\rangle\cdot Z_{extra}$. Here, $\left\langle 1\right\rangle$ is the pure partition function we defined earlier, \eqref{5dintegral} without the factor $Z^{(k)}_{defect}$. We do not want to count possible UV degrees of freedom present in the ADHM descrption but not in the actual QFT; we have collected these extra unwanted contributions as the factor $Z_{extra}$. Let us expand this normalized partition function in the (exponentiated) defect fugacity $e^{M}$, and write the result as
	\begin{equation}
	\label{generalcurveqqh}
	\widetilde{Z}(M) = \cT(M; \{e_i^{\epsilon_1, \epsilon_2}\}) \; .
	\end{equation}
	Here,  $\cT(M; \{e_i^{\epsilon_1, \epsilon_2}\})$ is a Laurent series in the  Wilson loop (exponentiated) fugacity $e^{M}$. The parameters  $\{e_i^{\epsilon_1, \epsilon_2}\}$ encode the classical Coulomb moduli  $\{a_i\}$, the various masses $\{m_d\}$, the $\Omega$-background parameters $\epsilon_1, \epsilon_2$, as well as the instanton counting parameter $\widetilde{\fq}$.\\
	
	Our second result is the following: we conjecture that $T$ is a \emph{finite} Laurent series in $e^{M}$, meaning that the number $i$ of parameters $\{e_i^{\epsilon_1, \epsilon_2}\}$ is finite.
	
	Ultimately, this result comes about from expanding the normalized partition function in  $e^{M}$, and showing that the resulting expression is pole-free in that fugacity. It follows that a general proof of the finiteness of  $\cT(M; \{e_i^{\epsilon_1, \epsilon_2}\})$ is related to the explicit description of the set of poles  $\mathcal{M}_k\setminus \mathcal{M}^{pure}_k$, for all instantons $k$. The sets $\mathcal{M}_k$ and  $\mathcal{M}^{pure}_k$ happen to both be known exactly in the case $G=SU(N)$ (pure or with adjoint/fundamental matter), and a proof of the regularity of  $\cT(M; \{e_i^{\epsilon_1, \epsilon_2}\})$ is therefore possible. When $G$ is one of the other classical gauge groups, however, an explicit description of the poles is lacking, so we leave the general proof of the statement to future work. For now, we will content ourselves with a check of the conjecture following the JK-residue algorithm, which can be performed to arbitrarily high instanton number.\\
	
We therefore conclude that 
	\beq
	\label{Schwinger}
	\boxed{\frac{1}{\widetilde{\left\langle 1 \right\rangle}}\left[\left\langle Y(M) \right\rangle + \sum_{k=1}^{\infty}\, \sum_{j=1}^{\mathcal{M}_k\setminus \mathcal{M}^{pure}_k} \widetilde{\fq}^{j} \, F_{k,j}\big(\left\langle Y(M)\right\rangle,\{a_i\}, \{m_d\}, M , \epsilon_1, \epsilon_2, R\big)\right]  = \cT(M; \{e_i^{\epsilon_1, \epsilon_2}\}) \, ,}
	\eeq
	together with the requirement that $\cT(M; \{e_i^{\epsilon_1, \epsilon_2}\})$ be finite in $e^M$, can be thought of as non-perturbative Schwinger-Dyson equations for the theory. The defect $Y$-operator vevs we construct as \eqref{Yoperator} are explicit solutions of these equations. In five dimensions, the $Y$-operator is understood as a loop defect. Since instantons in 5d are particles, the $Y$-operator can correctly mediate a change in instanton number $k$, as long as the particles wrap the loop. This changes the topological sector of $T^{5d}$, and the left-hand side of \eqref{Schwinger} encodes a corresponding symmetry.
	
	Before we end this section, we point out that the above discussion straightforwardly generalizes to an arbitrary number of fermions, $N'>1$. In that case, we still obtain Schwinger-Dyson equations similar to the ones above, but involving correlation functions of a higher number of $Y$-operators.\\

	\subsection{The Doubly-Quantized Seiberg-Witten Geometry}

	We would like to propose that the above Schwinger-Dyson (SD) identities can be understood as a double quantization of Seiberg-Witten (SW) geometry, where the two quantum parameters are the spacetime  $\Omega$-background parameters $\epsilon_1$ and $\epsilon_2$. In this work, we test the idea by taking the flat space limit $\epsilon_1, \epsilon_2 \rightarrow 0$ in  \eqref{Schwinger}, and we argue that we recover the SW curve of the theory. The SD equations involve correlators of Wilson loop operators, which should now be thought of as a complex coordinate on an auxiliary cylinder  \cite{Nekrasov:2012xe},
	\beq
	\left\langle Y(M)\right\rangle \rightarrow  y\in \mathbb{C}^* \; .
	\eeq
	We argue that the resulting equation
	\beq
	\label{SW5d}
	y+ \widetilde{\fq} \ldots = \cT(M; \{e_i^{0,0}\}) \, ,
	\eeq
	is the SW curve of $T^{5d}$, with coordinates $(e^M, y)\in \mathbb{C}^*\times \mathbb{C}^*$. From our discussion in the last section, $\cT(M; \{e_i^{0,0}\})$ is a finite Laurent series in $e^M$, and the parameters $\{e_i^{0,0}\}$ define the vacuum.\\

	This result has highly nontrivial implications; for instance, the left-hand side of \eqref{Schwinger} typically is an infinite sum of correlators, since the set $\mathcal{M}_k\setminus \mathcal{M}^{pure}_k$ becomes increasingly bigger as $k$ grows. However, when $T^{5d}$ is a pure $G$-theory, we conjecture that all but a finite number of terms disappear in the flat space limit. More precisely, in those cases, we will argue that when the instanton number $k$ is sufficiently high\footnote{This typically happens for $k\geq 2$ or $k \geq 3$, depending on the gauge group and matter content. It can also happen that all higher instanton terms survive the limit, as is the case for instance for $SU(N)$ with adjoint matter.} in the sum \eqref{Schwinger}, all terms will cancel against each other at each order, resulting in a finite sum that reproduces the known SW curve of the model.\\

	Perhaps the presentation of the curves is more familiar in a four-dimensional context, so we deem it useful to rewrite our non-perturbative SD equations in four dimensions first. There are a priori many different ways to take a 4d limit, so we should be specific: here, we want $T^{5d}$ to become a four-dimensional gauge theory $T^{4d}$, with the same gauge group $G$ and flavor content as the higher-dimensional theory.

	\begin{figure}[h!]
		%\begin{center}
		\emph{}
		%\hspace{-20ex}
		\centering
		\includegraphics[trim={0 0 0 2cm},clip,width=0.8\textwidth]{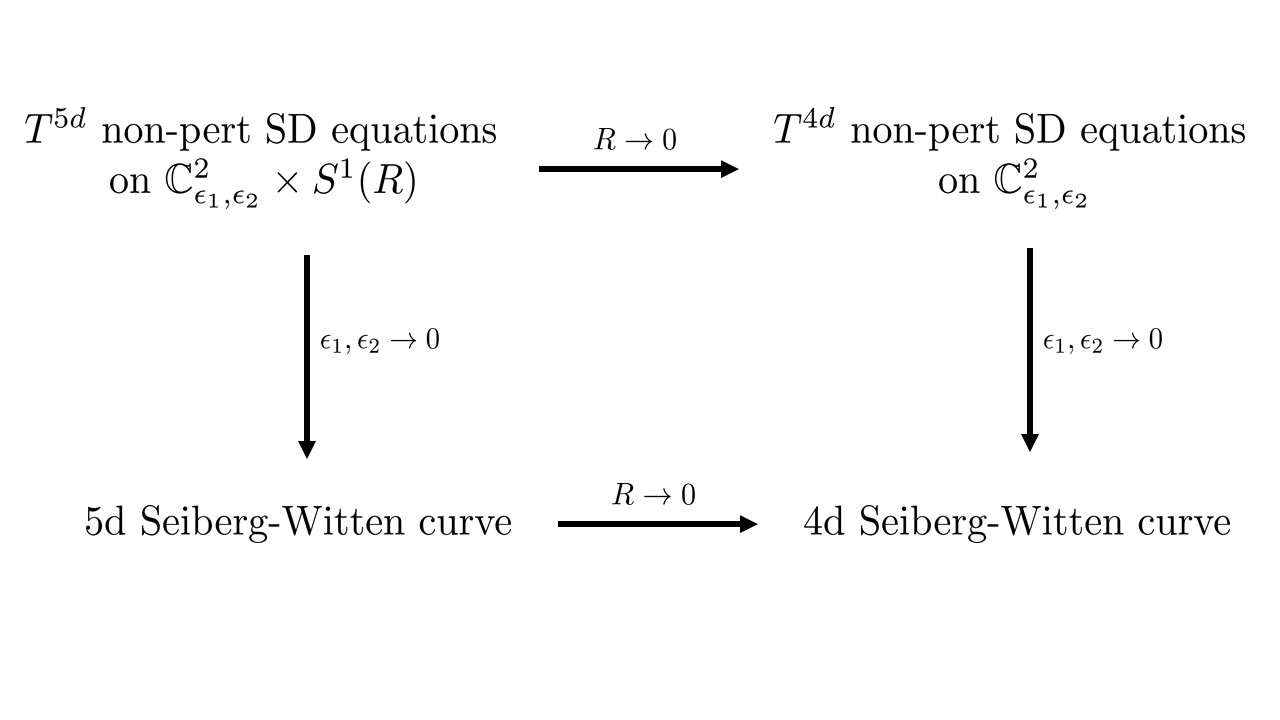}
		\vspace{-12pt}
		\caption{The various Schwinger-Dyson equations and limits to Seiberg-Witten curves, in 5d and in 4d.} 
		\label{fig:limits}
		%\end{center}
	\end{figure}

	Recall that $T^{5d}$ is defined on  the manifold $S^1(R)\times \mathbb{R}^4$, so we  will conveniently reduce the theory on the circle  $S^1(R)$ by sending $R\rightarrow 0$. At this point, it is necessary to reintroduce the explicit dependence of the various 5d fugacities on the radius $R$. We first rescale the gauge coupling as 
	\beq
	\label{RGflow}
	e^{-\frac{8 \pi^2 R}{g^2_{5d}}}=\left(-i R\right)^{2 h^\vee(G)-k(R)}\, \widetilde{\fq}\; ,
	\eeq

	and keep $\widetilde{\fq}$ fixed as we send $R$ to zero.	$h^\vee(G)$ is the dual Coxeter number of $G$ and  $k(R)$ the quadratic Casimir of the representation $R$. Furthermore, we require the $\Omega$-background parameters $\epsilon_1$, $\epsilon_2$, the Coulomb moduli $\{a_i\}$, the masses  $\{m_d\}$ and the defect fugacity $M$ to be kept fixed as we reduce the circle size. In practice, we simply rescale these fugacities by $R$, and then take the limit $R\rightarrow 0$. 
	The 4d gauge coupling of $T^{4d}$ is related to the 5d gauge coupling of $T^{5d}$ by
	\beq
	\frac{8 \pi^2 R}{g^2_{5d}}=\frac{8 \pi^2}{g^2_{4d}} \; .
	\eeq
	It follows that \eqref{RGflow} can be understood as an RG flow equation, which describes the running of the coupling $g_{4d}$ as the UV scale $R^{-1}$ is varied. An associated dynamical scale $\Lambda$ can therefore be defined:
	\beq
	\label{lambdadef}
	\widetilde{\fq}=\Lambda^{2 h^\vee(G)-k(R)}\; .
	\eeq
	We can now safely take the limit $R\rightarrow 0$ in the SD equations \eqref{Schwinger}, and obtain non-perturbative SD equations for $T^{4d}$, which have the same functional form as before:
	\beq
	\label{Schwinger4d}
	\frac{1}{\widetilde{\left\langle 1 \right\rangle}_{4d}}\left[\left\langle Y(M) \right\rangle_{4d} + \sum_{k=1}^{\infty} \sum_{j=1}^{\mathcal{M}_k\setminus \mathcal{M}^{pure}_k} \widetilde{\fq}^{j}\, F^{4d}_{k,j}\big(\left\langle Y(M)\right\rangle_{4d},\{a_i\}, \{m_d\},M, \epsilon_1, \epsilon_2\big)\right]  = {\cT}_{4d}(M; \{e_i^{\epsilon_1, \epsilon_2}\}) \, .
	\eeq
	Note that this time, the function ${\cT}_{4d}(M; \{e_i^{\epsilon_1, \epsilon_2}\})$ on the right-hand side is a finite polynomial in $M$, with a finite number $i$ of roots $\{e_i^{\epsilon_1, \epsilon_2}\}$ which characterize the vacuum. On the left-hand side, the $Y$-operator vev $ \left\langle Y\right\rangle_{4d}$ is now a point defect sitting at the origin of spacetime, since it is the reduction of the Wilson loop that used to wrap the circle $S^1(R)$.
	
	Finally, we can proceed to take the flat space limit $\epsilon_1, \epsilon_2 \rightarrow 0$ just as we did in 5d, to obtain
	\beq
	\label{SW4d}
	y+ \widetilde{\fq} \ldots = \cT_{4d}(M; \{e_i^{0,0}\}) \, .
	\eeq
	We will argue in the next section that in all examples we present, the above is nothing but the SW curve of $T^{4d}$, with coordinates $(M, y)\in \mathbb{C}\times \mathbb{C}^*$.\\

	Let us end this section with a remark on representation theory. In the 5d picture and for a pure $SU(N)$ gauge theory, the left-hand side of the SD equation \eqref{Schwinger} can be interpreted  as a deformation of the $q$-character of the fundamental representation of the quantum affine algebra $U_\hbar(\widehat{A_1})$ \cite{Nekrasov:2015wsu,Kimura:2015rgi}.  In four dimensions, the left-hand side is understood instead as a deformation of the $q$-character of the fundamental representation of the Yangian $Y(A_1)$, as we reviewed in the introduction. The terms on the left-hand side are built from an i-Weyl group action acting on the highest weight  $\left\langle Y(M) \right\rangle$. When $G\neq SU(N)$, the relation to the representation theory of Yangians (in 4d) or quantum affine algebras (in 5d) is lacking, and we will see that the i-Weyl group is no longer finite. Nevertheless, the regularity of the Laurent polynomial ${\cT}(M; \{e_i^{\epsilon_1, \epsilon_2}\})$ in $e^M$ suggests that an interpretation of the 5d partition function as a $qq$-character of some representation of a quantum affine algebra may still exist. A hint may come from the study of quantum integrable systems, such as the XXZ spin chain, which is known to exhibit quantum affine symmetry. when $G=SO(N), Sp(N)$, one needs to consider open boundaries, and the XXZ spin chain analysis requires the use of a reflection matrix \cite{Cherednik:1985vs,Sklyanin:1988yz}. The role of this reflection matrix will essentially be played by an orientifold plane in our string theory construction, as we will see in the next section.

	\subsection{Existence of the Theories}
	\label{ss:existence}
	
	Gauge theories in five dimensions are non-renormalizable. As such, $T^{5d}$ is not actually useful as a microscopic theory.  After taking the 4d limit, we should worry whether or not the resulting gauge theory $T^{4d}$ makes sense in the first place, as a conformal or asymptotically free theory. Therefore, we henceforth impose the condition $2 h^\vee(G)-k(R)\geq 0$. In particular, for a fixed gauge group rank, this translates to restrictions on the allowed flavor symmetry of the theory. For instance, as far as fundamental matter is concerned, if $G=SU(N)/SO(N)/Sp(N)$, the allowed flavor content must now satisfy $N_f\leq 2 N\; /\; N_f\leq N-2 \;/\; N_f\leq N+2$, respectively.\\
	
	There is a subtlety, however. Indeed, consider a 4d conformal or asymptotically free theory $T^{4d}$, and its 5d uplift $T^{5d}$. In the ADHM quantum mechanics description of $T^{5d}$, it is possible a continuum opens up in the 1d Coulomb branch, which manifests itself in 5d as extra decoupled states in the bulk. In practice, such a continuum can arise when there are new poles at $\infty$ for the 1d vector multiplet scalar in the integrand \eqref{5dintegral}. For example, consider the case $G=SO(N)$; the conformality/asymptotic freedom condition on fundamental matter is  $N_f\leq N-2$, but there are poles at infinity in the ADHM quantum mechanics of $T^{5d}$ when $N_f=N-4, N-3, N-2$. Our analysis in this paper is only strictly valid when $N_f\leq N-5$. For definiteness, then, we will only limit ourselves in the rest of this paper to the cases where the ADHM quantum mechanics does not develop such continua in the 1d Coulomb branch. We leave the detailed study of the remaining cases and their 4d limits to future work.

	\section{Explicit Derivation of the Defect Partition Function}
	
	We now come to the explicit derivation of the quantized Seiberg-Witten geometry, for all classical gauge groups and a wide variety of matter contents.

	\subsection{The $SU(N)$ Gauge Theory}

Recently, the quantized SW geometry of $SU(N)$ gauge theories has been an active topic of investigation. Before we review it in our formalism, let us give a brief overview of what has been done in the literature. The study of the instanton moduli space of 5d $SU(N)$ in the presence of a 1/2-BPS Wilson loop was first considered in \cite{Tong:2014cha}, in the same type IIA string background as us. Our presentation will follow closely the discussion in \cite{Kim:2016qqs}, where the instanton partition function of 5d $SU(N)$ SYM  with a single Wilson loop insertion was computed as the  Witten index of a $(0,4)$ quantum mechanics. 
	
The existence of non-perturbative SD identities was discovered in \cite{Nekrasov:2015wsu}. A type IIB realization is given there, which is one T-duality away from ours; this is consistent with the fact that the focus in that work is on a point defect in 4d $\cN=2$ $SU(N)$ SYM. Our presentation is simply the $K$-theoretic version of the cohomological results obtained there, as can be seen after reducing our 5d instanton partition on the circle wrapped by the Wilson loop.  \cite{Nekrasov:2015wsu} and \cite{Kim:2016qqs} both compute the defect partition function for pure $SU(N)$ SYM, as well as in the presence of fundamental and adjoint matter. The $S$-duality of the $SU(N)$ Wilson loop in 5d was studied with fundamental matter in \cite{Assel:2018rcw}. The $S$-duality of the Wilson loop in the presence of adjoint matter was the focus of \cite{Agarwal:2018tso}.
	
Yet another T-dual brane configuration was given in \cite{Tong:2014yna}, with the aim of giving a UV description to the holographic dual of $AdS_3\times S^3 \times S^3 \times S^1$. The quantization of the various open strings in the presence of $B$-fields was carried out in that setup in \cite{Nekrasov:2016gud}.
	
Finally, a $q$-CFT perspective was given in \cite{Kimura:2015rgi} in the context of so-called deformed $\cW$-algebras, see also \cite{Mironov:2016yue}; the non-perturbative SD equations arise there as a $q$-deformed Ward identities in two dimensions.\\

Our starting point is type IIA string theory in flat space. We compactify the $x_0$-direction and introduce the following branes:
	\begin{table}[H]\centering
		\begin{tabular}{|l|cccccccccc|} \hline
			&  0 &  1 &  2 &  3 &  4 &  5 &  6 &  7 &  8 &  9  \\
			\hline\hline
			$k$\,\,	D0 &\xTB&   &   &   &   &    &    &    &    &    \\
			\hline
			$N$	D4 &\xTB&\xTB&\xTB&\xTB&\xTB&    &    &    &    &     \\
			\hline
			$1$\,\, D$4'$ &\xTB&    &    &    &    &\xTB&\xTB&\xTB&\xTB& \\
			\hline
			\;	F1 &\xTB&    &    &    &    & & & & & \xTB  \\
			\hline
		\end{tabular}
		\caption{The directions of the various branes.}
		\label{table:BranesSUN}
	\end{table}

	The effective theory on the D4 branes is a 5d $U(N)$ gauge theory on $S^1(R)\times\mathbb{R}^4$, called $T^{5d}$. Separating the D4 branes along the $x_9$-direction correponds to giving a non-zero vector multiplet vev $\langle\Phi_9 \rangle\neq 0$, which describes the Coulomb branch of the theory.
	The D$4'$ brane realizes a 1/2-BPS  Wilson loop\footnote{A stringy realization of Wilson loops was first proposed in \cite{Maldacena:1998im, Rey:1998ik}, in the context of holography; there, a Wilson loop in the first fundamental representation of $SU(N)$ can be described as a fundamental string whose worldsheet ends at the loop, located at the boundary of an $AdS$ geometry. Later, a realization in terms of branes was given instead  \cite{Drukker:2005kx,Gomis:2006sb,Yamaguchi:2006tq}, allowing for loops in more general representations.  More precisely, a 1/2-BPS Wilson loop can be realized as $N'$ D$4'$ branes that are codimension 4 with respect to the original $N$ D4 branes; the representation in which the Wilson loop transforms is then determined by $N'$.} wrapping the $x_0$-circle and which sits at the origin of $\mathbb{R}^4$. Pulling the  D$4'$ brane a distance $M$ away in the $x_9$-direction, there are now open strings with nonzero tension between the D4 and D$4'$ branes. These are heavy fermion probes, with mass proportional to the distance $M$. The ADHM mechanics we are after is the quantum mechanics of D0 branes in this background of D4 and D$4'$ branes. The gauge group on the D0 branes is $\widehat{G}=U(k)$. We put  the theory on the 5d $\Omega$-background $S^1(R)\times\mathbb{R}^4_{\epsilon_1, \epsilon_2}$. Finally, we find it convenient to introduce the following fugacities, 
	\ie
	\epsilon_+\equiv{\epsilon_1 + \epsilon_2\over 2},~~~ \epsilon_-\equiv{\epsilon_1 - \epsilon_2\over 2},
	~~~ m \equiv {\epsilon_3 - \epsilon_4\over 2},
	\fe
	where $\epsilon_1$, $\epsilon_2$, $\epsilon_3$, and $\epsilon_4$ are the chemical potentials associated to the rotations of the $\bR^2_{\text{\tiny{12}}}$, $\bR^2_{\text{\tiny{34}}}$, $\bR^2_{\text{\tiny{56}}}$, and $\bR^2_{\text{\tiny{78}}}$ planes, respectively.

	The instanton partition function is the index of the D0 brane quantum mechanics. In our context, it reads\footnote{The partition function we will compute is for a $U(N)$ gauge theory. To obtain the $SU(N)$ result, one has to further set a constraint on the Coulomb parameters, $\sum_{i=1}^N a_i =0$.}:	
	\begin{align}
	\label{5dintegralSUN}
	&Z(M) = Z_{D4/D4'}(M)\sum_{k=0}^{\infty}\; \frac{\widetilde{\fq}^k}{|W(\widehat{G})|} \oint_{\mathcal{M}_k}  \left[ \prod_{I=1}^{k}\frac{d\phi_I}{2\pi i}\right]Z^{(k)}_{D0/D0}\cdot Z^{(k)}_{D0/D4}\cdot Z^{(k)}_{D0/D4'}(M)  \; .
	\end{align}
For the sake of brevity, we only made the dependence on the defect fugacity $M$ explicit in writing the partition function $Z(M)$.
	$|W|$ is the order of the Weyl group of $\widehat{G}$, and the various factors are given by the 1-loop determinants of the quantum mechanical fields, coming from the quantization of the strings stretching between the various branes.	
	\begin{table}[H]\centering
		\begin{tabular}{|c|c|c|c|}
			\hline
			strings &Multiplets & fields & {\footnotesize $SO(4)\times SO(4)_R\times U(k)\times U(N)\times U(N')$ }%& O($k$)$\times$Sp(1)$\times$ SO($2N_f$)
			\\ \hline
			\multirow{8}{*}{D0-D0}& \multirow{3}{*}{vector} & gauge field &$({\bf 1},{\bf 1},{\bf 1},{\bf 1},{\bf adj},{\bf 1},{\bf 1})$
			\\ \cline{3-4}
			& & scalar & $({\bf 1},{\bf 1},{\bf 1},{\bf 1},{\bf adj},{\bf 1},{\bf 1})$
			\\ \cline{3-4}
			&  & fermions &$({\bf 1},{\bf 2},{\bf 1},{\bf 2},{\bf adj},{\bf 1},{\bf 1})$
			\\ \cline{2-4}
			&Fermi & fermions & $({\bf 2},{\bf 1},{\bf 2},{\bf 1},{\bf adj},{\bf 1},{\bf 1})$
			\\ \cline{2-4}
			&\multirow{2}{*}{t-hyper}  & scalars & $({\bf 1},{\bf 1},{\bf 2},{\bf 2},{\bf adj},{\bf 1},{\bf 1})$
			\\ \cline{3-4}
			&& fermions & $({\bf 1},{\bf 2},{\bf 2},{\bf 1},{\bf adj},{\bf 1},{\bf 1})$
			\\ \cline{2-4}
			&\multirow{2}{*}{hyper} & scalars & $({\bf 2},{\bf 2},{\bf 1},{\bf 1},{\bf adj},{\bf 1},{\bf 1})$
			\\ \cline{3-4}
			&& fermions & $({\bf 2},{\bf 1},{\bf 1},{\bf 2},{\bf adj},{\bf 1},{\bf 1})$
			\\ \hline
			\multirow{3}{*}{D0-D4}& \multirow{2}{*}{hyper} & scalars & $({\bf 2},{\bf 1},{\bf 1},{\bf 1},{\bf k},{\bf \bar N},{\bf 1})$
			\\ \cline{3-4}
			& & fermions & $({\bf 1},{\bf 1},{\bf 1},{\bf 2},{\bf k},{\bf\bar N},{\bf 1})$
			\\ \cline{2-4}
			& Fermi & fermions & $({\bf 1},{\bf 1},{\bf 2},{\bf 1},{\bf k},{\bf\bar N},{\bf 1})$
			\\ \hline
			\multirow{3}{*}{D0-D$4'$}& \multirow{2}{*}{t-hyper} & scalars & $({\bf 2},{\bf 1},{\bf 1},{\bf 1},{\bf k},{\bf 1},{\bf\bar N'})$
			\\ \cline{3-4}
			& & fermions & $({\bf 1},{\bf 1},{\bf 1},{\bf 2},{\bf k},{\bf 1},{\bf\bar N'})$
			\\ \cline{2-4}
			& Fermi & fermions & $({\bf 1},{\bf 1},{\bf 2},{\bf 1},{\bf k},{\bf 1},{\bf\bar N'})$
			\\ \hline
			\multirow{1}{*}{D4-D$4'$}& \multirow{1}{*}{Fermi} & \multirow{1}{*}{fermions} & $({\bf 1},{\bf 1},{\bf 1},{\bf 1},{\bf 1},{\bf \bar N},{\bf N'})$
			\\ \hline
		\end{tabular}
		\caption{The field content of the quantum mechanics. $U(k)=\widehat{G}$ is the D0 brane group, $U(N)=G$ is the D4 brane group and  $U(N')=G'$ is the D$4'$ brane group. $SO(4)=SU(2)_-\times SU(2)_+$, $SO(4)_R=SU(2)^R_-\times SU(2)^R_+$. In this paper, we set $N'=1$ throughout.  t-hyper denotes a $(0,4)$ twisted hypermultiplet. All multiplets in the second column are  $(0,4)$ multiplets, with the exception of the last row, which is an honest  $(0,2)$ Fermi multiplet; still, it can be made compatible with $(0,4)$ supersymmetry \cite{Tong:2014yna}, as is the case here.}
		\label{table:fieldcontentSUN}
	\end{table}
	
	The 1-loop determinants for the D0/D0 strings give
	\ie
	{Z}^{(k)}_{D0/D0}=\prod_{\substack{I\neq J\\ I,J=1}}^k\sh(\phi_I-\phi_J)\prod_{I, J=1}^k\frac{\sh(\phi_I-\phi_J+2\E_+)}{\sh(\phi_I-\phi_J+\E_+\pm\E_-)}\prod_{I, J=1}^k\frac{\sh(\phi_I-\phi_J\pm m-\E_-)}{\sh(\phi_I-\phi_J\pm m-\E_+)} \; .
	\fe
	
	The 1-loop determinants for the D0/D4 strings give
	\ie
	{Z}^{(k)}_{D0/D4}=\prod_{I=1}^k\prod_{i=1}^N\frac{\sh(\pm(\phi_I-a_i)+m)}{\sh(\pm(\phi_I-a_i)+\E_+)} \; .
	\fe

	In order to deduce the contribution of D0/D$4'$ strings, we can make use of a symmetry of the brane configuration \tabref{table:BranesSUN}. Namely, under the exchange of the coordinates $x_{1,2,3,4}  \leftrightarrow x_{5,6,7,8}$, D4 and D4' branes get swapped, while the D0 branes are unaffected.  Hence,  we can write the $D0/D4'$ contribution from the $D0/D4$ one, after simply exchanging $\E_1,\E_2$ with $\E_3,\E_4$, meaning
	\ie
	a_i\leftrightarrow M,\quad m\leftrightarrow \E_-,\quad \E_+\leftrightarrow -\E_+ \; .
	\fe
	We obtain
	\ie
	{Z}^{(k)}_{D0/D4'}=\prod_{I=1}^k\frac{\sh(\pm(\phi_I-M)+\E_-)}{\sh(\pm(\phi_I-M)-\E_+)} \; .
	\fe

	The quantization of the D4/D$4'$ string was first carried out in our context in \cite{Gomis:2006sb} (see also the work \cite{Banks:1997zs}, which considered a T-dual setup of D0/D8 branes). The corresponding multiplet is Fermi, in a bifundamental representation of $G\times G'$, and we obtain  
	\ie
	Z_{D4/D4'}=\prod_{i=1}^N\sh(M-a_i) \; .
	\fe
	
In the above, we defined $\sh(x)\equiv 2\sinh(x/2)$ and products over all signs inside an argument should be considered; for example, $\sh(a\pm b)\equiv\sh(a+b)\,\sh(a-b)$. 

\newpage

For completeness, let us also write the index in terms of field theory quantities, as $Z^{(k)}_{D0/D0}\cdot Z^{(k)}_{D0/D4} = Z^{(k)}_{vec}\cdot Z^{(k)}_{adjoint}$ and  $Z^{(k)}_{D0/D4'}\cdot Z_{D4/D4'} = Z^{(k)}_{defect}$. We obtain
	\begin{align}
	\label{5dintegralA1}
	& Z(M)  =\sum_{k=0}^{\infty}\;\frac{\widetilde{\fq}^{k}}{k!} \, \oint_{\mathcal{M}_k} \left[\frac{d\phi_I}{2\pi i}\right]Z^{(k)}_{vec}\cdot Z^{(k)}_{adjoint}\cdot  Z^{(k)}_{defect}(M)  \; , \\
	&Z^{(k)}_{vec} =\prod_{\substack{I\neq J\\ I,J=1}}^k\sh(\phi_I-\phi_J)\prod_{I, J=1}^k\frac{\sh(\phi_I-\phi_J+2\E_+)}{\sh(\phi_I-\phi_J+\E_+\pm\E_-)}\prod_{I=1}^k\prod_{i=1}^N\frac{1}{\sh(\pm(\phi_I-a_i)+\E_+)}\; ,\\
	&Z^{(k)}_{adjoint} =\prod_{I, J=1}^k\frac{\sh(\phi_I-\phi_J\pm m-\E_-)}{\sh(\phi_I-\phi_J\pm m-\E_+)}\prod_{I=1}^k\prod_{i=1}^N\sh(\pm(\phi_I-a_i)+m)\; ,\\
	&Z^{(k)}_{defect} =\prod_{i=1}^N\sh(M-a_i) \prod_{I=1}^k\frac{\sh(\pm(\phi_I-M)+\E_-)}{\sh(\pm(\phi_I-M)-\E_+)}\, .
	\end{align}
This identification has a potential caveat, though. The D0 brane  quantum mechanical index typically counts states that are present in the UV complete string theory but not necessarily part of the low energy QFT, which was pointed out in \cite{Hwang:2014uwa}. This manifests itself as extra spurious contributions to the index, which we denote collectively as $Z_{extra}$. We will need to normalize the partition function by this factor when it is nontrivial to get sensible results. We will be more explicit in what follows.\\

\vspace{10mm}
	
------\; {\bf Pure case}\; ------\\
	
	Let us fist study the case of a pure $SU(N)$ gauge theory. This can be done by decoupling the adjoint mass $m\rightarrow \infty$. In the present case, integration commutes with the limit (as long as we properly rescale the instanton counting parameter with $m$), so we can simply take the limit inside the integrand, which amounts to setting $Z^{(k)}_{adjoint}\rightarrow 1$:
	\begin{align}
	\label{5dintegralA1pure}
	& Z(M)  =\sum_{k=0}^{\infty}\;\frac{\widetilde{\fq}^{k}}{k!} \, \oint  \left[\frac{d\phi_I}{2\pi i}\right]Z^{(k)}_{vec}\cdot  Z^{(k)}_{defect}(M)  \; . 
	\end{align}
	
	In order to derive non-perturbative SD equations, we apply our program and  study the pole structure of \eqref{5dintegralA1pure} in the defect fugacity $M$, as we argued in detail in section \ref{ssec:Yoperator}. Namely, for a fixed instanton number $k$, let $\mathcal{M}_k$ be the set of poles selected by the JK-residue prescription in the defect partition function $Z(M)$. Meanwhile, let $\mathcal{M}^{pure}_k$ be the set poles selected by the JK-residue prescription in the partition function  $\langle 1 \rangle$, that is to say in the absence of the factor $Z^{(k)}_{defect}$. The set $\mathcal{M}_k$ happens to contain exactly one element depending on $M$, for all $k$. To see this, first note that $\mathcal{M}_k$  contains at least one element depending on $M$, since one of the $k$ contours, say the $I$-th one, is required to pick up exactly one pole coming from the defect factor $Z^{(k)}_{defect}$, namely
	\begin{equation}
	\label{onlyonepole}
	\phi_I-M-\epsilon_+=0 \;\;\mbox{for some} \;  I=1,\ldots, k \; .
	\end{equation}    
	In order to show that  $\mathcal{M}_k$  contains at most one element depending on $M$, one argues as follows: for some $1\leq I\leq k$, consider the pole at $\phi_I=M+\epsilon_+$. First, one could think there exists another pole at $\phi_J-M-\epsilon_+=0$, with  $J \neq I$. But that cannot be so, because the numerator has a zero $\phi_I - \phi_J=0$ at this locus. Second, there is another source of potential poles in $M$; indeed, the JK prescription requires us to include the hyperplanes $\phi_{J}-\phi_{I}+\epsilon_1=0$ and $\phi_{J}-\phi_{I}+\epsilon_2=0$. But the numerators of $Z^{(k)}_{defect}$ cancel these potential pole locations, so we have succeeded in showing that there is exactly one $M$-dependent pole in $\mathcal{M}_k$. Put differently,  $|\mathcal{M}_k|=|\mathcal{M}^{pure}_k|+1$, for all $k$. In conclusion, for a given $k$, we choose one of the contours to pick up the unique $M$-dependent pole in $\mathcal{M}_k$, namely \eqref{onlyonepole}, and the $k-1$ other poles are to be chosen in the set $\mathcal{M}^{pure}_{k-1}$\footnote{Though this fact is not needed in our analysis, we mention here in passing that the set of poles $\mathcal{M}^{pure}_k$ which do not depend on $M$ have a famous classification in terms of $N$-colored Young diagrams $\overrightarrow{\boldsymbol{\mu}}=\{\boldsymbol{\mu}_1, \boldsymbol{\mu}_2, \ldots, \boldsymbol{\mu}_N\}$ such that $\left|\overrightarrow{\boldsymbol{\mu}}\right|=k$. In practice, this assigns one Young tableau per $U(1)$ Coulomb modulus. Explicitly, the $k$ integration variables $\phi_1, \ldots, \phi_k$ are chosen such that
		\beq
		\label{A1youngtuples}
		\phi_I=a_i -\epsilon_+ -(s_1-1)\, \epsilon_1 - (s_2-1)\, \epsilon_2 , \text{with}\; (s_1, s_2)\in \boldsymbol{\mu}_i \; .
		\eeq
		See \cite{Nekrasov:2002qd} and \cite{Nekrasov:2003rj} for details.}
	Having analyzed the pole structure in the fugacity $M$, we are ready to derive SD equations. All we need is to define a Wilson loop operator vev:
	\begin{align}
	\label{YoperatorA1}
	\left\langle \left[Y(M)\right]^{\pm 1}\right\rangle =
	\sum_{k=0}^{\infty}\frac{\widetilde{\fq}^{k}}{k!} \, \oint_{\mathcal{M}^{pure}_k}  \left[\frac{d\phi_I}{2\pi i}\right]Z^{(k)}_{vec}\cdot \left[Z^{(k)}_{defect}(M)\right]^{\pm 1}\, . 
	\end{align}
	Crucially, we take the contour prescription in the definition of the $Y$-operator vev to only enclose poles in the set $\mathcal{M}^{pure}_k$, meaning it ignores the pole at  $\phi_I=M+\epsilon_+$.
	
	Then, we find that the partition function can be expressed in terms of these $Y$-operators, as a sum of exactly two terms:
	\begin{align}
	\label{A1pure}
	Z(M) = \left\langle Y(M) \right\rangle + \widetilde{\fq} \; \left\langle\frac{1}{Y(M+2\,\epsilon_+)}\right\rangle\; .
	\end{align}
	The meaning of this expression is as follows: The first term on the right-hand side encloses almost all the ``correct" poles in the partition function integrand, but we are missing exactly one: the extra pole at $\phi_I-M-\epsilon_+=0$. The second term on the right-hand side makes up for this missing pole, and relies on a key observation: one can trade a contour enclosing this extra pole for a contour which does not enclose it, at the expense of inserting the operator $Y(M + 2\, \epsilon_+)^{-1}$ inside the vev. This result is derived at once from the integral expression \eqref{5dintegralA1pure}, and the $Y$-operator integral definition \eqref{YoperatorA1}. Finally, note the presence of the parameter $\widetilde{\fq}$ in the second term; it counts exactly one instanton, to make up for the missing $M$-pole.

	After normalizing the partition function and expanding it in the (exponentiated) defect fugacity $e^{M}$, we find 
	\beq
	\label{SchwingerA1pure}
	\frac{1}{\left\langle 1 \right\rangle}\left[	\left\langle Y(M) \right\rangle + \widetilde{\fq} \; \left\langle\frac{1}{Y(M+2\,\epsilon_+)}\right\rangle\right]  = \cT(M; \{e_i^{\epsilon_1, \epsilon_2}\}) \; .
	\eeq
	Above, $\cT(M; \{e_i^{\epsilon_1, \epsilon_2}\})$ is a finite Laurent series in $e^M$.  To argue that this is the case, one can simply show that the left-hand side is pole-free in $M$; for details, see for instance \cite{Kim:2016qqs}. Furthermore, the asymptotics of $Z(M)$ at $M\rightarrow \pm\infty$  tell us that the number of terms in the $e^M$--series is $N+1$. Therefore, \eqref{SchwingerA1pure} can be thought of as non-perturbative SD equations for the $SU(N)$ gauge theory, solved explicitly by the $Y$-operator vev \eqref{YoperatorA1}. Taking the flat space limit $\epsilon_1, \epsilon_2 \rightarrow 0$, we recover the SW curve of 5d pure $SU(N)$ super Yang-Mills.\\
	
	Let us study the more familiar four-dimensional limit in detail. Namely, we reintroduce the radius $R$ explicitly in the partition function, and take the limit $R\rightarrow 0$, leaving all fugacities fixed in the process (essentially, all $\text{sinh}(x)$ functions simply become $x$, inside \eqref{YoperatorA1} and \eqref{5dintegralA1pure}). The SD equations become
	\beq
	\label{Schwinger4dA1pure}
	\frac{1}{\left\langle 1 \right\rangle_{4d}}\left[	\left\langle Y(M) \right\rangle_{4d} + \widetilde{\fq} \; \left\langle\frac{1}{Y(M+2\,\epsilon_+)}\right\rangle_{4d}\right] = {\cT}_{4d}(M; \{e_i^{\epsilon_1, \epsilon_2}\}) \, ,
	\eeq
	where
	\beq
	\label{RHSA1pure}
	{\cT}_{4d}(M; \{e_i^{\epsilon_1, \epsilon_2}\})=\prod_{i=1}^{N}(M-e_i^{\epsilon_1, \epsilon_2}) \; .
	\eeq 
	The roots $\{e_i^{\epsilon_1, \epsilon_2}\}$ characterize the vacuum, and are determined straight from the expansion of the left-hand side of \eqref{Schwinger4dA1pure} in the fugacity $M$.
	
	As an example, here is ${\cT}_{4d}$ when  $N=2$, meaning $G=SU(2)$, with Coulomb parameter $a\equiv a_1=-a_2$:
	\begin{align}
	\label{SUNpurepolyNis2}
	{\cT}_{4d}(M; \{e_i^{\epsilon_1, \epsilon_2}\}) = M^2 + d_{(0)}^{\epsilon_1, \epsilon_2}
	\end{align}
	with $d_{(0)}^{\epsilon_1, \epsilon_2} = \left(e_1^{\epsilon_1, \epsilon_2}\right)^2$, and
	\begin{align}
	\label{SUNpurepolycoef}
	d_{(0)}^{\epsilon_1, \epsilon_2} &= - a^2 - \widetilde{\fq}\,\frac{1}{2(a^2-\epsilon_+^2)}+\widetilde{\fq}^2\; \ldots
	\end{align}
	
	Taking the flat space limit $\epsilon_1, \epsilon_2 \rightarrow 0$ of the SD equations, we obtain at once
	\beq
	\label{SUNpureSW4d}
	y(M) + \frac{\widetilde{\fq}}{y(M)}  = \prod_{i=1}^{N}(M-e_i^{0,0})\, .
	\eeq
	The parameters $\{e_i^{0,0}\}$ are simply the $\epsilon_1, \epsilon_2\rightarrow 0$ limit of the roots \eqref{RHSA1pure}.  Rewriting the instanton counting parameter as the dynamical scale  $\widetilde{\fq}\equiv \Lambda^{2 N}$, we recover the familiar SW curve of 4d pure $SU(N)$:
	\beq
	\label{SUNSWpure}
	y(M) + \frac{\Lambda^{2 N}}{y(M)} = \prod_{i=1}^{N}(M-e_i^{0,0})\; .
	\eeq

	As an aside, it is straightforward to generalize our discussion to an arbitrary number $N'>1$ of D$4'$ branes. In this case, the JK residue prescription dictates that for each value of $\rho$ in the set $\{1, 2, \ldots, N'\}$, the contours of the partition function $Z(M_{\rho})$ have to enclose exactly one pole at
\begin{equation}
\phi_I-M_\rho-\epsilon_+=0 \;  , \;\;\; I=1, \ldots, k.
\end{equation}
$Z(M_{\rho})$ can then be expressed as a $qq$-character of the spin-$N'/2$ representation of the quantum affine algebra $U_\hbar(\widehat{A_1})$. From this character point of view, it is clear that only the spin-$1/2$ representation has a relation to Seiberg-Witten geometry.
Lastly, unlike the $N'=1$ case, this time around the normalized partition function is not a finite Laurent series in any of the exponentiated defect fugacities $e^{M_{\rho}}$.\\	
	
\vspace{10mm}

------\; {\bf Fundamental matter}\; ------\\
	
We can add $N_f$ fundamental hypermultiplets by introducing $N_f$ D8 branes in the 012345678-directions. This results in a new sector of D0/D8 strings, and consequently new fermionic zero modes arising from Fermi multiplets in the bifundamental representation of $U(k)\times U(N_f)$. We therefore need to modify the index to account for these new Fermi multiplets in the quantum mechanics:
	\begin{align}
	\label{5dintegralA1fund}
	& Z(M)  =\sum_{k=0}^{\infty}\;\frac{\widetilde{\fq}^{k}}{k!} \, \oint_{\mathcal{M}_k}  \left[\frac{d\phi_I}{2\pi i}\right]Z^{(k)}_{vec}\cdot Z^{(k)}_{fund}\cdot  Z^{(k)}_{defect}  \; , \\
	&Z^{(k)}_{fund} =\prod_{I=1}^{k} \prod_{d=1}^{N_f} \sh\left(\phi_I-m_d\right) \equiv\prod_{I=1}^{k} Q(\phi_I)\; ,\label{fundQA1}
	\end{align}
As we explained in section \ref{ss:existence}, we will only restrict ourselves to a  number of hypermultiplets where the 4d limit is a conformal or asymptotically free theory, and where $Z^{(k)}_{fund}$ does not introduce new poles at $\phi_I \rightarrow \infty$.\\

Then, the matter factor does not contribute new poles, and we compute the partition function to be
\begin{align}
\label{A1fund}
Z(M) = \left\langle Y(M) \right\rangle + \widetilde{\fq} \; Q(M+\epsilon_+) \left\langle\frac{1}{Y(M+2 \, \epsilon_+)}\right\rangle\; .
\end{align}
Notice the argument of the matter factor $Q(M+\epsilon_+)$ is precisely the locus of the $M$-dependent pole we did not consider in the first term.

	After normalizing the partition function and expanding it in the (exponentiated) defect fugacity $e^{M}$, we find the SD equation solved by the $Y$-operators:
	\beq
	\label{SchwingerA1fund}
	\frac{1}{\left\langle 1 \right\rangle}\left[\left\langle Y(M) \right\rangle + \widetilde{\fq} \;  Q(M+\epsilon_+) \left\langle\frac{1}{Y(M+2\,\epsilon_+)}\right\rangle\right]  = \cT(M; \{e_i^{\epsilon_1, \epsilon_2}\}) \; .
	\eeq
	Above, $\cT(M; \{e_i^{\epsilon_1, \epsilon_2}\})$ is a finite Laurent series in $e^M$, with $N+1$ terms.
	Taking the flat space limit $\epsilon_1, \epsilon_2 \rightarrow 0$, we recover the SW curve of 5d $SU(N)$ super Yang-Mills with fundamental matter.\\
	
	Let us study the more familiar four-dimensional limit in detail. Namely, we reintroduce the radius $R$ explicitly in the partition function, and take the limit $R\rightarrow 0$, leaving all fugacities fixed. The SD equations become
	\beq
	\label{Schwinger4dA1fund}
	\frac{1}{\left\langle 1 \right\rangle_{4d}}\left[	\left\langle Y(M) \right\rangle_{4d} + \widetilde{\fq} \; \prod_{d=1}^{N_f} \left(M+\E_+ -m_d\right) \left\langle\frac{1}{Y(M+2\,\epsilon_+)}\right\rangle_{4d}\right] = {\cT}_{4d}(M; \{e_i^{\epsilon_1, \epsilon_2}\}) \, ,
	\eeq
	where
	\beq
	\label{RHSA1fund}
	{\cT}_{4d}(M; \{e_i^{\epsilon_1, \epsilon_2}\})=\prod_{i=1}^{N}(M-e_i^{\epsilon_1, \epsilon_2}) \; .
	\eeq 
	
	As an example, we write ${\cT}_{4d}$ when  $N=2$ and $N_f=2$, meaning $G=SU(2)$ with 2 fundamental hypermultiplets. If the Coulomb parameter is denoted as $a\equiv a_1=-a_2$ and the two masses as $m_1$ and $m_2$, we get
	\begin{align}
	\label{SUNfundpolyNis2}
	{\cT}_{4d}(M; \{e_i^{\epsilon_1, \epsilon_2}\}) = M^2 + d_{(0)}^{\epsilon_1, \epsilon_2}
	\end{align}
	with $d_{(0)}^{\epsilon_1, \epsilon_2} = \left(e_1^{\epsilon_1, \epsilon_2}\right)^2$, and
	\begin{align}
	\label{SUNfundpolycoef}
	d_{(0)}^{\epsilon_1, \epsilon_2} &= - a^2 - \widetilde{\fq}\,\frac{1}{2\, m_1^{1/2}m_2^{1/2}(a^2-\epsilon_+^2)}+  \widetilde{\fq}^2\, \ldots
	\end{align}
	
	Taking the flat space limit $\epsilon_1, \epsilon_2 \rightarrow 0$ and  reintroducing the dynamical scale $\Lambda$, we recover the familiar SW curve of 4d $SU(N)$ with $N_f$ fundamental hypermultiplets:
	\beq
	\label{SUNSWfund}
	y(M) + \frac{\Lambda^{2 N-N_f}\prod_{d=1}^{N_f} \left(M-m_d\right)}{y(M)} = \prod_{i=1}^{N}(M-e_i^{0,0})\; .
	\eeq

\vspace{10mm}
	
------\; {\bf Adjoint matter}\; ------\\
	
We come back to our initial quantum mechanics index, \eqref{5dintegralA1}, which we rewrite here for convenience:
\beq
\label{5dintegralA1again}
Z(M)  =\sum_{k=0}^{\infty}\;\frac{\widetilde{\fq}^{k}}{k!} \, \oint_{\mathcal{M}_k}  \left[\frac{d\phi_I}{2\pi i}\right]Z^{(k)}_{vec}\cdot Z^{(k)}_{adjoint}\cdot  Z^{(k)}_{defect}(M)  \; .
\eeq
Then, we compute the partition function to be
\begin{align}
\label{A1adjoint}
Z(M) = &\left\langle Y(M) \right\rangle\\
&+\widetilde{\fq} \; c_{(1)}^{\epsilon_1, \epsilon_2} \left\langle\frac{Y(M+m+ \epsilon_+)\cdot Y(M-m + \epsilon_+)}{Y(M+2 \, \epsilon_+)} \right\rangle\nonumber\\
&+\widetilde{\fq}^2 \ldots \nonumber
\end{align}
The above ``$\ldots$" stands for an infinite series in the instanton counting parameter $\widetilde{\fq}$, because there are now an infinite number of new poles depending on the defect mass $M$. Namely, each term corresponds to an element of the set $\mathcal{M}_k\setminus \mathcal{M}^{pure}_k$. In particular, the $k=1$ term stands for the residue at the pole $\phi_I=M+\epsilon_+$, the only element of $\mathcal{M}_1\setminus \mathcal{M}^{pure}_1$. It turns out that the infinite sum can  be written combinatorially as a sum over partitions, but we will not need this fact here, and refer instead to \cite{Nekrasov:2015wsu, Kim:2016qqs, Agarwal:2018tso} for details.  The coefficient $c_{(1)}^{\epsilon_1, \epsilon_2}$ is a function of the adjoint mass $m$, the defect fugacity $M$, and the $\Omega$-background parameters only. We will give precise expressions in four-dimensional variables momentarily.\\

After normalizing the partition function and expanding it in the (exponentiated) defect fugacity $e^{M}$, we find the SD equation solved by the $Y$-operators:
\beq
\label{SchwingerA1adjoint}
\frac{1}{\widetilde{\left\langle 1 \right\rangle}}\left[\left\langle Y(M) \right\rangle
+\widetilde{\fq} \; c_{(1)}^{\epsilon_1, \epsilon_2} \left\langle\frac{Y(M+m+ \epsilon_+)\cdot Y(M-m + \epsilon_+)}{Y(M+2 \, \epsilon_+)} \right\rangle
+\widetilde{\fq}^2 \ldots \right]  = \cT(M; \{e_i^{\epsilon_1, \epsilon_2}\}) \; .
\eeq
Above, $\cT(M; \{e_i^{\epsilon_1, \epsilon_2}\})$ is a finite Laurent series in $e^M$, with $N+1$ terms. The normalization is delicate here, as the index is counting extra states in the UV not present in the QFT, $Z_{extra}\neq 1$.  Namely, we have defined $\widetilde{\left\langle 1 \right\rangle}\equiv \left\langle 1 \right\rangle\cdot Z_{extra}$, where $\left\langle 1 \right\rangle=\sum_{k=0}^{\infty} \widetilde{\fq}^k \oint_{\mathcal{M}_k} Z^{(k)}_{D0/D0} Z^{(k)}_{D0/D4}$ is the usual partition function in the absence of defect, and $Z_{extra}\equiv\sum_{k=0}^{\infty} \widetilde{\fq}^k \oint_{\mathcal{M}_k} Z^{(k)}_{D0/D0} Z^{(k)}_{D0/D4'}$.
Taking the flat space limit $\epsilon_1, \epsilon_2 \rightarrow 0$, the above equation describes the SW geometry of 5d $SU(N)$ super Yang-Mills with adjoint matter.\\

Let us study the more familiar four-dimensional limit in detail. Namely, we reintroduce the radius $R$ explicitly in the partition function, and take the limit $R\rightarrow 0$, leaving all fugacities fixed. The SD equations become
\beq
\label{Schwinger4dA1adjoint}
	\resizebox{.96\hsize}{!}{$\frac{1}{\widetilde{\left\langle 1 \right\rangle}_{4d}}\left[\left\langle Y(M) \right\rangle_{4d}
+\widetilde{\fq} \; c_{(1), 4d}^{\epsilon_1, \epsilon_2} \left\langle\frac{Y(M+m+ \epsilon_+)\cdot Y(M-m + \epsilon_+)}{Y(M+2 \, \epsilon_+)} \right\rangle_{4d}
+\widetilde{\fq}^2 \ldots \right]  = \cT_{4d}(M; \{e_i^{\epsilon_1, \epsilon_2}\}) \, ,$}
\eeq
where 
\beq
c_{(1), 4d}^{\epsilon_1, \epsilon_2}=\frac{(m-\epsilon_-)(m+\epsilon_-)}{(m-\epsilon_+)(m+\epsilon_+)}
\eeq
and
\beq
\label{RHSA1adjoint}
{\cT}_{4d}(M; \{e_i^{\epsilon_1, \epsilon_2}\})=\prod_{i=1}^{N}(M-e_i^{\epsilon_1, \epsilon_2}) \; .
\eeq 

As an example, let us write ${\cT}_{4d}$ when  $N=2$, meaning $G=SU(2)$. We find
\begin{align}
\label{SUNadjointpolyNis2}
{\cT}_{4d}(M; \{e_i^{\epsilon_1, \epsilon_2}\}) = M^2 + d_{(0)}^{\epsilon_1, \epsilon_2}
\end{align}
with $d_{(0)}^{\epsilon_1, \epsilon_2} = \left(e_1^{\epsilon_1, \epsilon_2}\right)^2$, and
\begin{align}
\label{SUNadjointpolycoef}
d_{(0)}^{\epsilon_1, \epsilon_2} &= - a^2 - \widetilde{\fq}\,\frac{(m-\epsilon_-)(m+\epsilon_-)(m-\epsilon_+)(m+\epsilon_+)}{2\, (a^2-\epsilon_+^2)}+ \widetilde{\fq}^2 \, \ldots
\end{align}

Taking the flat space limit $\epsilon_1, \epsilon_2 \rightarrow 0$, we recover the SW geometry of 4d $SU(N)$ with adjoint matter (see for instance section 5 of \cite{DHoker:1997hut}):
\beq
\label{SUNSWadjoint}
y(M)  +\widetilde{\fq}\, \frac{\prod_{i=1}^{N} (M-e^{0,0}_i+m)(M-e^{0,0}_i-m)}{y(M)}+\ldots = \prod_{i=1}^{N}(M-e_i^{0,0})\; .
\eeq

Having reviewed what has been studied in the literature, we will now apply our techniques to study new matter content and gauge groups.\\

\vspace{10mm}

------\; {\bf Symmetric matter}\; ------\\

Here, we provide a definition of the Witten index that does not originate from a D0 brane quantum mechanics in type IIA. Namely, we introduce symmetric matter following the field theory analysis performed in \cite{Shadchin:2004yx}. We propose the following index:

\begin{align}
\label{5dintegralA1symmetric}
	& Z(M)  =\sum_{k=0}^{\infty}\;\frac{\widetilde{\fq}^{k}}{k!} \, \oint_{\mathcal{M}_k}  \left[\frac{d\phi_I}{2\pi i}\right]Z^{(k)}_{vec}\cdot Z^{(k)}_{sym}\cdot  Z^{(k)}_{defect}  \; , \\
&	Z^{(k)}_{sym} =\prod_{I=1}^{k} \sh(2\phi_I+m-\epsilon_+\pm \epsilon_-) \prod_{i=1}^{N}\sh(\phi_I+a_i+m-\epsilon_+)\;\times\\ &\qquad\qquad\times\prod_{J>I}^{k}\frac{\sh(\phi_I+\phi_J+m-\epsilon_+\pm\epsilon_-)}{\sh(\phi_I+\phi_J+m)\sh(\phi_I+\phi_J+m-2\epsilon_+)}\; ,\nonumber
\end{align}
Then, we compute the partition function and find
\begin{align}
\label{A1symmetric}
Z(M) = &\left\langle Y(M) \right\rangle\\
&+\widetilde{\fq} \; c_{(1)}^{\epsilon_1, \epsilon_2}\prod_{i=1}^{N}\sh(M+m+a_i) \left\langle\frac{1}{Y(M+2 \, \epsilon_+)} \right\rangle\nonumber\\
&+\widetilde{\fq}^2 \; \bigg[c_{(2,1)}^{\epsilon_1, \epsilon_2}\left\langle\frac{Y(M)}{Y(M-2 \, \epsilon_+)\,Y(-M-m)}\right\rangle +c_{(2,2)}^{\epsilon_1, \epsilon_2}\left\langle\frac{1}{Y(-M-m-2\epsilon_+)} \right\rangle\nonumber\\
&\qquad\;\;\;+c_{(2,3)}^{\epsilon_1, \epsilon_2}\left\langle\frac{1}{Y(-M-m+4\epsilon_+)} \right\rangle 
+c_{(2,4)}^{\epsilon_1, \epsilon_2}\left\langle\frac{Y(M)}{Y(M+2 \, \epsilon_+)\,Y(-M-m+2\epsilon_+)} \right\rangle \bigg]\nonumber\\
&+\widetilde{\fq}^3 \; \ldots\nonumber
\end{align}
The above ``$\ldots$" is an infinite series in the instanton counting parameter $\widetilde{\fq}$. As usual, each term represents an element of the set $\mathcal{M}_k\setminus \mathcal{M}^{pure}_k$. Following the JK residue prescription, here are the first poles: 

The $k=1$ term is the residue at $\phi_I=M+\epsilon_+$. 

The $(2,1)$ term is the residue at  $\phi_I= -\phi_J-m ,\; \phi_J= -M-m+\epsilon_+$. 

The $(2,2)$  term is the residue at $\phi_I= -\phi_J-m ,\; \phi_J= M+\epsilon_+$. 

The $(2,3)$  term is the residue at  $\phi_I= -\phi_J-m+2\epsilon_+ ,\; \phi_J= -M-m+3\epsilon_+$. 

The $(2,4)$  term is the residue at $\phi_I= -\phi_J-m+2\epsilon_+  ,\; \phi_J= M+\epsilon_+$. 

This is a priori puzzling because the SW curve of $SU(N)$ with symmetric matter is a cubic curve. As we will see momentarily, the resolution of this paradox is that an infinite number of terms will elegantly  cancel out against each other once we turn off the $\Omega$-background.\\

After normalizing the partition function and expanding it in the (exponentiated) defect fugacity $e^{M}$, we find the SD equation solved by the $Y$-operators:
\beq
\label{SchwingerA1symmetric}
\frac{1}{\widetilde{\left\langle 1 \right\rangle}}\left[\left\langle Y(M) \right\rangle
+\widetilde{\fq} \; \ldots\right]  = \cT(M; \{e_i^{\epsilon_1, \epsilon_2}\}) \; .
\eeq
Above, $\cT(M; \{e_i^{\epsilon_1, \epsilon_2}\})$ is a finite Laurent series in $e^M$, with $N+1$ terms.\\

Let us study the more familiar four-dimensional limit in detail. Namely, we reintroduce the radius $R$ explicitly in the partition function, and take the limit $R\rightarrow 0$, leaving all fugacities fixed. The SD equations become
\beq
\label{Schwinger4dA1symmetric}
\frac{1}{\widetilde{\left\langle 1 \right\rangle}_{4d}}\left[\left\langle Y(M) \right\rangle_{4d}
+\widetilde{\fq} \; \ldots\right]  = \cT_{4d}(M; \{e_i^{\epsilon_1, \epsilon_2}\}) \, .
\eeq
The coefficients in \eqref{A1symmetric} become
\begin{align}
\label{SUNsymmetriccoefs}
c_{(1), 4d}^{\epsilon_1, \epsilon_2} &= (2M+m+\epsilon_+\pm\epsilon_-) \\
c_{(2,1), 4d}^{\epsilon_1, \epsilon_2} &= \frac{-1}{2}(2M+m-2\epsilon_+)(2M+m-4\epsilon_+) \nonumber\\
c_{(2,2), 4d}^{\epsilon_1, \epsilon_2} &=\frac{-1}{2}(2M+m+\epsilon_+\pm\epsilon_-)(2M+m+2\epsilon_+) (2M+m+4\epsilon_+) \nonumber\\
c_{(2,3), 4d}^{\epsilon_1, \epsilon_2} &=\frac{1}{2}(2M+m-3\epsilon_+\pm\epsilon_-)(2M+m-4\epsilon_+) (2M+m-6\epsilon_+) \nonumber\\
c_{(2,4), 4d}^{\epsilon_1, \epsilon_2} &=\frac{1}{2}(2M+m-3\epsilon_+\pm\epsilon_-)(2M+m-4\epsilon_+) (2M+m-6\epsilon_+) \nonumber
\end{align}
and the right-hand side becomes
\beq
\label{RHSA1symmetric}
{\cT}_{4d}(M; \{e_i^{\epsilon_1, \epsilon_2}\})=\prod_{i=1}^{N}(M-e_i^{\epsilon_1, \epsilon_2}) \; .
\eeq 

Taking  the flat space limit $\epsilon_1, \epsilon_2 \rightarrow 0$, we see some remarkable simplifications: 
\begin{align}
\label{SUNsymmetricSW}
Z(M) = &\left\langle Y(M) \right\rangle\\
&+\widetilde{\fq} \; c_{(1), 4d}^{0,0}\; {\cT}_{4d}(-M-m) \left\langle\frac{1}{Y(M)} \right\rangle_{4d}\nonumber\\
&+\widetilde{\fq}^2 \left[c_{(2,1), 4d}^{0,0}+c_{(2,2), 4d}^{0,0}+c_{(2,3), 4d}^{0,0}+c_{(2,4), 4d}^{0,0}\right] \left\langle\frac{1}{Y(-M)} \right\rangle_{4d}\nonumber\\
&+ \widetilde{\fq}^3 \left[c_{(3,1), 4d}^{0,0}+c_{(3,2), 4d}^{0,0}+c_{(3,3), 4d}^{0,0}+c_{(3,4), 4d}^{0,0}+c_{(3,5), 4d}^{0,0}+c_{(3,6), 4d}^{0,0}\right] \left\langle\frac{1}{Y(M)^2} \right\rangle_{4d}\nonumber\\
&+\ldots\nonumber
\end{align}
Explicitly, we find 
\begin{align}
\label{SUNsymmetriccoefsSW}
& c_{(1), 4d}^{0,0}  =  (2M+m)^2\\
& c_{(2,1), 4d}^{0,0} = c_{(2,2), 4d}^{0,0} =  \frac{-(2M+m)^4}{2}  \; ,\;\;\;\;  c_{(2,3), 4d}^{0,0} = c_{(2,4), 4d}^{0,0} =  \frac{+(2M+m)^4}{2}\; .\nonumber
\end{align}
All terms at order 2 cancel out. We further find that six terms at order 3 survive the limit. Those are the ones corresponding to the following JK-poles:
\begin{align}
\label{JKpolesorder3SUNsym}
 \phi_I&=M-\E_- -2\E_+ \, ,  &\phi_J&=M-\E_+  \, , &\phi_K&=-M-m+\E_+   \\
 \phi_I&=M+\E_- -2\E_+ \, , &\phi_J&=M-\E_+ \, , &\phi_K&=-M-m+\E_+  \nonumber \\
 \phi_I&=M-\E_+ \, ,  &\phi_J&=M-\E_- -2\E_+  \, , &\phi_K&=-M-m+\E_+   \nonumber\\
 \phi_I&=M-\E_+ \, ,  &\phi_J&=M+\E_- -2\E_+  \, ,  &\phi_K&=-M-m+\E_+  \nonumber \\
 \phi_I&=M+\E_- +2\E_+ \, ,   &\phi_J&=-M-m-\E_- -2\E_+  \, ,  &\phi_K&=M+\E_+  \nonumber \\
\phi_I&=M-\E_- +2\E_+ \, ,   &\phi_J&=-M-m+\E_- -2\E_+  \, ,  &\phi_K&=M+\E_+  \; . \nonumber
\end{align}
The sum of the corresponding coefficients at order 3 gives
\beq
\label{SUNsymmetriccoefsSWorder3}
\sum_{p=1}^6 c_{(3,p), 4d}^{0,0} = \frac{+(2M+m)^6}{2} \; .
\eeq
Looking at instanton corrections beyond order 3, we found numerically that all terms cancel in a similar fashion to the terms at order 2. We conjecture that this is a generic feature of all higher instantons contributions. We therefore conjecture the following flat space limit for the SD equations \eqref{Schwinger4dA1symmetric}:
\begin{align}
\label{SUNsymmetricSW2}
&\left\langle Y(M) \right\rangle +\widetilde{\fq} \;(2M+m)^2\; {\cT}_{4d}(-M-m; \{e_i^{0,0}\}) \left\langle\frac{1}{Y(M)} \right\rangle_{4d}\\
&\qquad\qquad\;\;\; + \widetilde{\fq}^3 \; \frac{(2M+m)^6}{2} \left\langle\frac{1}{Y(M)^2} \right\rangle_{4d} = {\cT}_{4d}(M ; \{e_i^{0,0}\})\; .\nonumber
\end{align}
After introducing the dynamical scale $\widetilde{\fq}\equiv \Lambda^{N-2}$, multiplying both sides by $y(M)^2$ and rescaling,  this is precisely the SW curve of $T^{4d}$ (see for instance \cite{Ennes:1998ve,Ennes:1998gh} or more recently \cite{Shadchin:2005cc}):

\begin{align}
\label{SUNSWsymmetric}
&y^3(M) \;+\; {\cT}_{4d}(M ; \{e_i^{0,0}\})\, y^2(M)\\
&  \qquad\;\;\;\; +\;\Lambda^{N-2} \; \left(M+\frac{m}{2}\right)^2\; {\cT}_{4d}(-M-m; \{e_i^{0,0}\})\, y(M) \; +\; \Lambda^{3N-6} \; \left(M+\frac{m}{2}\right)^6=0 \, .\nonumber
\end{align}

\vspace{10mm}

------\; {\bf Symmetric and Fundamental matter}\; ------\\

As another example, we consider a $SU(N)$ gauge theory with both a symmetric  and $N_f$ fundamental hypermultiplets together. Deriving the quantized SW geometry simply amounts to performing the JK-residue prescription on the following integral:  
\begin{align}
\label{5dintegralA1symmetricandfund}
& Z(M)  =\sum_{k=0}^{\infty}\;\frac{\widetilde{\fq}^{k}}{k!} \, \oint_{\mathcal{M}_k} \left[\frac{d\phi_I}{2\pi i}\right]Z^{(k)}_{vec}\cdot Z^{(k)}_{sym}\cdot  Z^{(k)}_{fund}\cdot  Z^{(k)}_{defect}  \; .
\end{align}
Evaluating the integrals, we find once again an infinite $\widetilde{\fq}$-series in defect $Y$-operator vevs:
\begin{align}
\label{A1symmetricplusfund}
Z(M) = &\left\langle Y(M) \right\rangle\\
&+\widetilde{\fq} \; c_{(1)}^{\epsilon_1, \epsilon_2}\, Q(M+\epsilon_+)\prod_{i=1}^{N}\sh(M+m+a_i) \left\langle\frac{1}{Y(M+2 \, \epsilon_+)} \right\rangle\nonumber\\
&+\widetilde{\fq}^2 \; \ldots\nonumber
\end{align}
The various coefficients $c_{(i)}^{\epsilon_1, \epsilon_2}$ are the same as we found before, in the case with symmetric matter only. We will not repeat here the full analysis, as it is identical to that case.  We will however deduce the SW geometry directly from the observation that the fundamental matter factors do not contribute new poles. Therefore, after evaluation of the residues, the argument of these factors will simply record the various $M$-dependent poles we enclosed. For instance, note the presence of the fundamental matter factor $Q(M+\epsilon_+)$ at $k=1$, signaling that the $k=1$ term has a pole at $\phi_I=M+\epsilon_+$.

Then, to deduce the SW curve of the theory, we simply have to keep track of the poles picked up by JK at order 1 and at order 3, and take the flat space limit. The $k=3$ poles were recorded above \eqref{JKpolesorder3SUNsym}.  Every set of pole there is of the form $\phi_I= M+\ldots$, $\phi_J= M+\ldots$, and $\phi_K= -M-m+\ldots$. Remarkably, this is exactly the required pole structure to produce  the SW curve of the corresponding four-dimensional theory:
\begin{align}
\label{SUNSWsymmetricandfund}
&y^3(M) + {\cT}_{4d}(M ; \{e_i^{0,0}\})\, y^2(M)  +\Lambda^{N-2} \; \prod_{d=1}^{N_f}\left(M-m_d\right) \left(M + \frac{m}{2}\right)^2\; {\cT}_{4d}(-M-m; \{e_i^{0,0}\})\, y(M)\nonumber\\
&\qquad\;\;\;\;  + \Lambda^{3N-6} \; \left[\prod_{d=1}^{N_f}\left(M-m_d\right)\right]^2\prod_{d=1}^{N_f}\left(-M-m-m_d\right) \left(M+\frac{m}{2}\right)^6=0 \, .
\end{align}

\vspace{10mm}
	
------\; {\bf Antisymmetric matter}\; ------\\
	
As a final example, We briefly mention here how to proceed with antisymmetric matter. The field theory ADHM analysis in the absence of defects was carried out in \cite{Shadchin:2004yx}. We will consider the 5d uplift in the partition function integrand:
\begin{align}
\label{5dintegralA1antisymmetric}
& Z(M)  =\sum_{k=0}^{\infty}\;\frac{\widetilde{\fq}^{k}}{k!} \, \oint_{\mathcal{M}_k}  \left[\frac{d\phi_I}{2\pi i}\right]Z^{(k)}_{vec}\cdot Z^{(k)}_{antisym}\cdot  Z^{(k)}_{defect}  \; , \\
&Z^{(k)}_{antisym} =\prod_{I=1}^{k} \frac{1}{\sh(2\phi_I+m)\sh(2\phi_I+m-2\epsilon_+)} \prod_{i=1}^{N}\sh(\phi_I+a_i+m-\epsilon_+)\;\times\\ &\qquad\;\;\;\times\prod_{J>I}^{k}\frac{\sh(\phi_I+\phi_J+m-\epsilon_+\pm\epsilon_-)}{\sh(\phi_I+\phi_J+m)\sh(\phi_I+\phi_J+m-2\epsilon_+)}\; ,\nonumber
\end{align}	
The partition function is again an infinite	$\widetilde{\fq}$-series when expressed in terms of defect $Y$-operator vevs:
\begin{align}
\label{A1antisymmetric}
Z(M) = &\left\langle Y(M) \right\rangle\\
&+\widetilde{\fq} \; c_{(1)}^{\epsilon_1, \epsilon_2}\prod_{i=1}^{N}\sh(M+m+a_i) \left\langle\frac{1}{Y(M+2 \, \epsilon_+)} \right\rangle\nonumber\\
&+\widetilde{\fq}^2 \; \ldots\nonumber
	\end{align}
The $k=1$ term originates from the pole at $\phi_I=M+\E_+$. In the 4d limit, the corresponding coefficient is given by
\beq
\label{SUNantisymmetriccoefs}
c_{(1), 4d}^{\epsilon_1, \epsilon_2} = \frac{1}{(2M+m) (2M+m+2\epsilon_+)} \; .
\eeq
Taking the flat space limit, this is indeed the one-instanton correction to $SU(N)$ with antisymmetric matter as obtained from usual SW theory \cite{Naculich:1998rh}. In order to obtain the quantized geometry, we simply expand \eqref{A1antisymmetric} to arbitrarily high order, following the JK-residue prescription. We then take the four-dimensional and flat space limits. An infinite number of terms should disappear in the limit, to yield a cubic SW curve in the end. We will skip the details here as the analysis is similar to the previous example.

\vspace{5mm}

\subsection{The $SO(2 N)$ Gauge Theory}

We now construct the quantized SW geometry of gauge theories with other classical groups. The ADHM D0-brane quantum mechanics of such theories has been worked out in \cite{Hwang:2014uwa} (see also the field theory perspective of \cite{Nekrasov:2004vw}, and \cite{Marino:2004cn} in a four-dimensional setup). The challenge is twofold: first, is it possible to incorporate a Wilson loop using branes? Second, how can we make sense of the D0 brane quantum mechanics in field theory terms? When $G=SO(2N)$ or $G=Sp(N)$, we claim that we are able to construct such a defect using O$8$ orientifold planes, by making use of a particular symmetry of the brane setup. We further claim that we will be able to derive non-perturbative SD equations for the low energy gauge theories, and further write down the SW geometry after taking the flat space limit.\\

Our starting point is the same setup we considered in the $SU(N)$ case, but with the addition of an O$8^+$ plane, as follows: 
\begin{table}[H]\centering
	\begin{tabular}{|l|cccccccccc|} \hline
		&  0 &  1 &  2 &  3 &  4 &  5 &  6 &  7 &  8 &  9 \\
		\hline\hline
		\;\;O$8^+$ &\xTB& \xTB  &\xTB   & \xTB  & \xTB  &  \xTB  &  \xTB  & \xTB   & \xTB   &    \\
		\hline
		$k$\,\,	D0 &\xTB&   &   &   &   &    &    &    &    &    \\
		\hline
		$N$	D4 &\xTB&\xTB&\xTB&\xTB&\xTB&    &    &    &    &     \\
		\hline
		$1$\,\, D$4'$ &\xTB&    &    &    &    &\xTB&\xTB&\xTB&\xTB&  \\
		\hline
		\;	F1 &\xTB&    &    &    &    & & & & & \xTB  \\
		\hline
	\end{tabular}
	\caption{The directions of the various branes.}
	\label{table:BranesSO2N}
\end{table}

The effective theory on the D4 branes is a 5d $SO(2N)$ gauge theory on $S^1(R)\times\mathbb{R}^4$, called $T^{5d}$. 
The D$4'$ brane realizes a 1/2-BPS  Wilson loop with symmetry group  $SO(2)$, wrapping the $x_0$-circle and which sits at the origin of $\mathbb{R}^4$. Pulling the  D$4'$ brane a distance $M$ away in the $x_9$-direction, there are now open strings with nonzero tension between the D4 and D$4'$ branes. These are heavy fermion probes, with mass proportional to the distance $M$. The gauge group on the D0 branes is $\widehat{G}=Sp(k)$. We put the theory on the 5d $\Omega$-background $S^1(R)\times\mathbb{R}^4_{\epsilon_1, \epsilon_2}$. Correspondingly, we introduce the same fugacities as in the $SU(N)$ case:
\ie
\epsilon_+\equiv{\epsilon_1 + \epsilon_2\over 2},~~~ \epsilon_-\equiv{\epsilon_1 - \epsilon_2\over 2},
~~~ m \equiv {\epsilon_3 - \epsilon_4\over 2},
\fe
where $\epsilon_1$, $\epsilon_2$, $\epsilon_3$, and $\epsilon_4$ are the chemical potentials associated to the rotations of the $\bR^2_{\text{\tiny{12}}}$, $\bR^2_{\text{\tiny{34}}}$, $\bR^2_{\text{\tiny{56}}}$, and $\bR^2_{\text{\tiny{78}}}$ planes, respectively.\\

Before we go further, it is important to point out that the  O$8^+$ plane has introduced a Romans mass in type IIA.  This will affect the D4 brane theory by introducing an effective Chern-Simons term interaction. As a result, the $U(1)$ topological charge, aka the instanton number $k$, receives an anomalous contribution and gets shifted. Thankfully, the quantum mechanics index turns out not to be sensitive to this shift, so we will safely proceed. Note that no CFT is expected to exist in the UV in this construction.\\

The instanton partition function is the index of the D0 brane quantum mechanics. In our context, it reads	
\begin{align}
\label{5dintegralSO2N}
&Z(M) = Z_{D4/D4'}(M)\sum_{k=0}^{\infty}\; \frac{\widetilde{\fq}^k}{|W(\widehat{G})|} \oint_{\mathcal{M}_k}  \left[ \prod_{I=1}^{k}\frac{d\phi_I}{2\pi i}\right]Z^{(k)}_{D0/D0}\cdot Z^{(k)}_{D0/D4}\cdot Z^{(k)}_{D0/D4'}(M)  \; .
\end{align}
For the sake of brevity, we only made the dependence on the defect fugacity $M$ explicit in writing the partition function $Z(M)$.
$|W|$ is the order of the Weyl group of $\widehat{G}=Sp(k)$, and the various factors are given by the 1-loop determinants of the quantum mechanical fields, coming from the quantization of the strings stretching between the various branes.

\newpage

The field content of the quantum mechanics on the D0 branes is:
\begin{table}[H]\centering
	\begin{tabular}{|c|c|c|c|}
		\hline
		strings &Multiplets & fields & {\footnotesize $SO(4)\times SO(4)_R\times Sp(k)\times SO(2N)\times SO(2N')$ }%& O($k$)$\times$Sp(1)$\times$ SO($2N_f$)
		\\ \hline
		\multirow{8}{*}{D0-D0}& \multirow{3}{*}{vector} & gauge field &$({\bf 1},{\bf 1},{\bf 1},{\bf 1},{\bf adj},{\bf 1},{\bf 1})$
		\\ \cline{3-4}
		& & scalar & $({\bf 1},{\bf 1},{\bf 1},{\bf 1},{\bf adj},{\bf 1},{\bf 1})$
		\\ \cline{3-4}
		&  & fermions &$({\bf 1},{\bf 2},{\bf 1},{\bf 2},{\bf adj},{\bf 1},{\bf 1})$
		\\ \cline{2-4}
		&Fermi & fermions & $({\bf 2},{\bf 1},{\bf 2},{\bf 1},{\bf adj},{\bf 1},{\bf 1})$
		\\ \cline{2-4}
		&\multirow{2}{*}{t-hyper}  & scalars & $({\bf 1},{\bf 1},{\bf 2},{\bf 2},{\bf \frac{k(k-1)}{2}},{\bf 1},{\bf 1})$
		\\ \cline{3-4}
		&& fermions & $({\bf 1},{\bf 2},{\bf 2},{\bf 1},{\bf \frac{k(k-1)}{2}},{\bf 1},{\bf 1})$
		\\ \cline{2-4}
		&\multirow{2}{*}{hyper} & scalars & $({\bf 2},{\bf 2},{\bf 1},{\bf 1},{\bf \frac{k(k-1)}{2}},{\bf 1},{\bf 1})$
		\\ \cline{3-4}
		&& fermions & $({\bf 2},{\bf 1},{\bf 1},{\bf 2},{\bf \frac{k(k-1)}{2}},{\bf 1},{\bf 1})$
		\\ \hline
		\multirow{3}{*}{D0-D4}& \multirow{2}{*}{hyper} & scalars & $({\bf 2},{\bf 1},{\bf 1},{\bf 1},{\bf 2k},{\bf \overline{ 2N}},{\bf 1})$
		\\ \cline{3-4}
		& & fermions & $({\bf 1},{\bf 1},{\bf 1},{\bf 2},{\bf 2k},{\bf \overline{ 2N}},{\bf 1})$
		\\ \cline{2-4}
		& Fermi & fermions & $({\bf 1},{\bf 1},{\bf 2},{\bf 1},{\bf 2k}{\bf \overline{ 2N}},{\bf 1})$
		\\ \hline
		\multirow{3}{*}{D0-D$4'$}& \multirow{2}{*}{t-hyper} & scalars & $({\bf 2},{\bf 1},{\bf 1},{\bf 1},{\bf 2k},{\bf 1},{\bf \overline{2 N'}})$
		\\ \cline{3-4}
		& & fermions & $({\bf 1},{\bf 1},{\bf 1},{\bf 2},{\bf 2k},{\bf 1},{\bf \overline{2 N'}})$
		\\ \cline{2-4}
		& Fermi & fermions & $({\bf 1},{\bf 1},{\bf 2},{\bf 1},{\bf 2k},{\bf 1},{\bf \overline{2 N'}})$
		\\ \hline
		\multirow{1}{*}{D4-D$4'$}& \multirow{1}{*}{Fermi} & \multirow{1}{*}{fermions} & $({\bf 1},{\bf 1},{\bf 1},{\bf 1},{\bf 1},{\bf \overline{2 N}},{\bf 2 N'})$
		\\ \hline
	\end{tabular}
	\caption{The field content of the quantum mechanics. $Sp(k)=\widehat{G}$ is the D0 brane group, $SO(2N)=G$ is the D4 brane group and  $SO(2N')=G'$ is the D$4'$ brane group. $SO(4)=SU(2)_-\times SU(2)_+$, $SO(4)_R=SU(2)^R_-\times SU(2)^R_+$. Of interest to us is the case of a single D$4'$ brane, $N'=1$.  t-hyper denotes a $(0,4)$ twisted hypermultiplet.}
	\label{table:fieldcontentSO2N}
\end{table}

The 1-loop determinants for the D0/D0 strings give
\ie
\hspace*{-1cm}{Z}_{D0/D0}^{k}=&  \prod_{I=1}^k\sh\left(\pm\phi_I\right)\sh\left(\pm\phi_I+\E_+\right)\,\prod_{I < J}^{k}\frac{\sh\left(\pm \phi_I \pm \phi_J\right) \;\sh\left(\pm \phi_{I} \pm \phi_{J} + 2\epsilon_+\right)}{\sh\left(\pm \phi_{I} \pm \phi_{J} \pm \epsilon_- + \epsilon_+\right)}\\
&\times\prod^k_{I=1}\frac{1}{\sh\left(\pm\phi_I\pm m-\E_+\right)}\prod^k_{I<J}\frac{\sh\left(\pm\phi_I\pm\phi_J\pm m-\E_-\right)}{\sh\left(\pm\phi_I\pm\phi_J\pm m-\E_+\right)}.
\fe

The 1-loop determinants for the D0/D4 strings give
\ie
\hspace*{-1cm}{Z}_{D0/D4}^{k}=\prod^k_{I=1}\prod^{N}_{i=1}\frac{\sh\left(\pm\phi_I\pm a_i+m\right)}{\sh\left(\pm\phi_I \pm a_i+\E_+\right)}
\fe

The hard part is to find the contribution of D0/D$4'$ strings, but thanks to a symmetry of our brane configuration \tabref{table:BranesSO2N}, we do not actually have to compute anything else. The symmetry in question is the same one we used in the $SU(N)$ analysis. Namely, under the exchange of the coordinates $x_{1,2,3,4}  \leftrightarrow x_{5,6,7,8}$, D4 and D$4'$ branes get swapped, while the D0 branes and O8 plane remain invariant. Hence,  we can write the D0/D$4'$ contribution from the D0/D4 one, after simply exchanging $\E_1,\E_2$ with $\E_3,\E_4$. This translates to
\ie
a_i\leftrightarrow M,\quad m\leftrightarrow \E_-,\quad \E_+\leftrightarrow -\E_+ \; ,
\fe
and we obtain
\ie
{Z}^{(k)}_{D0/D4'}=\prod^k_{I=1}\frac{\sh\left(\pm\phi_I\pm M+\E_-\right)}{\sh\left(\pm\phi_I \pm M-\E_+\right)} \; .
\fe

The quantization of the D4/D$4'$ string is responsible for a Fermi multiplet, in a bifundamental representation of $SO(2N)\times SO(2)$, which results in  
\ie
Z_{D4/D4'}=\prod_{i=1}^N\sh(M\pm a_i) \; .
\fe
In the above, we used the same notation as in the $SU(N)$ case: we defined $\sh(x)\equiv 2\sinh(x/2)$, and products over all signs inside an argument have to be considered; for example, $\sh(a\pm b)\equiv\sh(a+b)\,\sh(a-b)$.\\

The various factors in the integrand can be written in gauge theory terms, with the caveat of factoring out possible spurious contributions $Z_{extra}$ present in the UV. We  identify $Z^{(k)}_{D0/D0}\cdot Z^{(k)}_{D0/D4} = Z^{(k)}_{vec}\cdot Z^{(k)}_{adjoint}$ and  $Z^{(k)}_{D0/D4'}\cdot Z_{D4/D4'} = Z^{(k)}_{defect}$. Then the index can also be written as
\begin{align}
\label{5dintegralDN}
Z(M)  =&\sum_{k=0}^{\infty}\;\frac{\widetilde{\fq}^{k}}{2^k\,k!} \, \oint  \left[\frac{d\phi_I}{2\pi i}\right]Z^{(k)}_{vec}\cdot Z^{(k)}_{adjoint}\cdot  Z^{(k)}_{defect}(M)  \; , \\
Z^{(k)}_{vec} = &\prod_{I=1}^k\frac{\sh\left(\pm\phi_I\right)\sh\left(\pm\phi_I+\E_+\right)}{\prod^{N}_{i=1}\sh\left(\pm\phi_I \pm a_i+\E_+\right)}\,\prod_{I < J}^{k}\frac{\sh\left(\pm \phi_I \pm \phi_J\right) \;\sh\left(\pm \phi_{I} \pm \phi_{J} + 2\epsilon_+\right)}{\sh\left(\pm \phi_{I} \pm \phi_{J} \pm \epsilon_- + \epsilon_+\right)}\\
Z^{(k)}_{adjoint} =&\prod^k_{I=1}\frac{\prod^{N}_{i=1}{\sh\left(\pm\phi_I\pm a_i+m\right)}}{\sh\left(\pm\phi_I\pm m-\E_+\right)}\prod^k_{I<J}\frac{\sh\left(\pm\phi_I\pm\phi_J\pm m-\E_-\right)}{\sh\left(\pm\phi_I\pm\phi_J\pm m-\E_+\right)}\; ,\\
Z^{(k)}_{defect} =&\prod_{i=1}^{N}\sh\left(M\pm a_i\right)\prod^k_{I=1}\frac{\sh\left(\pm\phi_I\pm M+\E_-\right)}{\sh\left(\pm\phi_I \pm M-\E_+\right)}\, .
\end{align}

\vspace{10mm}

------\; {\bf Pure case}\; ------\\

Let us first study the case of a pure $SO(2N)$ gauge theory. This can be done by decoupling the adjoint mass $m\rightarrow \infty$. As long as we properly rescale the instanton counting parameter with $m$, we can take the limit inside the integrand, and study the following integral:
\begin{align}
\label{5dintegralDNpure}
& Z(M)  =\sum_{k=0}^{\infty}\;\frac{\widetilde{\fq}^{k}}{2^k\,k!} \, \oint_{\mathcal{M}_k}  \left[\frac{d\phi_I}{2\pi i}\right]Z^{(k)}_{vec}\cdot  Z^{(k)}_{defect}(M)  \; . 
\end{align}

In order to derive non-perturbative SD equations, we need to know the content of the set $\mathcal{M}_k\setminus \mathcal{M}^{pure}_k$, that is the set of ``new" poles due solely to $Z^{(k)}_{defect}(M)$. We find that the set grows unbounded as $k$ increases, which tells us that $Z(M)$ is an infinite $\widetilde{\fq}$-series in defect $Y$-operator vevs.
We present the first few terms here:
\begin{align}
\label{SO2Npure}
Z(M) = &\left\langle Y(M) \right\rangle\\
&+\widetilde{\fq} \; \bigg[c_{(1,1)}^{\epsilon_1, \epsilon_2} \left\langle\frac{1}{Y(M-2 \, \epsilon_+)} \right\rangle + c_{(1,2)}^{\epsilon_1, \epsilon_2} \left\langle\frac{1}{Y(M+2 \, \epsilon_+)}\right\rangle \bigg]\nonumber\\
&+\widetilde{\fq}^2 \bigg[c_{(2,1)}^{\epsilon_1, \epsilon_2} \left\langle\frac{1}{Y(M-2 \, \epsilon_+)Y(M-\epsilon_- -\epsilon_+)Y(M-\epsilon_- -3\epsilon_+)} \right\rangle\nonumber\\ 
&\qquad+ c_{(2,2)}^{\epsilon_1, \epsilon_2} \left\langle\frac{1}{Y(M+2 \, \epsilon_+)Y(M+\epsilon_- +\epsilon_+)Y(M+\epsilon_- +3\epsilon_+)} \right\rangle\nonumber\\
&\qquad  + c_{(2,3)}^{\epsilon_1, \epsilon_2} \left\langle\frac{1}{Y(M-2 \, \epsilon_+)Y(M+\epsilon_- -\epsilon_+)Y(M+\epsilon_- -3\epsilon_+)} \right\rangle \nonumber\\
&\qquad+ c_{(2,4)}^{\epsilon_1, \epsilon_2} \left\langle\frac{1}{Y(M+2 \, \epsilon_+)Y(M-\epsilon_- +\epsilon_+)Y(M-\epsilon_- +3\epsilon_+)} \right\rangle \bigg]\nonumber\\
&+\widetilde{\fq}^3\; \ldots\nonumber
\end{align}
The set $\mathcal{M}_1\setminus \mathcal{M}^{pure}_1$ has two elements, which means we expect two terms at $k=1$.  

The $(1,1)$ term is the residue at $\phi_I=-M+\epsilon_+$.

The $(1,2)$ term is the residue at $\phi_I=M+\epsilon_+$. 

Meanwhile, the set $\mathcal{M}_2\setminus \mathcal{M}^{pure}_2$ has four elements, so  we expect four terms at $k=2$.

The $(2,1)$ term is the residue at $\phi_I= M-\epsilon_- -2\epsilon_+, \phi_J= -M+\epsilon_+$. 

The $(2,2)$ term is the residue at $\phi_I= -M-\epsilon_- -2\epsilon_+, \phi_J= M+\epsilon_+$.

The $(2,3)$ term is the residue at $\phi_I= M+\epsilon_- -2\epsilon_+, \phi_J= -M+\epsilon_+$.

The $(2,4)$ term is the residue at $\phi_I= -M+\epsilon_- -2\epsilon_+, \phi_J= M+\epsilon_+$.\\

After normalizing the partition function and expanding it in the (exponentiated) defect fugacity $e^{M}$, we find the SD equation solved by the $Y$-operator vevs:
\beq
\label{SchwingerSO2Npure}
\frac{1}{\widetilde{\left\langle 1 \right\rangle}}\left[\left\langle Y(M) \right\rangle
+\widetilde{\fq} \; \ldots\right]  = \cT(M; \{e_i^{\epsilon_1, \epsilon_2}\}) \; .
\eeq
Above, $\cT(M; \{e_i^{\epsilon_1, \epsilon_2}\})$ is a finite Laurent series in $e^M$.  Taking the flat space limit $\epsilon_1, \epsilon_2 \rightarrow 0$, the above equation describes the SW geometry of 5d $SO(2N)$ super Yang-Mills.\\ 

Let us study the more familiar four-dimensional limit in detail. Namely, we reintroduce the radius $R$ explicitly in the partition function, and take the limit $R\rightarrow 0$, leaving all fugacities fixed. The SD equations become
\beq
\label{Schwinger4dSO2Npure}
\frac{1}{\widetilde{\left\langle 1 \right\rangle}_{4d}}\left[\left\langle Y(M) \right\rangle_{4d}
+\widetilde{\fq} \; \ldots\right]  = \cT_{4d}(M; \{e_i^{\epsilon_1, \epsilon_2}\}) \, .
\eeq

We compute the coefficients in \eqref{SO2Npure} to be 
\begin{align}
\label{SO2Npurecoefs}
c_{(1,1), 4d}^{\epsilon_1, \epsilon_2} &=2(M - \epsilon_+) (2 M + \epsilon_- - \epsilon_+) (2 M - \epsilon_- - \epsilon_+) (M - 2 \epsilon_+)\\
c_{(1,2), 4d}^{\epsilon_1, \epsilon_2} &=2(M + \epsilon_+) (2 M - \epsilon_- + \epsilon_+) (2 M + \epsilon_- + \epsilon_+) (M + 2 \epsilon_+)\nonumber\\
c_{(2,1), 4d}^{\epsilon_1, \epsilon_2} &=\frac{2\epsilon_+}{\epsilon_-}(M - \epsilon_+) (2 M + \epsilon_- - \epsilon_+) (2 M - \epsilon_- - \epsilon_+)\nonumber\\
&\qquad (M -  \epsilon_- -  \epsilon_+) (M -  \epsilon_- - 2 \epsilon_+) (2 M -  \epsilon_- - 3 \epsilon_+) (M -  \epsilon_- - 3 \epsilon_+) (2M -  \epsilon_- - 5 \epsilon_+)\nonumber\\
c_{(2,2), 4d}^{\epsilon_1, \epsilon_2} &=\frac{2\epsilon_+}{\epsilon_-}(M + \epsilon_+) (2 M - \epsilon_- + \epsilon_+) (2 M + \epsilon_- + \epsilon_+)\nonumber\\
&\qquad (M +  \epsilon_- +  \epsilon_+) (M +  \epsilon_- + 2 \epsilon_+) (2 M +  \epsilon_- + 3 \epsilon_+) (M +  \epsilon_- + 3 \epsilon_+) (2M +  \epsilon_- + 5 \epsilon_+)\nonumber\\
c_{(2,3), 4d}^{\epsilon_1, \epsilon_2} &=\frac{-2\epsilon_+}{\epsilon_-}(M - \epsilon_+) (2 M + \epsilon_- - \epsilon_+) (2 M - \epsilon_- - \epsilon_+)\nonumber\\
& (M +  \epsilon_- -  \epsilon_+) (M +  \epsilon_- - 2 \epsilon_+) (2 M +  \epsilon_- - 3 \epsilon_+) (M +  \epsilon_- - 3 \epsilon_+) (2M +  \epsilon_- - 5 \epsilon_+)\nonumber\\
c_{(2,4), 4d}^{\epsilon_1, \epsilon_2} &=\frac{-2\epsilon_+}{\epsilon_-}(M + \epsilon_+) (2 M - \epsilon_- + \epsilon_+) (2 M + \epsilon_- + \epsilon_+)\nonumber\\
&\qquad(M -  \epsilon_- +  \epsilon_+) (M -  \epsilon_- +2 \epsilon_+) (2 M -  \epsilon_- + 3 \epsilon_+) (M -  \epsilon_- + 3 \epsilon_+) (2M -  \epsilon_- + 5 \epsilon_+)\nonumber
\end{align}
and we find for the right-hand side
\beq
\label{RHSSO2Npure}
{\cT}_{4d}(M; \{e_i^{\epsilon_1, \epsilon_2}\})=\prod_{i=1}^{N}(M \pm e_i^{\epsilon_1, \epsilon_2}) \; .
\eeq 

As an example, here is ${\cT}_{4d}$ when  $N=2$, meaning $G=SO(4)$, with Coulomb parameters $a_1$ and $a_2$:
\begin{align}
\label{SO2NpurepolyNis2}
{\cT}_{4d}(M; \{e_i^{\epsilon_1, \epsilon_2}\}) = M^4 + d_{(2)}^{\epsilon_1, \epsilon_2} M^2 + d_{(0)}^{\epsilon_1, \epsilon_2} \; ,
\end{align}
where
\begin{align}
\label{SO2Npurepolycoef}
d_{(2)}^{\epsilon_1, \epsilon_2} &=-a_1^2-a_2^2 + \widetilde{\fq}\,\frac{8(4\epsilon_+^2 - a_1^2 -a_2^2)}{(a_1+a_2\pm 2\epsilon_+)(a_1-a_2\pm 2\epsilon_+)}+ \widetilde{\fq}^2\,\ldots\\
d_{(0)}^{\epsilon_1, \epsilon_2} &=a_1^2\,a_2^2 + \widetilde{\fq}\,\frac{4(-4\epsilon_+^2\epsilon_-^2 + 4\epsilon_+^4 -4 a_1^2 \, a_2^2 + (a_1^2+a_2^2) (\epsilon_-^2 - \epsilon_+^2))}{(a_1+a_2\pm 2\epsilon_+)(a_1-a_2\pm 2\epsilon_+)}+\widetilde{\fq}^2\,\ldots\nonumber
\end{align}
The roots $\{e_i^{\epsilon_1, \epsilon_2}\}$ are obtained from the $d_{(i)}^{\epsilon_1, \epsilon_2}$ from usual root/coefficient relations for a polynomial.

Taking  the flat space limit $\epsilon_1, \epsilon_2 \rightarrow 0$, the partition function simplifies drastically:
\begin{align}
\label{SO2NpureSW}
Z(M) = &\left\langle Y(M) \right\rangle\\
&+\widetilde{\fq} \; \left[c_{(1,1), 4d}^{0,0}+ c_{(1,2), 4d}^{0,0} \right] \left\langle\frac{1}{Y(M)} \right\rangle_{4d}\nonumber\\
&+\widetilde{\fq}^2 \left[c_{(2,1), 4d}^{0,0}+c_{(2,2), 4d}^{0,0}+c_{(2,3), 4d}^{0,0}+c_{(2,4), 4d}^{0,0}\right] \left\langle\frac{1}{Y(M)^3}_{4d} \right\rangle\nonumber\\
&+ \widetilde{\fq}^3\; \ldots\nonumber
\end{align}
The flat space limit of the coefficients \eqref{SO2Npurecoefs} yields 
\begin{align}
\label{SO2NpurecoefsSW}
& c_{(1,1), 4d}^{0,0} = c_{(1,2), 4d}^{0,0} = 8 M^4\\
& c_{(2,1), 4d}^{0,0} = c_{(2,2), 4d}^{0,0} =  -32 M^8 \; ,\;\;\;  c_{(2,3), 4d}^{0,0} = c_{(2,4), 4d}^{0,0} =  +32 M^8\; .\nonumber
\end{align}
Remarkably, the terms at order $k=1$ add up, while the ones at $k=2$ cancel out. There are 42 nonzero residues at $k=3$. We find that they also do not survive the flat space limit. We were able to show this phenomenon at $k=4$ as well. We conjecture that this pattern continues at every instanton order, and that only the first instanton contribution $k=1$ survives the flat space limit\footnote{An exact proof for all $k$ would require a classification of poles in the partition function, which is not known. In other words, as of today, there is no explicit description of the set $\mathcal{M}_k\setminus \mathcal{M}^{pure}_k$.}. In the end, the four-dimensional SD equations \eqref{Schwinger4dSO2Npure} take the following form:  
\begin{align}
\label{SO2NpureSW2}
\left\langle Y(M) \right\rangle_{4d} +\widetilde{\fq} \; 16\, M^4\left\langle\frac{1}{Y(M)} \right\rangle_{4d} =\prod_{i=1}^{N}(M \pm e_i^{0,0}) \; .
\end{align}
After introducing the dynamical scale $\widetilde{\fq}\equiv \Lambda^{4N-4}$ and rescaling,  this is nothing but the SW curve of the pure 4d $SO(2N)$ theory  (see for instance \cite{DHoker:1996kdj} or \cite{Argyres:1995fw}):
\beq
\label{SO2NSW}
y(M) +  M^4\, \frac{\Lambda^{4 N-4}}{y(M)} = \prod_{i=1}^{N}(M \pm e_i^{0,0}) \; .
\eeq

\vspace{10mm}

------\; {\bf Fundamental matter}\; ------\\

We cannot add D8 branes in the presence of an O$8^+$ plane. Then, we will abandon the string theory picture and introduce $N_f$ fundamental hypermultiplets following a purely field theoretic route. 
The inclusion of fundamental matter will introduce new fermionic zero modes arising from Fermi multiplets in the bifundamental representation of $Sp(k)\times Sp(N_f)$. We therefore need to modify the index to account for these new Fermi multiplets:
\begin{align}
\label{5dintegralSO2Nfund}
& Z(M)  =\sum_{k=0}^{\infty}\;\frac{\widetilde{\fq}^{k}}{2^k\,k!} \, \oint_{\mathcal{M}_k}  \left[\frac{d\phi_I}{2\pi i}\right]Z^{(k)}_{vec}\cdot Z^{(k)}_{fund}\cdot  Z^{(k)}_{defect}  \; , \\
&Z^{(k)}_{fund} =\prod_{I=1}^{k} \prod_{d=1}^{N_f} \sh\left(\pm\phi_I+ m_d\right) \equiv\prod_{I=1}^{k} Q(\phi_I)\; ,\label{fundQDN}
\end{align}
The set $\mathcal{M}_k\setminus \mathcal{M}^{pure}_k$ is the same whether we consider a theory with or without fundamental matter  (as long as it does not introduce new poles at infinity, which was our assumption to begin with). The matter function $Q$  simply keep track of the various JK-poles in its argument. In our case, we compute at once:
\begin{align}
\label{SO2Nfund}
Z(M) = &\left\langle Y(M) \right\rangle\\
&+\widetilde{\fq} \; \left[c_{(1,1)}^{\epsilon_1, \epsilon_2}\, Q(-M+\epsilon_+) \left\langle\frac{1}{Y(M-2 \, \epsilon_+)} \right\rangle + c_{(1,2)}^{\epsilon_1, \epsilon_2}\, Q(M+\epsilon_+) \left\langle\frac{1}{Y(M+2 \, \epsilon_+)}\right\rangle \right]\nonumber\\
&+\widetilde{\fq}^2 \;\ldots\nonumber 
\end{align}
The coefficients $c_{(i, j)}^{\epsilon_1, \epsilon_2}$ are the same as in the pure case, and the derivation of non-perturbative SD equations is identical. Taking the four-dimensional and flat space limits, we find once again that all terms at $k\geq 2$ cancel out against each other, for every $k$ we tested. 

Notice that the function $Q(M)$ is even. It follows immediately that at order $k=1$, the two  terms above combine into one. After introducing the dynamical scale $\widetilde{\fq}\equiv \Lambda^{4 N-4-2N_f}$, we therefore obtain the SW curve of $SO(2N)$ with $N_f$ fundamental flavors:
\beq
\label{SO2NSWfund}
y(M) + M^4\; \prod_{d=1}^{N_f} \left(\pm M+ m_d\right) \frac{\Lambda^{4 N-4-2N_f}}{y(M)} = \prod_{i=1}^{N}(M \pm e_i^{0,0}) \; .
\eeq

\vspace{10mm}

------\; {\bf Adjoint matter}\; ------\\

We come back to our initial quantum mechanics index, \eqref{5dintegralDN}, which we rewrite here for convenience:
\beq
\label{5dintegralDNagain}
Z(M)  =\sum_{k=0}^{\infty}\;\frac{\widetilde{\fq}^{k}}{2^k\, k!} \, \oint_{\mathcal{M}_k}  \left[\frac{d\phi_I}{2\pi i}\right]Z^{(k)}_{vec}\cdot Z^{(k)}_{adjoint}\cdot  Z^{(k)}_{defect}(M)  \; .
\eeq
We compute the partition function
\begin{align}
\label{SO2Nadjoint}
Z(M) = &\left\langle Y(M) \right\rangle\\
&+\widetilde{\fq} \; \bigg[c_{(1,1)}^{\epsilon_1, \epsilon_2} \left\langle\frac{Y(M+m- \epsilon_+)\, Y(M-m - \epsilon_+)}{Y(M-2 \, \epsilon_+)} \right\rangle\nonumber\\ 
&\qquad+ c_{(1,2)}^{\epsilon_1, \epsilon_2} \left\langle\frac{Y(M+m + \epsilon_+)\, Y(M-m+ \epsilon_+)\,}{Y(M+2 \, \epsilon_+)}\right\rangle \bigg]\\
&+\widetilde{\fq}^2\; \ldots \nonumber
\end{align}
The above ``$\ldots$" indicates that there are an infinite number of terms in the $\widetilde{\fq}$-series. Each term stands for an element of the set $\mathcal{M}_k\setminus \mathcal{M}^{pure}_k$. In particular, the $k=1$ term above is the  residue at the pole $\phi_I=-M+\epsilon_+$ and at the pole $\phi_I=M+\epsilon_+$, respectively. These are the only two elements of the set  $\mathcal{M}_1\setminus \mathcal{M}^{pure}_1$. 
After normalizing the partition function and expanding it in the (exponentiated) defect fugacity $e^{M}$, we are able to write the SD equation solved by the $Y$-operators:
\beq
\label{SchwingerSO2Nadjoint}
\frac{1}{\widetilde{\left\langle 1 \right\rangle}}\left[\left\langle Y(M) \right\rangle
+\widetilde{\fq} \;  \ldots \right]  = \cT(M; \{e_i^{\epsilon_1, \epsilon_2}\}) \; .
\eeq
Above, $\cT(M; \{e_i^{\epsilon_1, \epsilon_2}\})$ is a finite Laurent series in $e^M$.
Taking the flat space limit $\epsilon_1, \epsilon_2 \rightarrow 0$, we conjecture that the above equation describes the SW geometry of 5d $SO(2N)$ super Yang-Mills with adjoint matter.\\

Let us study the more familiar four-dimensional limit in detail. Namely, we reintroduce the radius $R$ explicitly in the partition function, and take the limit $R\rightarrow 0$, leaving all fugacities fixed. The SD equations become
\beq
\label{Schwinger4dSO2Nadjoint}
\frac{1}{\widetilde{\left\langle 1 \right\rangle}_{4d}}\left[\left\langle Y(M) \right\rangle_{4d}
+\widetilde{\fq} \;  \ldots \right]  = \cT_{4d}(M; \{e_i^{\epsilon_1, \epsilon_2}\}) \, ,
\eeq
where 
\begin{align}
\label{SO2Nadjointcoefs}
c_{(1,1), 4d}^{\epsilon_1, \epsilon_2} &= \frac{(m - \epsilon_-) (m + \epsilon_-) (M - 2 \epsilon_+) (M - \epsilon_+) (2 M + \epsilon_- - \epsilon_+) (2 M - \epsilon_- - \epsilon_+)}{2  (m - \epsilon_+) (m + \epsilon_+) (2 M+m - \epsilon_+) (- 
	2 M+m + \epsilon_+) (2 M +m- 3 \epsilon_+) (- 2 M+m + 3 \epsilon_+)}\\
c_{(1,2), 4d}^{\epsilon_1, \epsilon_2} &= \frac{(m - \epsilon_-) (m + \epsilon_-) (M + 2 \epsilon_+) (M + \epsilon_+) (2 M - \epsilon_- + \epsilon_+) (2 M + \epsilon_- + \epsilon_+)}{2 (m - \epsilon_+)(m + \epsilon_+) (m - 2 M - \epsilon_+)  (- 2 M+m - 3 \epsilon_+) (m + 2 M + \epsilon_+) (m + 2 M + 3 \epsilon_+)}\nonumber
\end{align}
and
\beq
\label{RHSSO2Nadjoint}
{\cT}_{4d}(M; \{e_i^{\epsilon_1, \epsilon_2}\})=\prod_{i=1}^{N}(M \pm e_i^{\epsilon_1, \epsilon_2}) \; .
\eeq 
Taking the flat space limit $\epsilon_1, \epsilon_2 \rightarrow 0$, we conjecture that we recover the SW geometry of 4d $SO(2N)$ with adjoint matter. In particular, the $k=1$ term \eqref{SO2Nadjoint} we presented here is in perfect agreement with the first instanton correction computed in \cite{Ennes:1999fb}. The higher corrections are identified exactly as we did above, following the JK prescription. 

\vspace{5mm}

\subsection{The $SO(2 N + 1)$ Gauge Theory}
	
	Engineering a $SO(2 N + 1)$ gauge theory from string theory can usually be done with the help of a different type of orientifold, sometimes called $\widetilde{\text{O}p}^+$ plane. For $p\leq 5$, such orientiolds  can be interpreted as an O$p$ orientifold plane with a D$p$ brane stuck on it, and their construction relies on the existence of a $\mathbb{Z}_2$ discrete torsion associated with the RR fields of the theory. However, when $p\geq 6$, no such $\mathbb{Z}_2$ torsion exists, and the existence of $\widetilde{\text{O}p}$ planes becomes a subtle question. It turns out in particular that $\widetilde{\text{O}8}$ planes do not exist \cite{Hyakutake:2000mr}. Therefore, we will not rely on a stringy construction here, and provide only a field theory definition of the defect partition function. The 5d $SO(2 N + 1)$ instanton partition function in the absence of Wilson loop was constructed in \cite{Nekrasov:2004vw, Hwang:2014uwa}. The field content of the quantum mechanics is the same as  that of the $SO(2N)$ case, but there is a change in the 1-loop determinants.
	 To simplify notations, let us denote by $Z^{(k,2N)}$ the $SO(2N)$ 1-loop determinants written down in the previous section, and let us denote by $Z^{(k,2N+1)}$ the $SO(2N+1)$ 1-loop determinants.\\

Without an underlying brane construction, it is a priori unclear how to construct $Z^{(k, 2N+1)}_{defect}$. We propose that the defect group should be $G'=SO(2N')$. In particular, this implies that the set of poles $\mathcal{M}_k\setminus \mathcal{M}^{pure}_k$ depending on the defect fugacity $M$ will be exactly the same as in the previous section\footnote{The defect contribution is no longer obtained from the other 1-loop determinants after exchanging fugacities (the groups $G=SO(2N+1)$ and $G'=SO(2N')$ no longer even agree). In our attempts, the naive strategy of building a $SO(2N'+1)$ defect group from field theory seems to fail to reproduce the correct $SO(2N+1)$ SW geometry. It is somewhat intriguing that this seems consistent with the lack of a $\widetilde{\text{O}8}$ plane construction in string theory.}.  We will see that this choice enables us to make contact with $SO(2N+1)$ SW geometry. Even though the defect group $G'$ will be the same as we previously encountered, the 1-loop determinants are not exactly identical: we must remember that there is a classical contribution (what we previously called D4/D$4'$ strings) resulting in a Fermi multiplet in the bifundamental representation of $G \times G'$. In our case, we can readily write the 1-loop determinant for such a $SO(2N+1)\times SO(2N')$ Fermi multiplet:
\ie
\sh(M)\,\prod_{i=1}^N\sh(M\pm a_i) \; .
\fe

All in all, we therefore propose the following defect partition function
	\begin{align}
	\label{5dintegralBN}
	& Z(M)  =\sum_{k=0}^{\infty}\;\frac{\widetilde{\fq}^{k}}{2^k\,k!} \, \oint_{\mathcal{M}_k}  \left[\frac{d\phi_I}{2\pi i}\right]Z^{(k,2N+1)}_{vec}\cdot Z^{(k,2N+1)}_{adjoint}\cdot  Z^{(k,2N+1)}_{defect}(M)  \; , \\
	&Z^{(k,2N+1)}_{vec} =Z^{(k,2N)}_{vec}\times\prod^k_{I=1}\frac{1}{\sh(\pm\phi_I+\E_+)}\; ,\quad Z^{(k,2N+1)}_{adjoint} =Z^{(k,2N)}_{adjoint}\times\prod^k_{I=1}\sh(\pm\phi_I+m)\; ,\\
	&Z^{(k,2N+1)}_{defect} =Z^{(k,2N)}_{defect}\times\sh(M)\, .
	\end{align}
	
\vspace{10mm}
	
------\; {\bf Pure case}\; ------\\
	
We decouple the adjoint mass by taking the limit $m\rightarrow\infty$, which once again commutes with integration. Then, consider the integral
\begin{align}
\label{5dintegralBNpure}
Z(M)  =\sum_{k=0}^{\infty}\;\frac{\widetilde{\fq}^{k}}{2^k\,k!} \, \oint_{\mathcal{M}_k}  \left[\frac{d\phi_I}{2\pi i}\right]Z^{(k,2N+1)}_{vec}\cdot  Z^{(k,2N+1)}_{defect}(M)  \; .
\end{align}

The partition function organizes itself as an infinite $\widetilde{\fq}$-series of defect $Y$-operators, as follows:
\begin{align}
\label{SO2NPLUS1pure}
Z(M) = &\left\langle Y(M) \right\rangle\\
&+\widetilde{\fq} \; \left[c_{(1,1)}^{\epsilon_1, \epsilon_2} \left\langle\frac{1}{Y(M-2 \, \epsilon_+)} \right\rangle + c_{(1,2)}^{\epsilon_1, \epsilon_2} \left\langle\frac{1}{Y(M+2 \, \epsilon_+)}\right\rangle \right]\nonumber\\
&+\widetilde{\fq}^2 \; \ldots\nonumber
\end{align}
The JK residue prescription tells us that the set $\mathcal{M}_1\setminus \mathcal{M}^{pure}_1$ has two elements, which means we expect two terms at $k=1$. We wrote them explicitly above.

The $(1,1)$ term is the residue at $\phi_I=-M+\epsilon_+$.

The $(1,2)$ term is the residue at $\phi_I=M+\epsilon_+$.\\

After normalizing the partition function and expanding it in the (exponentiated) defect fugacity $e^{M}$, we derive the following SD equation:
\beq
\label{SchwingerSO2NPLUS1pure}
\frac{1}{\widetilde{\left\langle 1 \right\rangle}}\left[\left\langle Y(M) \right\rangle
+\widetilde{\fq} \; \ldots\right]  = \cT(M; \{e_i^{\epsilon_1, \epsilon_2}\}) \; .
\eeq
Above, $\cT(M; \{e_i^{\epsilon_1, \epsilon_2}\})$ is a finite Laurent series in $e^M$. Taking the flat space limit $\epsilon_1, \epsilon_2 \rightarrow 0$, the above equation describes the SW geometry of 5d $SO(2N+1)$ super Yang-Mills.\\ 

Let us study the more familiar four-dimensional limit in detail. Namely, we reintroduce the radius $R$ explicitly in the partition function, and take the limit $R\rightarrow 0$, leaving all fugacities fixed. The SD equations become
\beq
\label{Schwinger4dSO2NPLUS1pure}
\frac{1}{\widetilde{\left\langle 1 \right\rangle}_{4d}}\left[\left\langle Y(M) \right\rangle_{4d}
+\widetilde{\fq} \; \ldots\right]  = \cT_{4d}(M; \{e_i^{\epsilon_1, \epsilon_2}\}) \, .
\eeq

We compute the coefficients in \eqref{SO2NPLUS1pure} to be 
\begin{align}
\label{SO2NPLUS1purecoefs}
c_{(1,1), 4d}^{\epsilon_1, \epsilon_2} &=2 (M - \epsilon_+) (2 M + \epsilon_- - \epsilon_+) (2 M - \epsilon_- - \epsilon_+) (M - 2 \epsilon_+)\\
c_{(1,2), 4d}^{\epsilon_1, \epsilon_2} &=2 (M + \epsilon_+) (2 M - \epsilon_- + \epsilon_+) (2 M + \epsilon_- + \epsilon_+) (M + 2 \epsilon_+)\nonumber
\end{align}
and the right-hand side comes out to be
\beq
\label{RHSSO2NPLUS1pure}
{\cT}_{4d}(M; \{e_i^{\epsilon_1, \epsilon_2}\})=M\,\prod_{i=1}^{N}(M \pm e_i^{\epsilon_1, \epsilon_2}) \; .
\eeq 

As an example, here is ${\cT}_{4d}$ when  $N=2$, meaning $G=SO(5)$, with Coulomb parameters $a_1$ and $a_2$:
\begin{align}
\label{SO2NPLUS1purepolyNis2}
{\cT}_{4d}(M; \{e_i^{\epsilon_1, \epsilon_2}\}) = M^5 + d_{(3)}^{\epsilon_1, \epsilon_2} M^3 + d_{(1)}^{\epsilon_1, \epsilon_2} M\; ,
\end{align}
where
\begin{align}
\label{SO2NPLUS1purepolycoef}
d_{(3)}^{\epsilon_1, \epsilon_2} &=-a_1^2-a_2^2 + \widetilde{\fq}\,\frac{16}{(a_1+a_2\pm 2\epsilon_+)(a_1-a_2\pm 2\epsilon_+)}+\widetilde{\fq}^2\ldots\\
d_{(1)}^{\epsilon_1, \epsilon_2} &=a_1^2\, a_2^2 + \widetilde{\fq}\,\frac{8(-3\epsilon_+^2-\epsilon_-^2 +a_1^2+a_2^2)}{(a_1+a_2\pm 2\epsilon_+)(a_1-a_2\pm 2\epsilon_+)}+\widetilde{\fq}^2\ldots
\end{align}
The roots $\{e_i^{\epsilon_1, \epsilon_2}\}$ are obtained from the $d_{(i)}^{\epsilon_1, \epsilon_2}$ from usual root/coefficient relations for a polynomial.\\

Taking  the flat space limit $\epsilon_1, \epsilon_2 \rightarrow 0$, the partition function simplifies drastically:
\begin{align}
\label{SO2NPLUS1pureSW}
Z(M) = &\left\langle Y(M) \right\rangle_{4d}\\
&+\widetilde{\fq} \; \left[c_{(1,1), 4d}^{0,0}+ c_{(1,2), 4d}^{0,0} \right] \left\langle\frac{1}{Y(M)} \right\rangle_{4d}\nonumber\\
&+\widetilde{\fq}^2\; \ldots\nonumber
\end{align}
The flat space limit of the coefficients \eqref{SO2NPLUS1purecoefs} yields 
\begin{align}
\label{SO2NPLUS1purecoefsSW}
& c_{(1,1), 4d}^{0,0} = c_{(1,2), 4d}^{0,0} = 8 M^4
\end{align}
We find once again that $k=1$ terms add up, while the terms at $k\geq 2$ cancel out among each other (we checked this up to $k=4$). We conjecture that this pattern continues at every instanton order, and that only the first instanton contribution $k=1$ survives the flat space limit. In the end, the four-dimensional SD equations \eqref{Schwinger4dSO2NPLUS1pure} take the following form:  
\begin{align}
\label{SO2NPLUS1pureSW2}
\left\langle Y(M) \right\rangle_{4d} +\widetilde{\fq} \; 16\, M^4\left\langle\frac{1}{Y(M)} \right\rangle_{4d} =M\, \prod_{i=1}^{N}(M \pm e_i^{0,0}) \; .
\end{align}
After introducing the dynamical scale $\widetilde{\fq}\equiv \Lambda^{4N-2}/16$ and rescaling the $Y$-operators by $M$,  this is nothing but the SW curve of the pure 4d $SO(2N+1)$ theory (see for instance \cite{DHoker:1996kdj,Danielsson:1995is,Argyres:1995fw}):
\beq
\label{SO2NPLUS1SW}
y(M) + M^2\,\frac{\Lambda^{4 N-2}}{y(M)} = \prod_{i=1}^{N}(M \pm e_i^{0,0}) \; .
\eeq

\vspace{10mm}

------\; {\bf Fundamental matter}\; ------\\

Introducing $N_f$ fundamental hypermultiplets amounts to considering new fermionic zero modes. These will arise from Fermi multiplets in the bifundamental representation of $Sp(k)\times Sp(N_f)$. We therefore need to modify the index accordingly:
\begin{align}
\label{5dintegralSO2NPLUS1fund}
& Z(M)  =\sum_{k=0}^{\infty}\;\frac{\widetilde{\fq}^{k}}{2^k\,k!} \, \oint_{\mathcal{M}_k}  \left[\frac{d\phi_I}{2\pi i}\right]Z^{(k)}_{vec}\cdot Z^{(k)}_{fund}\cdot  Z^{(k)}_{defect}  \; , \\
&Z^{(k)}_{fund} =\prod_{I=1}^{k} \prod_{d=1}^{N_f} \sh\left(\pm\phi_I+ m_d\right) \equiv\prod_{I=1}^{k} Q(\phi_I)\; ,\label{fundQBN}
\end{align}
The set $\mathcal{M}_k\setminus \mathcal{M}^{pure}_k$ is the same whether we consider a theory with or without fundamental matter  (as long as it does not introduce new poles at infinity, which was our assumption to begin with). The matter function $Q$  simply keeps track of the various JK-poles in its argument. In our case, we compute at once:
\begin{align}
\label{SO2NPLUS1fund}
Z(M) = &\left\langle Y(M) \right\rangle\\
&+\widetilde{\fq} \; \left[c_{(1,1)}^{\epsilon_1, \epsilon_2}\, Q(-M+\epsilon_+) \left\langle\frac{1}{Y(M-2 \, \epsilon_+)} \right\rangle + c_{(1,2)}^{\epsilon_1, \epsilon_2}\, Q(M+\epsilon_+) \left\langle\frac{1}{Y(M+2 \, \epsilon_+)}\right\rangle \right]\nonumber\\
&+\widetilde{\fq}^2 \; \ldots \nonumber
\end{align}
The coefficients $c_{(i, j)}^{\epsilon_1, \epsilon_2}$ are the same as in the pure case, and the derivation of non-perturbative SD equations is identical.\\

 Taking the four-dimensional and flat space limits, we find once again that all terms at $k\geq 2$ cancel out against each other, for every $k$ we tested.
The function $Q(M)$ is even. It follows immediately that at order $k=1$, the two  terms above combine into one. After introducing the dynamical scale $\widetilde{\fq}\equiv \Lambda^{4 N-2-2N_f}$ and rescaling, we land on the SW curve of $SO(2N+1)$ with $N_f$ fundamental flavors:
\beq
\label{SO2NPLUS1SWfund}
y(M) + M^2\; \prod_{d=1}^{N_f} \left(\pm M+ m_d\right) \frac{\Lambda^{4 N-4-2N_f}}{y(M)} = \prod_{i=1}^{N}(M \pm e_i^{0,0}) \; .
\eeq

\vspace{10mm}
	
------\; {\bf Adjoint matter}\; ------\\

We now consider the addition of adjoint matter:
\begin{align}
\label{5dintegralBNagain}
Z(M)  =\sum_{k=0}^{\infty}\;\frac{\widetilde{\fq}^{k}}{2^k\,k!} \, \oint_{\mathcal{M}_k}  \left[\frac{d\phi_I}{2\pi i}\right]Z^{(k,2N+1)}_{vec}\cdot Z^{(k,2N+1)}_{adjoint}\cdot  Z^{(k,2N+1)}_{defect}(M)  \; .
\end{align}
Following the JK-residue prescription, we compute the partition function
\begin{align}
\label{SO2NPLUS1adjoint}
Z(M) = &\left\langle Y(M) \right\rangle\\
&+\widetilde{\fq} \; \bigg[c_{(1,1)}^{\epsilon_1, \epsilon_2} \left\langle\frac{Y(M+m- \epsilon_+)\, Y(M-m - \epsilon_+)}{Y(M-2 \, \epsilon_+)} \right\rangle\nonumber\\ 
&\qquad + c_{(1,2)}^{\epsilon_1, \epsilon_2} \left\langle\frac{Y(M+m + \epsilon_+)\, Y(M-m+ \epsilon_+)\,}{Y(M+2 \, \epsilon_+)}\right\rangle \bigg]\nonumber\\
&+\widetilde{\fq}^2 \; \ldots \nonumber
\end{align}
The above ``$\ldots$" stand for an infinite series in the instanton counting parameter $\widetilde{\fq}$. Each term stands for an element of the set $\mathcal{M}_k\setminus \mathcal{M}^{pure}_k$. In particular, the $k=1$ term above is the  residue at the pole $\phi_I=-M+\epsilon_+$ and at the pole $\phi_I=M+\epsilon_+$, respectively. These are the only two elements of the set  $\mathcal{M}_1\setminus \mathcal{M}^{pure}_1$. 
After normalizing the partition function and expanding it in the (exponentiated) defect fugacity $e^{M}$, we derive the SD equation solved by the $Y$-operators:
\beq
\label{SchwingerSO2NPLUS1adjoint}
\frac{1}{\widetilde{\left\langle 1 \right\rangle}}\left[\left\langle Y(M) \right\rangle
+\widetilde{\fq} \;  \ldots \right]  = \cT(M; \{e_i^{\epsilon_1, \epsilon_2}\}) \; .
\eeq
Above, $\cT(M; \{e_i^{\epsilon_1, \epsilon_2}\})$ is a finite Laurent series in $e^M$.
Taking the flat space limit $\epsilon_1, \epsilon_2 \rightarrow 0$, we conjecture that the above equation describes the SW geometry of 5d $SO(2N+1)$ super Yang-Mills with adjoint matter.\\

Let us study the more familiar four-dimensional limit in detail. Namely, we reintroduce the radius $R$ explicitly in the partition function, and take the limit $R\rightarrow 0$, leaving all fugacities fixed. The SD equations become
\beq
\label{Schwinger4dSO2NPLUS1adjoint}
\frac{1}{\widetilde{\left\langle 1 \right\rangle}_{4d}}\left[\left\langle Y(M) \right\rangle_{4d}
+\widetilde{\fq} \;  \ldots \right]  = \cT_{4d}(M; \{e_i^{\epsilon_1, \epsilon_2}\}) \, ,
\eeq
where 
\begin{align}
\label{SO2NPLUS1adjointcoefs}
c_{(1,1), 4d}^{\epsilon_1, \epsilon_2} &= \frac{2(m - \epsilon_-) (m + \epsilon_-) (M - 2 \epsilon_+) (M - \epsilon_+) (2 M + \epsilon_- - \epsilon_+) (2 M - \epsilon_- - \epsilon_+)}{(m - \epsilon_+) (m + \epsilon_+) (2 M+m - \epsilon_+) (- 2 M+m + \epsilon_+) (2 M +m- 3 \epsilon_+) (- 2 M+m + 3 \epsilon_+)}\\
c_{(1,2), 4d}^{\epsilon_1, \epsilon_2} &= \frac{2(m - \epsilon_-) (m + \epsilon_-) (M + 2 \epsilon_+) (M + \epsilon_+) (2 M - \epsilon_- + \epsilon_+) (2 M + \epsilon_- + \epsilon_+)}{(m - \epsilon_+)(m + \epsilon_+) (m - 2 M - \epsilon_+)  (- 2 M+m - 3 \epsilon_+) (m + 2 M + \epsilon_+) (m + 2 M + 3 \epsilon_+)}\nonumber
\end{align}
and
\beq
\label{RHSSO2NPLUS1adjoint}
{\cT}_{4d}(M; \{e_i^{\epsilon_1, \epsilon_2}\})=M\, \prod_{i=1}^{N}(M \pm e_i^{\epsilon_1, \epsilon_2}) \; .
\eeq 
Taking the flat space limit $\epsilon_1, \epsilon_2 \rightarrow 0$, we conjecture that we recover the SW geometry of 4d $SO(2N)$ with adjoint matter. In particular, the $k=1$ term \eqref{SO2Nadjoint} we presented here is in perfect agreement with the first instanton correction computed in \cite{Ennes:1999fb}. The higher corrections are identified exactly as we did above, following the JK prescription.

\vspace{5mm}

\subsection{The $Sp(N)$ Gauge Theory}

We now move on to our last class of examples, the $G=Sp(N)$ gauge theory. We once again engineer the defect partition function as the index of a D0 brane quantum mechanics, this time in the presence of an O$8^-$ orientifold plane\footnote{This construction was first proposed recently  in \cite{Chang:2016iji} in the case $N=1$, in order to study the superconformal index in the presence of a Wilson loop. The resulting $Sp(1)$ ray operators constructed there captured the enhancement of the global symmetry at the CFT point to $E_{N_f+1}$ \cite{Seiberg:1996bd}.}:

\begin{table}[H]\centering
	\begin{tabular}{|l|cccccccccc|} \hline
		&  0 &  1 &  2 &  3 &  4 &  5 &  6 &  7 &  8 &  9\\
		\hline\hline
		\;\;O$8^-$ &\xTB& \xTB  &\xTB   & \xTB  & \xTB  &  \xTB  &  \xTB  & \xTB   & \xTB   &    \\
		\hline
		$k$\,\,	D0 &\xTB&   &   &   &   &    &    &    &    &    \\
		\hline
		$N$	D4 &\xTB&\xTB&\xTB&\xTB&\xTB&    &    &    &    &     \\
		\hline
		$1$\,\, D$4'$ &\xTB&    &    &    &    &\xTB&\xTB&\xTB&\xTB&  \\
		\hline
		\;	F1 &\xTB&    &    &    &    & & & & & \xTB   \\
		\hline
	\end{tabular}
	\caption{The directions of the various branes.}
	\label{table:BranesSpN}
\end{table}

The effective theory on the D4 branes is a 5d $Sp(N)$ gauge theory on $S^1(R)\times\mathbb{R}^4$, called $T^{5d}$. 
The D$4'$ brane realizes a 1/2-BPS  Wilson loop with symmetry group  $Sp(1)$, wrapping the $x_0$-circle and which sits at the origin of $\mathbb{R}^4$. Pulling the  D$4'$ brane a distance $M$ away in the $x_9$-direction, there are now open strings with nonzero tension between the D4 and D$4'$ branes. These are heavy fermion probes, with mass proportional to the distance $M$.  The gauge group on the D0 branes is $\widehat{G}=O(k)$.
We put the theory on the 5d $\Omega$-background $S^1(R)\times\mathbb{R}^4_{\epsilon_1, \epsilon_2}$. We will rely on the same fugacity notation we have used throughout this paper:
\ie
\epsilon_+\equiv{\epsilon_1 + \epsilon_2\over 2},~~~ \epsilon_-\equiv{\epsilon_1 - \epsilon_2\over 2},
~~~ m \equiv {\epsilon_3 - \epsilon_4\over 2},
\fe
where $\epsilon_1$, $\epsilon_2$, $\epsilon_3$, and $\epsilon_4$ are the chemical potentials associated to the rotations of the $\bR^2_{\text{\tiny{12}}}$, $\bR^2_{\text{\tiny{34}}}$, $\bR^2_{\text{\tiny{56}}}$, and $\bR^2_{\text{\tiny{78}}}$ planes, respectively.\\

It is worth pointing out a new feature due to the gauge group on the D0 branes, $\widehat{G}=O(k)$. Indeed, $O(k)$ has a $\mathbb{Z}_2$ center, and as a set, it is $O(k)^+\cup O(k)^-$, where the superscript is the charge under this $\mathbb{Z}_2$. Hence, one needs to treat the cases $O(k)^+=SO(k)$ and $O(k)^-$ separately when computing the 1-loop determinants of the quantum mechanical fields, and sum over each contribution at the end. Correspondingly, let us denote the 1-loop determinants related to $\widehat{G}=O(k)^+$ as ${Z}^{(k,+)}$, and those related to $\hat{G}=O(k)^-$ as ${Z}^{(k,-)}$.

The field content of the quantum mechanics on the D0 branes is 
\begin{table}[H]\centering
	\begin{tabular}{|c|c|c|c|}
		\hline
		strings &Multiplets & fields & {\footnotesize $SO(4)\times SO(4)_R\times O(k)\times Sp(N)\times Sp(N')$ }%& O($k$)$\times$Sp(1)$\times$ SO($2N_f$)
		\\ \hline
		\multirow{8}{*}{D0-D0}& \multirow{3}{*}{vector} & gauge field &$({\bf 1},{\bf 1},{\bf 1},{\bf 1},{\bf adj},{\bf 1},{\bf 1})$
		\\ \cline{3-4}
		& & scalar & $({\bf 1},{\bf 1},{\bf 1},{\bf 1},{\bf adj},{\bf 1},{\bf 1})$
		\\ \cline{3-4}
		&  & fermions &$({\bf 1},{\bf 2},{\bf 1},{\bf 2},{\bf adj},{\bf 1},{\bf 1})$
		\\ \cline{2-4}
		&Fermi & fermions & $({\bf 2},{\bf 1},{\bf 2},{\bf 1},{\bf adj},{\bf 1},{\bf 1})$
		\\ \cline{2-4}
		&\multirow{2}{*}{t-hyper}  & scalars & $({\bf 1},{\bf 1},{\bf 2},{\bf 2},{\bf \frac{k(k+1)}{2}},{\bf 1},{\bf 1})$
		\\ \cline{3-4}
		&& fermions & $({\bf 1},{\bf 2},{\bf 2},{\bf 1},{\bf \frac{k(k+1)}{2}},{\bf 1},{\bf 1})$
		\\ \cline{2-4}
		&\multirow{2}{*}{hyper} & scalars & $({\bf 2},{\bf 2},{\bf 1},{\bf 1},{\bf \frac{k(k+1)}{2}},{\bf 1},{\bf 1})$
		\\ \cline{3-4}
		&& fermions & $({\bf 2},{\bf 1},{\bf 1},{\bf 2},{\bf \frac{k(k+1)}{2}},{\bf 1},{\bf 1})$
		\\ \hline
		\multirow{3}{*}{D0-D4}& \multirow{2}{*}{hyper} & scalars & $({\bf 2},{\bf 1},{\bf 1},{\bf 1},{\bf \bar k},{\bf 2N},{\bf 1})$
		\\ \cline{3-4}
		& & fermions & $({\bf 1},{\bf 1},{\bf 1},{\bf 2},{\bf \bar k},{\bf 2N},{\bf 1})$
		\\ \cline{2-4}
		& Fermi & fermions & $({\bf 1},{\bf 1},{\bf 2},{\bf 1},{\bf \bar k},{\bf 2N},{\bf 1})$
		\\ \hline
		\multirow{3}{*}{D0-D$4'$}& \multirow{2}{*}{t-hyper} & scalars & $({\bf 2},{\bf 1},{\bf 1},{\bf 1},{\bf \bar k},{\bf 1},{\bf 2N'})$
		\\ \cline{3-4}
		& & fermions & $({\bf 1},{\bf 1},{\bf 1},{\bf 2},{\bf \bar k},{\bf 1},{\bf 2N'})$
		\\ \cline{2-4}
		& Fermi & fermions & $({\bf 1},{\bf 1},{\bf 2},{\bf 1},{\bf \bar k},{\bf 1},{\bf 2N'})$
		\\ \hline
		\multirow{1}{*}{D4-D$4'$}& \multirow{1}{*}{Fermi} & \multirow{1}{*}{fermions} & $({\bf 1},{\bf 1},{\bf 1},{\bf 1},{\bf 1},{\bf 2N},{\bf 2N'})$
		\\ \hline
	\end{tabular}
	\caption{The field content of the quantum mechanics. $O(k)=\widehat{G}$ is the D0 brane group, $Sp(N)=G$ is the D4 brane group and  $Sp(N')=G'$ is the D$4'$ brane group. $SO(4)=SU(2)_-\times SU(2)_+$, $SO(4)_R=SU(2)^R_-\times SU(2)^R_+$. Of interest to us is the case of a single D$4'$ brane, $N'=1$.  t-hyper denotes a $(0,4)$ twisted hypermultiplet.}
	\label{table:fieldcontentSpN}
\end{table}

The instanton partition function is the index  of the D0 brane quantum mechanics. In our context, it reads\footnote{The component $Z^{(k,-)}$ can be understood as inserting a parity operator $(-1)^P$ in the Witten index, with $P$ the element $-1$ of $O(k)^-$. The partition function we wrote down is the projection to parity-even states with $\frac{1}{2}(1+(-1)^P)$. Another choice is possible, which corresponds to a projection to parity-odd states with $\frac{1}{2}(1-(-1)^P)$. That latter partition function corresponds to a choice of theta angle $\theta=\pi$ for $Sp(N)$. Indeed, such an angle exists in five dimensions, since $\pi_4(Sp(N))=\mathbb{Z}_2$. The analysis in that case is carried out in \cite{Haouzi:2020zls}.}:
\begin{align}
\label{5dintegralSpN}
&Z(M) = Z_{D4/D4'}(M)\sum_{k=0}^{\infty}\; \widetilde{\fq}^k\;\frac{1}{2} \left(Z^{(k,+)}(M) + Z^{(k,-)}(M)\right) \; ,
\end{align}
with
\begin{align}
{Z}^{(k,+)}(M) &=\frac{1}{|O(k)^+|^\chi}\oint_{\mathcal{M}_{k,+}}  \left[ \prod_{I=1}^{k}\frac{d\phi_I}{2\pi i}\right]Z^{(k,+)}_{D0/D0}\cdot Z^{(k,+)}_{D0/D4}\cdot Z^{(k,+)}_{D0/D4'}(M) \; ,\\
{Z}^{(k,-)}(M) &=\frac{1}{|O(k)^-|^\chi}\oint_{\mathcal{M}_{k,-}}  \left[ \prod_{I=1}^{k}\frac{d\phi_I}{2\pi i}\right]Z^{(k,-)}_{D0/D0}\cdot Z^{(k,-)}_{D0/D4}\cdot Z^{(k,-)}_{D0/D4'}(M)\; .
\end{align}
In the above, for the sake of brevity, we only made the dependence on the defect fugacity $M$ explicit in writing the partition function $Z(M)$.
$|O(k)^+|$ is the order of the Weyl group of $SO(k)$, and $|O(k)^-|$ is the order of the Weyl group of $O(k)^-$. Throughout the rest of this section, we will make use of the following convenient notation for the instanton number, 
\beq
k=2n+\chi \;\; ,\;\;\;\; \chi=0, 1 \; .
\eeq 	
We collected all the 1-loop determinants in the appendix (see also the work \cite{Kim:2012gu}).\\

We can write the index in terms of field theory quantities, as ${Z}^{(k,\pm)}_{D0/D0}\cdot {Z}^{(k,\pm)}_{D0/D4} \equiv Z^{(k,\pm)}_{vec}\cdot Z^{(k,\pm)}_{antisym}$, and ${Z}^{(k,\pm)}_{D0/D4'}\cdot {Z}_{D4/D4'} \equiv Z^{(k,\pm)}_{defect}$. We obtain
\begin{align}
\label{5dintegralCN}
& Z(M)  =\sum_{k=0}^{\infty}\;\widetilde{\fq}^{k} \, \frac{1}{2}\bigg[\frac{1}{|O(k)^+|^\chi}\oint_{\mathcal{M}_{k,+}}  \left[\frac{d\phi_I}{2\pi i}\right]Z^{(k,+)}_{vec}\cdot Z^{(k,+)}_{antisym}\cdot  Z^{(k,+)}_{defect}(M)\\
&\qquad\qquad\qquad\;\;\;\,+\frac{1}{|O(k)^-|^\chi}\oint_{\mathcal{M}_{k,-}}  \left[\frac{d\phi_I}{2\pi i}\right]Z^{(k,-)}_{vec}\cdot Z^{(k,-)}_{antisym}\cdot  Z^{(k,-)}_{defect}(M)\bigg]  \; .\nonumber
\end{align}

An important comment is in order. Note that in the absence of D8 branes, we do not cancel the charge of the O8 plane, which means the dilaton runs in the direction $x_9$. In fact, this will be true even when we consider fundamental matter, as we are only interested in theories that have a well-defined four-dimensional limit (conformal or asymptotically free), implying  $N_f \leq 4$; the number of D8 branes is still too low to cancel the orientifold charge. As we already mentioned, the $U(1)$ instanton charge will then receive an anomalous contribution, which causes a (fractional) shift to $k$. The quantum mechanics index we compute is not sensitive to this shift\footnote{This is in contrast to other physical quantities, such as the superconformal index, for example; see for instance the work \cite{Chang:2016iji}, where the index is highly sensitive to this anomalous shift.}, so we will safely proceed.\\

\vspace{10mm}

------\; {\bf Antisymmetric matter}\; ------\\

We will start with the case of a $Sp(N)$ gauge theory with one antisymmetric matter multiplet, of corresponding mass $m$. We find it easier to study the pure gauge later, by sending the mass $m\rightarrow \infty$ only at the very end\footnote{We suspect there may be some spurious contributions $Z_{extra}$ that do not decouple properly if we take the limit inside the integrand form the start, as we did for the other gauge groups.}.\\

First, we note a new feature of this index: the $k$-th term in the integral \eqref{5dintegralCN} only requires $\lfloor k/2 \rfloor$ integrals to be performed. For instance, the first instanton correction $k=1$ is just the integrand itself. So far in this paper, every term in the Schwinger-Dyson identities stood for a residue at some pole in the set $\mathcal{M}_k\setminus \mathcal{M}^{pure}_k$. However, we are claiming here that the sets $\mathcal{M}_{1,\pm}$ and $\mathcal{M}^{pure}_{1,\pm}$, for example, are empty to begin with. Then, we need to slightly modify our prescription to write down our partition function as a Laurent series in $Y$-operator vevs. We think that the odd-$k$ instanton corrections should contribute a constant term to the SD equations. For instance, the $k=1$ term comes out to be
\begin{align}
\label{SpNantisym1}
&\widetilde{\fq} \; \bigg[c_{(1), 4d}^{\epsilon_1, \epsilon_2}\; \left\langle 1\right\rangle_{4d} \bigg]\; ,
\end{align}
with
\beq
c_{(1), 4d}^{\epsilon_1, \epsilon_2}=\frac{(M-\epsilon_-)(M+\epsilon_-)}{(M-\epsilon_+)(M+\epsilon_+)}\; .
\eeq

The first integral we have to perform is found at $k=2$. The JK-residue prescription tells us that $\mathcal{M}_{2,+}\setminus \mathcal{M}^{pure}_{2,+}$ has exactly two elements; they are the pole at $\phi_I=-M+\epsilon_+$ and the pole at $\phi_I=M+\epsilon_+$, respectively. Meanwhile,  $\mathcal{M}_{2,-}$ and $\mathcal{M}^{pure}_{2,-}$ are the empty set, so the $O(2)^-$ sector only contributes at most a constant $\widetilde{\fq}^2$ to the SD equations.
Explicit formulas get involved very quickly, so let us focus on the four-dimensional limit $R\rightarrow 0$. Note that $Z^{(k,+)}$ and $Z^{(k,-)}$ scale differently in the limit, and only the $O(k)^+$ sector contribution survives. We therefore identify the $k=2$ term in the defect partition function to be
\begin{align}
\label{SpNantisym}
&\widetilde{\fq}^2 \; \bigg[c_{(2,1), 4d}^{\epsilon_1, \epsilon_2} \left\langle\frac{Y^+(M+m- \epsilon_+)\, Y^+(M-m - \epsilon_+)}{Y^+(M-2 \, \epsilon_+)} \right\rangle_{4d}\\
&\qquad + c_{(2,2), 4d}^{\epsilon_1, \epsilon_2} \left\langle\frac{Y^+(M+m + \epsilon_+)\, Y^+(M-m+ \epsilon_+)}{Y^+(M+2 \, \epsilon_+)}\right\rangle_{4d} \bigg]\; ,\nonumber
\end{align}
with
\begin{align}
\label{SpNantisymcoefs}
c_{(2,1), 4d}^{\epsilon_1, \epsilon_2} &= \frac{(m \pm \epsilon_-)}{4M(m \pm \epsilon_+)(M - \epsilon_+) (2 M\pm m - \epsilon_+)  (2 M \pm m- 3 \epsilon_+)(2 M\pm \epsilon_- - 3 \epsilon_+)}\\
c_{(2,2), 4d}^{\epsilon_1, \epsilon_2} &= \frac{(m \pm \epsilon_-)}{4M(m \pm \epsilon_+)(M + \epsilon_+) (2 M\pm m + \epsilon_+) (2 M \pm m+ 3 \epsilon_+) (2 M\pm\epsilon_- + 3 \epsilon_+)}\; .\nonumber
\end{align}
We used superscripts on the $Y$-operators to make explicit which $O(k)$ sector they are defined with respect to.
After normalizing the partition function and expanding it in the defect fugacity $M$, we derive the following four-dimensional SD equation:
\beq
\label{Schwinger4dSpNantisym}
\frac{1}{\widetilde{\left\langle 1 \right\rangle}_{4d}}\left[\left\langle Y(M) \right\rangle_{4d}
+\widetilde{\fq} \;  \ldots \right]  = \cT_{4d}(M; \{e_i^{\epsilon_1, \epsilon_2}\}) \, ,
\eeq
where we checked up to $k=4$ that
\beq
\label{RHSSpNantisym}
{\cT}_{4d}(M; \{e_i^{\epsilon_1, \epsilon_2}\})=\prod_{i=1}^{N}(M\pm e_i^{\epsilon_1, \epsilon_2}) \; .
\eeq 
The normalization by $Z_{extra}$ is crucial here in obtaining this result, as it decouples contributions that are not part of the QFT and makes ${\cT}_{4d}$ a finite polynomial in $M$.  Here, we find that this amounts to subtracting $\left.-Z(M)\right|_{\{a_i\}\rightarrow -\infty}$ from the partition function, with $\{a_i\}_{i=1, \ldots, N}$ the $N$ Coulomb parameters. This is consistent with the string theory expectation, since spurious contributions are expected from the  D0/D$4'$/O8 string sector. Then, we should decouple the D4 branes to identify the unwanted UV degrees of freedom.\\

Taking the flat space limit $\epsilon_1, \epsilon_2 \rightarrow 0$, we conjecture that we recover the SW geometry of 4d $Sp(N)$ with adjoint matter. In particular, the $k=1$ and $k=2$ terms are in perfect agreement with the literature in that limit \cite{Ennes:1999fb}. The higher corrections are similarly computed  as we did above, following the JK prescription.\\

\vspace{10mm}

------\; {\bf Pure case}\; ------\\
	
The pure $Sp(N)$ quantized geometry can now easily be deduced from the above results, by taking the limit $m\rightarrow \infty$ in the evaluated integrals. The $k=1$ term is as before,
\begin{align}
\label{SpNpure1}
&\widetilde{\fq} \; \bigg[c_{(1), 4d}^{\epsilon_1, \epsilon_2}\; \left\langle 1\right\rangle_{4d} \bigg]\; ,
\end{align}
with
\beq
c_{(1), 4d}^{\epsilon_1, \epsilon_2}=\frac{(M-\epsilon_-)(M+\epsilon_-)}{(M-\epsilon_+)(M+\epsilon_+)}\; .
\eeq

The $k=2$ term we previously derived becomes
 \begin{align}
 \label{SpNpure}
 &\widetilde{\fq}^2 \; \left[c_{(2,1), 4d}^{\epsilon_1, \epsilon_2} \left\langle\frac{1}{Y^+(M-2 \, \epsilon_+)} \right\rangle_{4d} + c_{(2,2), 4d}^{\epsilon_1, \epsilon_2} \left\langle\frac{1}{Y^+(M+2 \, \epsilon_+)}\right\rangle_{4d} \right] \; , 
 \end{align}
 with
 \begin{align}
 \label{SpNpurecoefs}
 c_{(2,1), 4d}^{\epsilon_1, \epsilon_2} &= \frac{1}{4M(M - \epsilon_+)(2 M\pm \epsilon_- - 3 \epsilon_+)}\\
 c_{(2,2), 4d}^{\epsilon_1, \epsilon_2} &= \frac{1}{4M(M + \epsilon_+)(2 M\pm \epsilon_- + 3 \epsilon_+)}\; .\nonumber
 \end{align}
The right-hand side of the SD equation,  $\cT_{4d}(M; \{e_i^{\epsilon_1, \epsilon_2}\}) $,  is again a finite polynomial  in $M$, which we checked up to $k=4$. To obtain this result, we had to remove extra spurious contributions present in the UV. As we pointed out in the previous section, this amounts to subtracting $\left.-Z(M)\right|_{\{a_i\}\rightarrow -\infty}$ from the partition function.

As an example, here is ${\cT}_{4d}$ when  $N=1$, meaning $G=Sp(1)$, with Coulomb parameter $a$:
\begin{align}
\label{SpNpurepolyNis2}
{\cT}_{4d}(M; \{e_i^{\epsilon_1, \epsilon_2}\}) = M^2 + d_{(0)}^{\epsilon_1, \epsilon_2} \; ,
\end{align}
with $d_{(0)}^{\epsilon_1, \epsilon_2} = \left(\{e_i^{\epsilon_1, \epsilon_2}\}\right)^2$, and
\begin{align}
\label{SpNpurepolycoef}
d_{(0)}^{\epsilon_1, \epsilon_2} &= - a^2 - \widetilde{\fq}\,\frac{1}{2(a^2-\epsilon_+^2)}+\widetilde{\fq}^2\; \ldots
\end{align}
Note this is exactly what we had found for  $G=SU(2)$, see \eqref{SUNpurepolycoef}. Above, we had to subtract an extra spurious contribution, 
\beq
\left.-Z(M)\right|_{a\rightarrow -\infty}=\frac{\widetilde{\fq}}{2(M^2-\epsilon_+^2)} \; ,
\eeq
which only affected the first instanton correction.\\

 We now take the flat space limit $\epsilon_1, \epsilon_2 \rightarrow 0$.
Following encouraging computer experiments, we conjecture that only the $\widetilde{\fq}$ and $\widetilde{\fq}^2$ terms in the SD equation survive this limit. As we argued in the previous example, the $\widetilde{\fq}$ term should simply be a constant in the SD equations, since there are no poles associated to it. Meanwhile, the $\widetilde{\fq}^2$ term \eqref{SpNpure} greatly simplifies in the limit. After introducing the dynamical scale $\widetilde{\fq}\equiv \Lambda^{2 N+2}$, we multiply the equation by $M^2$ and rescale the $Y$-operators. All in all, we recover the SW curve of the pure $Sp(N)$ gauge theory  (see \cite{DHoker:1996kdj} or \cite{Argyres:1995fw}):
\beq
\label{SpNSWpure}
y(M) + \Lambda^{2 N+2}+ \frac{\Lambda^{4 N+4}}{y(M)} =M^2\, \prod_{i=1}^{N}(M \pm e_i^{0,0}) \; .
\eeq

\vspace{10mm}

------\; {\bf Fundamental matter}\; ------\\
	
Finally, we consider adding $N_f$ fundamental hypermultiplets. In our brane setup, this can be done by adding $N_f$ D8 branes. As usual, we limit ourselves to a number of D8 branes such that in the four-dimensional limit, the resulting low energy gauge theory $T^{4d}$ is conformal or asymptotically free. In the presence of antisymmetric matter, this translates to $N_f \leq 4$, while in the absence of antisymmetric matter, this means $N_f \leq N+2$. In fact, we impose the stricter condition that the amount of fundamental matter should not introduce new poles at $\infty$ in the $\phi_I$ integrals.\\

In the appendix, we wrote down the various D0/D8 1-loop determinants. In particular, in the $O(k)^+$ sector,
\beq
{Z}^{(k=2n+\chi, +)}_{D0/D8}=\prod^{N_f}_{d=1}\sh(m_d)^\chi \prod^n_{I=1}\sh(\pm\phi_I + m_d)
\eeq
Then, it is not hard to see that for $k$ even, the argument of the matter factor is the locus of the various JK poles, while a single factor is picked up for $k$ odd. 
In the context of our previous discussion,  the $k=2$ term \eqref{SpNpure}  (in 4d)  is then modified as
 \begin{align}
\label{SpNfund}
\widetilde{\fq}^2 \; \left[c_{(2,1), 4d}^{\epsilon_1, \epsilon_2} \left\langle\frac{\prod^{N_f}_{d=1}(\pm(-M+\E_+) + m_d)}{Y^+(M-2 \, \epsilon_+)} \right\rangle_{4d} + c_{(2,2), 4d}^{\epsilon_1, \epsilon_2} \left\langle\frac{\prod^{N_f}_{d=1}(\pm(M+\E_+) + m_d)}{Y^+(M+2 \, \epsilon_+)}\right\rangle_{4d} \right] \; .
\end{align}
Meanwhile, the $k=1$ term picks up a factor $\prod^{N_f}_{d=1}\sh(m_d)$. All higher $k$ corrections follow this pattern. Taking the flat space limit, we conjecture (and checked up to $k=4$) that only the $k=1$ and $k=2$ terms survive in the SD equation. After introducing the dynamical scale $\widetilde{\fq}\equiv \Lambda^{2 N+2-N_f}$ and rescaling the $Y$-operators,  we recognize the SW curve of $Sp(N)$ with $N_f$ fundamental hypermultiplets (see \cite{DHoker:1996kdj} or \cite{Argyres:1995fw}):
\beq
\label{SpNSWfund}
y(M) + \left[\prod^{N_f}_{d=1}(m_d)\right]\Lambda^{2 N+2-N_f}+ \left[\prod^{N_f}_{d=1}(\pm M + m_d)\right]\frac{\Lambda^{4 N+4-2N_f}}{y(M)} =M^2\, \prod_{i=1}^{N}(M \pm e_i^{0,0}) \; .
\eeq
The case with both antisymmetric and fundamental matter can be treated the same way, by including the D0/D8 1-loop we just discussed determinants in the integrand  \eqref{5dintegralSpN}.

\section{On the Uniqueness of the Defects}

We end this paper with an open question. When $G=SO(N), Sp(N)$, there is some evidence that the codimension 4 defect we studied may not be unique. Indeed, already string theory suggests at least two different UV completions of the theories: instead of using O8 planes as we did, a natural construction is to rely on the use of O4 planes, aligned in the same direction as the D4 branes; an immediate advantage is that the case $G=SO(2N+1)$ would now be realized in a stringy picture, since $\widetilde{\text{O4}}$ planes do exist. However, quantizing the various strings may prove subtle; as we saw in the examples, the use of O8 planes enabled us to exploit a particular symmetry of the brane system, from which we easily deduced the new D0/D$4'$ 1-loop determinants due to the Wilson loop. With an O4 plane instead, such a symmetry is broken, since the D$4'$ branes are now orthogonal to the O4 plane, while the D4 branes sit on top of it. Therefore, one would need to do more  work to quantize the D0/D$4'$ strings. \\

In the O8 plane setup, the number of Dirichlet-Neumann directions was equal to 4 for the D$4'$/O8 configuration, the same as for the D4/O8 configuration. In conclusion, we saw an enhancement of both the  gauge and defect groups to the same classical group type: the groups $G$ and $G'$ were either both orthogonal or both symplectic. We can expect a similar enhancement when using an O4 plane: now, the number of Dirichlet-Neumann directions is equal to 8 for the D$4'$/O4 configuration, while it is 0 for the D4/O4 configuration. The implication is once again that $G$ and $G'$ would both see a symmetry enhancement to either an orthogonal group, or both to a symplectic group.\\

Even though we were able to derive the SW geometries in the O8 construction, it would then be interesting to further define (and compute, if possible) the index of the quantum mechanics on D0 branes using O4 planes, and find out if the resulting quantum geometry is distinct from ours or not. 
	
\section*{Acknowledgments}
We thank Mina Aganagic, Chi-Ming Chang, Ori Ganor, Amihay Hanany, Petr Horava, Saebyeok Jeong, Hee-Cheol Kim, Joonho Kim, Nikita Nekrasov, Kantaro Ohmori, Vivek Saxena, and Luigi Tizzano for useful discussions and comments at various stages of this project. JO is grateful to Chi-Ming Chang and Ori Ganor, who developed together a JK integral package in Mathematica, which was crucial for high instanton computations in this project. The research of JO is supported in part by Kwanjeong Educational Foundation and in part by the Berkeley Center of Theoretical Physics. The research of NH is supported by the Simons Center for Geometry and Physics.

\pagebreak

	\begin{appendix}
		
		\section{1-loop determinants for $G=Sp(N)$ and $\widehat{G}=O(k)$}
		
		We use the notation $\sh(x)\equiv 2 \sinh(x/2)$ and $\ch(x)\equiv 2 \cosh(x/2)$.  Products over all signs inside an argument have to be considered, and $\chi=0, 1$.
		
		The 1-loop determinants of the D0/D0 strings are given by
		\ie\label{Zplus}
		\hspace*{-1cm}{Z}_{D0/D0}^{(k=2n+\chi, +)}=& \bigg[\Big(\prod_{I=1}^{n} \sh(\pm \phi_I)\Big)^{\chi }\prod_{I<J}^{n} \sh(\pm \phi_I \pm \phi_J) \bigg]\sh(2\epsilon_+)^n \Big(\prod_{I=1}^{n}\sh(\pm \phi_I + 2\epsilon_+)\Big)^{\chi}\prod_{I < J}^{n}\sh(\pm \phi_{I} \pm \phi_{J} + 2\epsilon_+)
		\\
		&\hspace*{-.5cm}\times\sh(\pm m - \epsilon_-)^n \Big(\prod_{I=1}^{n}\sh(\pm\phi_I \pm m - \epsilon_-)\Big)^\chi \prod_{I<J}^{n}\sh(\pm\phi_I \pm \phi_J \pm m - \epsilon_-) 
		\\
		&\hspace*{-.5cm}\times {1\over \sh(\pm m - \epsilon_+)^{n+\chi}}\Big(\prod_{I=1}^{n} \frac{1}{\sh(\pm\phi_I \pm m - \epsilon_+)}\Big)^{\chi}\prod_{I=1}^{n} \frac{1}{\sh(\pm 2\phi_I \pm m - \epsilon_+) } \prod_{I<J}^{n} \frac{1}{\sh(\pm\phi_I \pm \phi_J \pm m - \epsilon_+)}
		\\
		&\hspace*{-.5cm}\times{1\over \sh(\pm \epsilon_- + \epsilon_+)^{n+\chi}} \Big( \prod_{I=1}^{n} \frac{ 1} {\sh(\pm \phi_I \pm \epsilon_- + \epsilon_+)}\Big)^{\chi}
		\prod_{I=1}^{n} \frac{1}{\sh(\pm 2\phi_{I} \pm \epsilon_- + \epsilon_+)} \prod_{I < J}^{n} \frac{1 }{\sh(\pm \phi_{I} \pm \phi_{J} \pm \epsilon_- + \epsilon_+)},
		\fe
		and
		\ie\label{Zminusodd}
		\hspace*{-1cm}{Z}_{D0/D0}^{(k=2n+1, -)}=&\Big( \prod_{I=1}^{n} \ch(\pm \phi_I) \prod_{I<J}^{n} \sh(\pm \phi_I \pm \phi_J)\Big) \sh(2\epsilon_+)^n\prod_{I=1}^{n} \ch(\pm \phi_I + 2\epsilon_+)\prod_{I < J}^{n}\sh(\pm \phi_{I} \pm \phi_{J} + 2\epsilon_+)
		\\
		&\hspace*{-0.5cm}\times\sh(\pm m - \epsilon_-)^n \prod_{I=1}^{n}\ch(\pm\phi_I \pm m - \epsilon_-) \prod_{I < J}^{n}\sh(\pm\phi_I \pm \phi_J \pm m - \epsilon_-)
		\\
		&\hspace*{-0.5cm}\times\frac{1}{\sh(\pm m - \epsilon_+)^{n+1}}  \prod_{I=1}^{n} \frac{1}{\ch(\pm\phi_I \pm m - \epsilon_+)\sh(\pm 2\phi_I \pm m - \epsilon_+) }  \prod_{I<J}^{n} \frac{1}{\sh(\pm\phi_I \pm \phi_J \pm m - \epsilon_+)}
		\\
		&\hspace*{-0.5cm}\times \frac{1}{\sh(\pm \epsilon_- + \epsilon_+)^{n+1} }  \prod_{I=1}^{n} \frac{ 1} {\ch(\pm \phi_I \pm \epsilon_- + \epsilon_+)  \sh(\pm 2\phi_{I} \pm \epsilon_- + \epsilon_+)} \prod_{I < J}^{n} \frac{ 1}{\sh(\pm \phi_{I} \pm \phi_{J} \pm \epsilon_- + \epsilon_+)},
		\fe
		and
		\ie\label{Zminuseven}
		\hspace*{-2cm}{Z}_{D0/D0}^{(k=2n, -)}=&\Big(\prod_{I<J}^{n-1} \sh(\pm \phi_I \pm \phi_J) \prod_{I=1}^{n-1}\sh(\pm2 \phi_I) \Big) \ch(2\epsilon_+)(\sh{\epsilon_+} )^{n-1}\prod_{I=1}^{n-1} \sh(\pm 2\phi_I + 4\epsilon_+)   \prod_{I < J}^{n-1}  \sh(\pm \phi_{I} \pm \phi_{J} + 2\epsilon_+)
		\\
		&\hspace*{-0.5cm}\times \ch(\pm m - \epsilon_-)\, \sh(\pm m - \epsilon_-)^{n-1}\prod_{I=1}^{n-1}\sh(\pm 2\phi_I \pm 2 m -2 \epsilon_-) \prod_{I<J}^{n-1}\sh(\pm\phi_I \pm \phi_J \pm m - \epsilon_-)
		\\
		&\hspace*{-0.5cm}\times \frac{1}{\sh(\pm m - \epsilon_+)^n \sh(\pm 2 m -2 \epsilon_+)} \prod_{I=1}^{n-1} \frac{1}{\sh(\pm 2\phi_I \pm 2 m - 2 \epsilon_+)\sh(\pm 2\phi_I \pm m - \epsilon_+)} 
		\prod_{I<J}^{n-1} \frac{1}{\sh(\pm\phi_I \pm \phi_J \pm m - \epsilon_+)}
		\\
		&\hspace*{-0.5cm}\times \frac{1}{\sh(\pm \epsilon_- + \epsilon_+)^n \sh(\pm 2 \epsilon_- + 2\epsilon_+)}  \prod_{I=1}^{n-1} \frac{1 }{\sh(\pm 2\phi_I \pm 2 \epsilon_- + 2 \epsilon_+) \sh(\pm 2\phi_{I} \pm \epsilon_- + \epsilon_+)}\prod_{I < J}^{n-1} \frac{1}{\sh(\pm \phi_{I} \pm \phi_{J} \pm \epsilon_- + \epsilon_+)}.
		\fe
		The first to the forth lines of the equations \eqref{Zplus}, \eqref{Zminusodd} and\eqref{Zminuseven}, are the 1-loop determinants of the $(0,4)$ vector multiplet, Fermi multiplet, twisted hypermultiplet and hypermultiplet, respectively.  The 1-loop determinants of the D0/D4 strings are given by
		\ie\label{D0-D4Integrand}
		&{Z}^{(k=2n+\chi, +)}_{D0/D4}=\prod_{i=1}^N\Big(\frac{\sh(m\pm a_i)}{\sh(\pm a_i+\epsilon_+)}\Big)^\chi\prod^n_{I=1}\frac{\sh(\pm\phi_I\pm a_i-m)}{\sh(\pm\phi_I\pm a_i+\epsilon_+)},
		\\
		&{Z}^{(k=2n+1, -)}_{D0/D4}=\prod_{i=1}^N\frac{\ch(m\pm a_i)}{\ch(\pm a_i+\epsilon_+)}\prod^{n}_{I=1}\frac{\sh(\pm\phi_I\pm a_i-m)}{\sh(\pm\phi_I\pm a_i+\epsilon_+)},
		\\
		&{Z}^{(k=2n, -)}_{D0/D4}
		%=Z^{-,\,k=2n+1}_{\text{\tiny D0-D4}}\Big|_{\phi_n=2\pi i}
		=\prod_{i=1}^N\frac{\sh(2m\pm2 a_i)}{\sh(\pm 2 a_i+2\epsilon_+)}\prod^{n-1}_{I=1}\frac{\sh(\pm\phi_I\pm a_i-m)}{\sh(\pm\phi_I\pm a_i+\epsilon_+)}.
		\fe
		In order to deduce the contribution of D0/D$4'$ strings, we once again make use of a symmetry of the brane setup \ref{table:BranesSpN}. Namely, under the exchange of the coordinates $x_{1,2,3,4}  \leftrightarrow x_{5,6,7,8}$, D4 and D$4'$ branes get swapped, while the D0 branes and the O$8^-$ orientifold plane are invariant.  Hence,  we can write the D0/D$4'$ contribution from the D0/D4 one, after simply exchanging $\E_1,\E_2$ with $\E_3,\E_4$. This translates to
		\ie
		a_i\leftrightarrow M,\quad m\leftrightarrow \E_-,\quad \E_+\leftrightarrow -\E_+ \; ,
		\fe
		and we obtain
		\ie\label{D0-D4pIntegrand}
		&{Z}^{(k=2n+\chi, +)}_{D0/D4'}=\Big(\frac{\sh(\pm M-\E_-)}{\sh(\pm M-\epsilon_+)}\Big)^\chi\prod^n_{I=1}\frac{\sh(\pm\phi_I\pm M-\E_-)}{\sh(\pm\phi_I\pm M-\epsilon_+)},
		\\
		&{Z}^{(k=2n+1, -)}_{D0/D4'}=\frac{\ch(\pm M-\E_-)}{\ch(\pm M-\epsilon_+)}\prod^{n}_{I=1}\frac{\sh(\pm\phi_I\pm M-\E_-)}{\sh(\pm\phi_I\pm M-\epsilon_+)},
		\\
		&{Z}^{(k=2n, -)}_{D0/D4'}
		%=Z^{-,\,k=2n+1}_{\text{\tiny D0-D4}}\Big|_{\phi_n=2\pi i}
		=\frac{\sh(\pm 2 M- 2\E_-)}{\sh(\pm 2 M-2\epsilon_+)}\prod^{n-1}_{I=1}\frac{\sh(\pm\phi_I\pm M-\E_-)}{\sh(\pm\phi_I\pm M-\epsilon_+)}.
		\fe
		The 1-loop determinant of the D4/D$4'$ strings arises from a Fermi multiplet transforming in the bifundamental representation of $G\times G'= Sp(N)\times Sp(1)$. We obtain
		\ie
		Z_{D4/D4'}=\prod_{i=1}^N\sh(M\pm a_i) \; .
		\fe
		The 1-loop determinants of D0/D8 strings are given by
		\ie\label{D0-D8integrand}
		&{Z}^{(k=2n+\chi, +)}_{D0/D8}=\prod^{N_f}_{d=1}\sh(m_d)^\chi \prod^n_{I=1}\sh(\pm\phi_I + m_d),
		\\
		&{Z}^{(k=2n+1, -)}_{D0/D8}=\prod^{N_f}_{d=1}\ch(m_d)\prod^n_{I=1}\sh(\pm\phi_I + m_d),
		\\
		&{Z}^{(k=2n, -)}_{D0/D8}=\prod^{N_f}_{d=1}\sh(2 m_d) \prod^{n-1}_{I=1}\sh(\pm\phi_I + m_d).
		\fe
	Finally, the Weyl factors of the $O(k)_+$ and $O(k)_-$ components  are given by
		\begin{align}
		\hspace{-1cm}|O(k)^+|^{\chi=0} = \frac{1}{2^{n-1} n!} ,\;\; |O(k)^+|^{\chi=1} = \frac{1}{2^n n!} ,\;\; |O(k)^-|^{\chi=0} =  \frac{1}{2^{n-1}(n-1)!},\;\; |O(k)^-|^{\chi=1} = \frac{1}{2^n n!}.
		\end{align}	
		
		\vspace{10mm}
		
		\section{Non-simple poles}

		Non-simple poles typically appear in our multi-dimensional integrals, and need to be treated with extra care. They are defined as follows: Consider a contour integral
		\ie
		\oint\prod_{i=1}^N\frac{d\phi_i}{2\pi i}\frac{f(\phi_1,\ldots,\phi_N)}{g(\phi_1,\ldots,\phi_N)} \; ,
		\fe
		where $f$ is regular at the zeroes of $g$.  We classify non-simple poles by specifying their degree. We define the degree $d$ of the pole $(\phi_i^\star)$ as
		\ie
		d=[\text{\# of vanishing factors in }g(\phi^\star_i)]-[\text{\# of vanishing factors in }f(\phi^\star_i)]-N
		\fe
		When $d=0$, dealing with such poles is fact benign\footnote{See also the page 36 of \cite{Hwang:2014uwa}}. As an example, suppose $N=2$, with $g(\phi_1,\phi_2)=(\phi_1-a)(\phi_1-\phi_2-b)$, and suppose $f(\phi_1,\phi_2)$ is regular at $\phi_1=a$ and $\phi_1-\phi_2=b$. Then
		\ie
		\oint\frac{d\phi_2}{2\pi i}\oint\frac{d\phi_1}{2\pi i}\frac{f(\phi_1,\phi_2)}{(\phi_1-a)(\phi_1-\phi_2-b)}&=\oint\frac{d\phi_2}{2\pi i}\left(\frac{f(a,\phi_2)}{a-b-\phi_2}+\frac{f(\phi_2+b,\phi_2)}{\phi_2+b-a}\right)\\
		&=-f(a,a-b)+f(a,a-b)\\
		&=0
		\fe
		In other words, there is a pair-wise cancellation of residues. Such a phenomenon is a generic feature of $d=0$ non-simple poles. Therefore, they do not contribute to the integral, even when singled out by the JK residue prescription, and vanish in the final instanton partition function formula. When $G\neq SU(N)$, non-simple poles with degree $d>0$ can appear at high instanton number. We do not make any claims on how to deal with them in full generality, and we treated them on a case-by-case basis when we encountered them at high instanton number in this work.

	\end{appendix}
  \newpage
	\bibliography{summarySOqq2}
	\bibliographystyle{JHEP}

\end{document}